\documentclass{aastex63}

\let\tablenum\relax
\usepackage{siunitx}
\usepackage{gensymb}
\usepackage{lineno}
\usepackage{hyperref}
\newcommand\chandra{{\it Chandra}}
\newcommand\xmm{{\it XMM-Newton}}
\newcommand\nustar{{\it NuSTAR}}
\newcommand\suzaku{{\it Suzaku}}
\newcommand\swift{{\it Swift}}
\newcommand\rosat{{\it ROSAT}}
\newcommand\iras{{\it IRAS}}
\newcommand{\uat}[2]{\href{http://astrothesaurus.org/uat/#2}{#1 (#2)}}

\begin{document}

\title{
NuSTAR Observations of AGN with Low Observed X-ray to [OIII] Luminosity Ratios:\\
Heavily Obscured AGN or Turned-Off AGN?
}

\correspondingauthor{M. Lynne Saade}
\email{mlsaade@astro.ucla.edu}

\author[0000-0001-7163-7015
]{M. Lynne Saade}
\affiliation{Department of Physics and Astronomy, University of California, 475 Portola Plaza, Los Angeles, CA 90095}

\author[0000-0002-8147-2602
]{Murray Brightman}
\affiliation{Cahill Center for Astronomy and Astrophysics, California Institute of Technology, 1216 E. California Blvd., Pasadena, CA 91125}

\author[0000-0003-2686-9241]{Daniel Stern}
\affiliation{Jet Propulsion Laboratory, California Institute of Technology, 4800 Oak Grove Drive, Pasadena,
CA 91109}

\author[0000-0001-6919-1237
]{Matthew A. Malkan}
\affiliation{Department of Physics and Astronomy, University of California, 475 Portola Plaza, Los Angeles, CA 90095}

\author[0000-0003-3828-2448]{
Javier~A.~Garc\'ia}
\affil{Cahill Center for Astronomy and Astrophysics, California Institute of Technology, 1216 E. California Blvd., Pasadena, CA 91125}
\affil{Dr. Karl Remeis-Observatory and Erlangen Centre for Astroparticle Physics, Sternwartstr.~7, 96049 Bamberg, Germany}

\begin{abstract}

Type~2 active galactic nuclei (AGN) show signatures of accretion onto a supermassive black hole through strong, high-ionization, narrow emission lines extended on scales of 100s to 1000s of parsecs, but they lack the broad emission lines from close in to the black hole that characterize type~1 AGN.  The lack of broad emission could indicate obscuration of the innermost nuclear regions, or could indicate that the black hole is no longer strongly accreting.  Since high-energy X-rays can penetrate thick obscuring columns, they have the power to distinguish these two scenarios. We present high-energy \nustar{} observations of 9 Seyfert~2 AGN from the \iras{} 12 $\mathrm{\mu}$m survey, supplemented with low-energy X-ray observations from \chandra, \xmm{}, and \swift{}. The galaxies were selected to have anomalously low observed 2-10 keV luminosities compared to their [\ion{O}{3}] optical luminosities, a traditional diagnostic of heavily obscured AGN, reaching into the Compton-thick regime for the highest hydrogren column densities ($N_{\rm H} > 1.5 \times 10^{24}\, {\rm cm}^{-2}$).  Based on updated [\ion{O}{3}] luminosities and intrinsic X-ray luminosities based on physical modeling of the hard X-ray spectra, we find that one galaxy was misclassified as type~2 (NGC~5005) and most of the remaining AGN are obscured, including three confirmed as Compton-thick (IC~3639, NGC~1386, and NGC~3982).  One galaxy, NGC~3627, appears to have recently deactivated. Compared to the original sample the 9 AGN were selected from, this is a rate of approximately 1\%.  We also find a new X-ray changing-look AGN in NGC~6890.

\end{abstract}


\section{Introduction} \label{sec:intro}

The presence of an actively accreting supermassive black hole (SMBH) in a galaxy is demonstrated through signatures of energetic processes near the central engine. In order of increasing distance from the black hole, the primary signs closest to the black hole are X-ray continuum from the hot corona \citep[found within a few Schwarzschild radii of the SMBH; e.g.,][]{2012MNRAS.422..129Z}, and ultraviolet and optical emission lines with widths greater than $\sim 1500\, {\rm km}\, {\rm s}^{-1}$ from the broad line region \citep[BLR -- found within 10s to 100s of light days from the SMBH; e.g.,][]{2000ApJ...533..631K, 2005ApJ...629...61K}. However, in heavily obscured active galactic nuclei (AGN) for which the line of sight hydrogen column density to the nucleus ($N_{\rm H}$) exceeds $10^{23}\: {\rm cm}^{-2}$, these signatures are not visible. For AGN with a characteristic luminosity of $\mathrm{10^{43}\: erg\,s^{-1}}$ (i.e., Seyfert galaxies), 60\% of sources are in this category \citep[e.g.,][]{2011ApJ...728...58B, 2015ApJ...815L..13R}.

Obscured AGN can still be identified through emission from further out from the central black hole, such as mid-infrared (MIR) thermal continuum from the dusty torus that is thought to surround the AGN accretion disk at distances of 0.1 pc $-$ 10s of pc \citep[e.g.,][]{2005ApJ...618L..17P, 2008ApJ...681..141R,
2010ApJ...715..736P,
2013A&A...558A.149B, 2016ApJ...822L..10I, 2016ApJ...823L..12G, 2016ApJ...829L...7G}, and the high ionization forbidden lines of the narrow line region (NLR) which occupies 100s to 1000s of pc scales \citep[e.g.,][]{1993ApJ...404L..51N, 2002ApJ...574L.105B, 2006A&A...456..953B,
2011ApJ...739...69M,
2017NatAs...1..679R}. However, because the torus and NLR are further away from the black hole, it is possible for accretion onto the SMBH to be recently shut off but still preserve the MIR and NLR emission \citep[e.g., within the last 10s to 100s of years;][]{2017ApJ...844...21I}, creating an AGN that looks like a classical Seyfert~2 with the BLR obscured, but that in truth intrinsically lacks a BLR. This could be related to a so-called `true' Seyfert~2 galaxy \citep[e.g.,][]{bianchi08}. While so far in the literature these sources have been assumed to be actively accreting, the lack of a BLR could also be due to an AGN that has recently deactivated.

X-rays with energies greater than 10~keV can penetrate thick obscuring columns and reveal the presence of an actively accreting central engine even in a heavily obscured Seyfert~2 galaxy.  As the first focusing X-ray telescope in orbit with sensitivity above 10~keV, the {\it Nuclear Spectroscopic Telescope Array} \citep[\nustar;][]{2013ApJ...770..103H} has identified and studied actively accreting SMBHs obscured by even Compton-thick levels of absorption \citep[$N_{\rm H} > 1.5 \times 10^{24}\: {\rm cm}^{-2}$; e.g.,][]{2015ApJ...815...36A, 2016ApJ...819....4R, 2016ApJ...833..245B}. \nustar\, thus gives us an opportunity to measure what fraction of the local Seyfert 2 population is currently accreting, and thereby constrain the AGN duty cycle.

To find Seyfert galaxies without a BLR requires a large sample of galaxies selected based on AGN signatures not blocked by obscuration, such as the warm dust from the torus. The most complete and brightest sample of such galaxies in the local universe are found in the 12~$\mathrm{\mu m}$ sample of galaxies \citep{1989ApJ...342...83S, 1993ApJS...89....1R}. This sample contains all galaxies in the second {\it Infrared Astronomical Satellite} \citep[{\it IRAS};][]{1984ApJ...278L...1N} point source catalogue that exceed 0.3~Jy in flux density at 12~$\mathrm{\mu m}$ that are also (a) brighter at 60 and 100~$\mathrm{\mu m}$ than at 12~$\mathrm{\mu m}$, and (b) located at declinations $\mathrm{|\delta|\:\geq}$  25\degree{}. \cite{2008MNRAS.390.1241B} investigated a subset of Seyfert 2 galaxies from this sample that appeared to be unabsorbed in the X-rays, finding two strong `true' Seyfert 2 candidates, NGC 3147 and NGC 3660. The X-ray spectral properties of the full galaxy sample with \xmm\ data were presented in \cite{2011MNRAS.413.1206B} and \cite{2011MNRAS.414.3084B}. Of the Seyfert 2 galaxies in this sample, 10 showed anomalously low observed (i.e., not absorption-corrected) 2-10 keV X-ray luminosities compared to their nuclear [\ion{O}{3}] luminosities. That is, these galaxies had observed 2-10~keV X-ray luminosities significantly less than expected for their observed [\ion{O}{3}] luminosities based on our fit to the $L_{2-10}$ to $L_{\rm [OIII]}$ relation for the {\bf39} Seyfert~1 galaxies in the {\it IRAS} 12~$\mathrm{\mu m}$ sample with X-ray observations:
\begin{equation} \label{eq:Sy1line}
\log(L_{2-10})=0.95\,\log(L_{\rm [OIII]})+3.89,
\end{equation}
where the luminosities are in units of ${\rm erg}\, {\rm s}^{-1}$.  The X-ray luminosities were derived directly from the observed 2-10~keV fluxes listed in \citet{2008A&A...483..151P}, \citet{2011MNRAS.413.1206B}, and \citet{2011MNRAS.414.3084B},  while the [\ion{O}{3}] luminosities were derived from fluxes listed in \citet{2017ApJ...846..102M}.

The 10 anomalously X-ray faint Seyfert 2 galaxies were an order of magnitude below this relation. This  made them candidate Compton-thick AGN, but  also potentially turned-off Seyfert~2 AGN if the central engines  were inactive. High-energy X-ray observations, as possible with \nustar, can distinguish  between these two scenarios. Of the 10 outlier galaxies, all but NGC 5953 had \nustar{} observations (Table \ref{tab:xraydata}) at the time of writing through a combination of archival data and dedicated observations from our Cycle~3 observing program (PID~3321). Three of the galaxies have already been reported as Compton-thick AGN in the literature based on these \nustar\ observations: NGC~1386 \citep{2015ApJ...805...41B}, NGC~4922 \citep{2017MNRAS.468.1273R}, and IC~3639 \citep{2016ApJ...833..245B}. However, since those publications, \citet{2018ApJ...854...42B} has released the BORUS X-ray spectral model which is designed for analyzing high-energy observations of heavily obscured AGN, allowing us to more accurately constrain the parameters of the obscuring torus. Therefore, we analyze all 9 outlier galaxies with \nustar\ data, including those that have already appeared in the literature.

\begin{figure}
    \centering
    \plotone{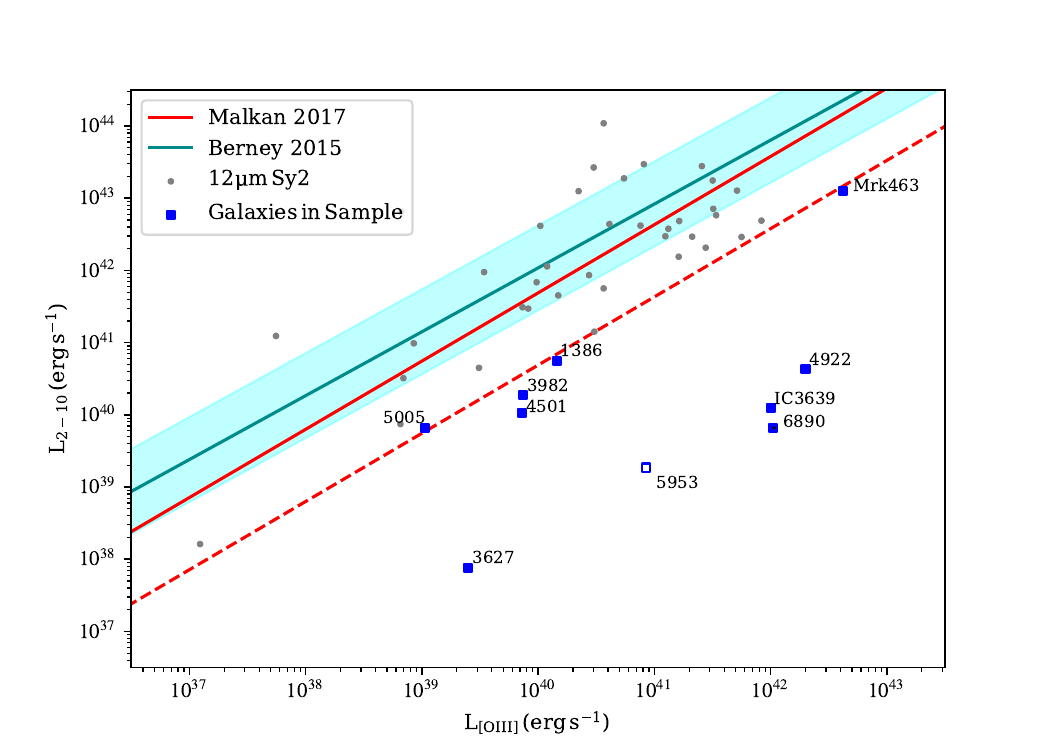}
    \caption{Observed 2-10~keV X-ray luminosity vs. [\ion{O}{3}] luminosity for the 9 Seyfert~2 galaxies in our sample. The 12 $\mathrm{\mu m}$ sample, based on data from \citet{2017ApJ...846..102M}. The solid red line shows the mean $L_{\rm{2-10}}$ vs $L_{\rm{[OIII]}}$ relation for the Seyfert~1 galaxies in the $\mathrm{12\,\mu m}$ sample (Eq.\ \ref{eq:Sy1line}). The dashed red line is the same line shifted by an order of magnitude down in observed 2-10 keV X-ray luminosity. The 9 galaxies  of the sample are labeled as blue squares. NGC 5953 is plotted as an open square as it had no \nustar{} data available, and so was left out of the final sample. The $L_{\rm{2-10}}$ vs $L_{\rm{[OIII]}}$ relation for Seyferts from \citet{2015MNRAS.454.3622B} is plotted in dark cyan for comparison, with the shaded region corresponding to its RMS scatter of 0.59 dex.}
    \label{fig:Proposal_Plot}
\end{figure}

 The 9 galaxies of the sample are plotted as the blue squares in Figure \ref{fig:Proposal_Plot} alongside the solid red line of Equation \ref{eq:Sy1line}. The dashed red line represents an order of magnitude below Equation \ref{eq:Sy1line}. We use updated $L_{\rm [OIII]}$ values from the literature instead of the original values from \citet{2017ApJ...846..102M} from this point on, though we did use the original \citet{2017ApJ...846..102M} values in the sample selection. For consistency's sake we use the reddening-corrected \citet{2011MNRAS.414.3084B} values for the [\ion{O}{3}] luminosities when available.  With the revised [\ion{O}{3}] luminosities, NGC~5005 no longer lies more than an order of magnitude below the mean relation from Equation \ref{eq:Sy1line}, though it did when using the \citet{2017ApJ...846..102M} values for the original sample selection.  Since the revised value is still very close to the selection line (see Figure \ref{fig:Proposal_Plot}), we keep this galaxy in the sample. NGC~5953 is plotted as an open blue square because it did not have \nustar{} observations at the time of writing and so did not end up in the final sample. The source of its [\ion{O}{3}] luminosity is \citet{2010ApJ...720..786L}. In addition to Equation \ref{eq:Sy1line} which is plotted in red, we also plot the intrinsic $L_{\rm{2-10}}$ vs intrinsic $L_{\rm{[OIII]}}$ relation for Seyfert galaxies from \citet{2015MNRAS.454.3622B} in dark cyan. This relation is derived from 340 Seyfert 1 and Seyfert 2 galaxies in the BAT AGN Spectroscopic Survey Data Release 1 \citep{2017ApJ...850...74K}, and its RMS scatter is 0.59 dex, shown as the light blue shaded region.

For calculating the distance scales on our images, we adopt the concordance cosmology, $\Omega_{\rm M} = 0.3$, $\Omega_\Lambda = 0.7$ and $H_0 = 70\, {\rm\,km\,s^{-1}\,Mpc^{-1}}$. For computing luminosities in XSPEC {\bf \citep{1996ASPC..101...17A}}, we use the default cosmology, which is $\Omega_{\rm M} = 0.27$, $\Omega_\Lambda = 0.73$ and $H_0 = 70\, {\rm\,km\,s^{-1}\,Mpc^{-1}}$. The 9 galaxies in this sample are very low redshift so the differences between the two cosmologies are negligible.

\smallskip

\section{X-Ray Observations and Analysis} \label{sec:obs}

The X-ray observations used in this paper are listed in Table~\ref{tab:xraydata}.  We include all available \nustar\ data for the 9 X-ray faint galaxies.  \nustar\ observes at 3-79~keV, though most sources are not detected out to the highest energies where the sensitivity of \nustar\ declines. Lower energy X-ray data are important for the spectral analysis, and several telescopes provide focused soft X-ray observations (0.5-10 keV).   Where available, we preferentially use archival \chandra{} observations, due to their sensitivity and high spatial resolution. With its 1\arcmin\ beam (half-power diameter), \nustar{} suffers confusion of off-nuclear point sources with the central AGN, which is particularly problematic for faint nuclei, as is the case for several of the galaxies analyzed here. When \chandra\ data were not available or were insufficient for understanding the true nature of some spectral features, we use archival \swift{} and/or \xmm{} data.

All X-ray spectra were grouped with a minimum of one count per bin. We fit the data in XSPEC (version 12.11.1). Due to the low number of counts for all sources, the C-statistic was used for fitting. We subtracted the background instead of modeling it separately, in which case XSPEC uses the modified W-statistic. We next describe the X-ray observations by each satellite in more detail.

\subsection{\nustar{}}
By design, the entire sample presented here has \nustar{} observations  obtained from the HEASARC archive\footnote{\url{https://heasarc.gsfc.nasa.gov/docs/archive.html}}. The \nustar{} data were reduced, filtered, and extracted using HEASOFT\footnote{ \url{https://heasarc.gsfc.nasa.gov/ftools/}}   \citep{2014ascl.soft08004N} version 6.28, the \nustar{} Data Analysis Software\footnote{
 \url{https://heasarc.gsfc.nasa.gov/docs/nustar/analysis/}} (NUSTARDAS; version 2.0.0), and the \nustar\ calibration database\footnote{ \url{https://heasarc.gsfc.nasa.gov/docs/heasarc/caldb/nustar/}} (CALDB; version 20200826). For the extractions, we used circular source regions 40\arcsec{} in radius centered on the galaxy nucleus positions and circular background regions 100\arcsec{} in radius. In the spectral fitting, we fixed the \nustar{} normalization constant to unity for FPMA and 1.04 for FPMB, where the latter is based on calibration observations of the bright source 3C~273 reported in \citet{2015ApJS..220....8M}. When multiple FPMA and FPMB observations were available, the normalization constants in the later observations were left as free parameters to account for variability. We used energies from 3 keV to 30 keV from the \nustar{} data for the spectral fitting. Above 30 keV background dominates over AGN emission for our sample.

\subsection{\chandra{}}
\chandra\ ACIS observations were available for 8 of the 9 galaxies  from the Chandra Data Archive\footnote{ \url{https://cxc.harvard.edu/cda/}}, with the exception being NGC~6890.  For most of this sample of X-ray faint, nearby galaxies, the sensitive, higher angular resolution \chandra\ observations identify multiple sources within the \nustar\ beam, primarily due to X-ray binaries within the target galaxies.  Using CIAO \citep{2006SPIE.6270E..1VF} version 4.12 and the \chandra\ CALDB\footnote{ \url{https://cxc.cfa.harvard.edu/caldb/}} version 4.9.1, we extracted \chandra\ spectra of all sources within a 40\arcsec{} radius circular aperture around the core of each galaxy, matching the \nustar{} beam. As discussed in the following discussion of individual sources, the \chandra\ aperture sizes varied depending on whether the source was unresolved and/or if the target was at a larger off-axis angle, for which the \chandra\ point spread function degrades. Sources in the \chandra{} images were identified by eye.  A circular background region 10\arcsec{} in radius was used for all \chandra{} data.  We used energies from 0.5 keV to 8.0 keV from the \chandra{} data for the spectral fitting, and ignored off-nuclear sources with less than 10\% the net count rate of the central AGN. We used all archival \chandra{} data available for these sources, with the exception of NGC~3627, which had a 1.3~ks observation (ObsID: 394) that was ultimately discarded in favor of a much deeper observation (50.3~ks; ObsID: 9548).

\subsection{\swift{}}
Because NGC~6890 lacked \chandra\ observations, we analyzed data from the X-Ray Telescope (XRT) on \swift{} for this galaxy. The data were obtained from the HEASARC archive. We extracted the data using HEASOFT version 6.28, the \swift{} XRT Data Analysis Software\footnote{ \url{https://swift.gsfc.nasa.gov/analysis/}} (SWXRTDAS; version 3.6.0), and the \swift\ CALDB\footnote{ \url{https://heasarc.gsfc.nasa.gov/docs/heasarc/caldb/swift/}} version 20200724. We used circular source regions of 25\arcsec{} radius and background regions of 50\arcsec{} radius for the spectral extraction. We used energies from 0.3 keV to 10.0 keV from the \swift{} data for the spectral fitting.

\subsection{\xmm{}}
We used \xmm{} data from the \xmm{} Science Archive\footnote{ \url{http://nxsa.esac.esa.int/nxsa-web}} for NGC 5005 and NGC 6890, the former to further investigate unusual spectral features found in the \nustar{} data, and the latter because no \chandra{} observations exist for the source. We used all three of the European Photon Imaging Camera (EPIC) CCDs --- i.e., pn, MOS1, and MOS2 --- in the spectral fitting.  We extracted the data using the \xmm\ Scientific Analysis System \citep[SAS, version 18.0.0;][]{2004ASPC..314..759G}.  Details on the \xmm{} spectral extractions are in the individual notes on each galaxy (\S3). We used energies from 0.2 keV to 10.0~keV from the \xmm{} data for the spectral fitting.

\subsection{X-ray Spectral Models}
For each galaxy spectrum we began fitting with a simple CONSTANT*TBABS*POWERLAW model in XSPEC. The constant is to account for source variability and cross-normalization differences between the different telescopes; in the text, we refer to this constant as either the cross-calibration coefficient or the normalization constant. The TBABS \citep{2000ApJ...542..914W} component models absorption of X-rays due to the interstellar medium of our own Milky Way galaxy, which we determined using the Galactic hydrogen column densities along the line of sight to each galaxy from \citet{2016A&A...594A.116H}. The POWERLAW\footnote{ \url{https://heasarc.gsfc.nasa.gov/xanadu/xspec/manual/node216.html}} component fits a simple powerlaw to the data with two parameters: the spectral index, $\Gamma$, and the normalization, defined as the number of $\mathrm{photons\, keV^{-1}\, cm^{-2}\, s^{-1}}$ at 1 keV in the source reference frame.  In luminous, unobscured AGN, Compton upscattering of thermal photons from the accretion disk by the SMBH corona generates a powerlaw X-ray spectrum across our observed range, and this component dominates the X-ray spectrum. In obscured AGN, this component is absorbed by gas, making the observed X-ray spectrum harder (i.e., a lower value of $\Gamma$).  For heavily absorbed, Compton-thick AGN, few photons from the intrinsic spectrum escape below 10~keV.  However, a small fraction of the intrinsic powerlaw generally always escapes \citep[e.g.,][]{2021MNRAS.504..428G}. This scattered, unabsorbed powerlaw component is typically just a few percent of the intrinsic spectrum.

In addition to this simple initial model, AGN, especially those with heavy absorption, may exhibit a soft excess in the 0.5-2~keV range that is thought due to thermal emission from hot gas along the line of sight. We account for this by adding an APEC \citep{2001ApJ...556L..91S} model, which simulates X-ray emission from a collisionally ionized plasma. Its parameters are the plasma temperature, elemental abundances, 
and normalization. The APEC normalization is defined as $\frac{10^{-14}}{4\pi [D_{A}(1+z)]^{2}} \int n_{e} N_{\rm H}dV$, where $D_{A}$ is the angular diameter distance to the source, and $n_{e}$ and $N_{\rm H}$ are the electron and hydrogen number densities, respectively. For this analysis, we set the elemental abundances to solar.

Obscured AGN also typically show a prominent neutral Fe~K-alpha line at 6.4~keV and a Compton hump at $\sim 20$~keV. These features arise from reflected emission and scattering off gas around the central engine.  The gas is believed to be toroidal in geometry and is presumed related to the cooler, more extended dusty torus that is responsible for AGN obscuration at visible wavelengths and AGN thermal emission at MIR wavelengths.  We fit the iron line and Compton hump by adding a BORUS model to the overall spectral model, which allows us to constrain the geometry of the torus. BORUS models torus reprocessing of an intrinsic SMBH corona powerlaw spectrum. Its free parameters are the spectral index of the intrinsic powerlaw ($\Gamma$), the high-energy cutoff, the torus hydrogen column density ($N_{\rm H}$), the torus covering factor (defined as the cosine of the opening angle of the torus), the inclination angle of the torus ($\theta_{\rm inc}$), the relative abundance of iron compared to the solar abundance,
and the normalization (which is defined the same as it is for the POWERLAW model). We consistently set the high energy cutoff to 500~keV and the iron abundance to solar. We also set the spectral indices of the BORUS model and the POWERLAW model to be the same in all cases except NGC~6890. In the case of an AGN with a BORUS component, the POWERLAW component represents the fraction of the intrinsic powerlaw that is scattered and transmitted through the torus, and so it should have the same spectral index as the BORUS component.

We also tried including a ZTBABS \citep{2000ApJ...542..914W} model in our fits. ZTBABS is similar to TBABS but represents absorption from hydrogen at the source, rather than from our Galaxy. However, though we investigated including a ZTBABS component for all of the AGN in this sample, none of the sources ultimately required it.  As noted below, a few of the extranuclear X-ray sources did find improved spectral fitting by including a ZTBABS component.

For NGC 5005 we tried a ZGAUSS\footnote{ \url{https://heasarc.gsfc.nasa.gov/xanadu/xspec/manual/node176.html}} component in addition to a BORUS component. This model represents a Gaussian emission line profile. Its free parameters are the source frame line energy in keV, the source frame line width in keV, the redshift to the source, and the normalization (which is defined as the total photons $\mathrm{cm^{-2}\, s^{-1}}$ in the emission line in the source frame). A ZGAUSS model was ultimately preferred over a BORUS model for this source.

Lastly, for NGC 3627 we used a CUTOFFPL\footnote{ \url{https://heasarc.gsfc.nasa.gov/xanadu/xspec/manual/node161.html}} instead of a POWERLAW component for the extranuclear point sources in the \nustar{} beam. This model component is the same as the POWERLAW component except it includes an exponential rolloff, $KE^{-\Gamma}\, \exp(-E/\beta)$, where $K$ is the normalization, $E$ is the energy, $\Gamma$ is the spectral index, and $\beta$ is the the e-folding energy of the rolloff.                                                              
\subsection{Measuring X-ray Luminosities}

 We measured the intrinsic X-ray luminosities from the BORUS normalization and $\mathrm{\Gamma}$ (which was fixed to the POWERLAW $\mathrm{\Gamma}$ for all but NGC 6890). We derived the errorbars on the intrinsic luminosities by turning the upper and lower errors on the BORUS $\Gamma$ and norm into fractional errors and then added fractional errors on each of the two parameters in quadrature to derive the fractional error on the luminosities. 

For NGC 3627 and NGC 5005 (for which BORUS components were not used), we added a CFLUX component to the POWERLAW components of their models. This component calculates the flux of the model component it is added to when the spectrum is fitted. We then converted the fluxes to luminosities using the Local-Group-corrected redshift distances listed in NED. The errors on intrinsic luminosity for these galaxies were derived from the 90\% confidence intervals reported by the CFLUX component.

\begin{deluxetable}{lcccccc}
\tablecaption{List of X-ray observations.}\label{tab:xraydata}
\tablewidth{0pt}
\tablehead{
\colhead{Target} & \colhead{R.A., Dec.} & \colhead{Observatory} & \colhead{ObsID} & \colhead{Date} & \colhead{Net Exposure Time} & \colhead{Net Count Rate}\\
\colhead{} & \colhead{(J2000)} & &\colhead{} & \colhead{} & \colhead{(ks)} & \colhead{(cts ${\rm ks}^{-1}$)}
}
\startdata
NGC 1386 & 03:36:46.18, $-$35:59:57.87 & \chandra{} & 4076 & 2003-11-19 & 19.6 & 52.5\\ 
{  } & {  } & - & 12289 & 2011-4-13  & 17.3 & 48.7\\ 
{  } & {  } & - & 13185 & 2011-4-13  & 29.7 & 45.4\\ 
{  } & {  } & - & 13257 & 2011-4-14  & 33.8 & 47.0\\ 
{  } & {  } & \nustar{} & 60001063002 & 2013-7-9  & 18.8/18.4 & 9.2/10.2\\ 
{  } & {  } & - & 60201024002 & 2016-5-11 &  25.4/25.8 & 9.9/9.2\\ 
NGC 3627 & 11:20:14.96, +12:59:29.54 & \chandra{} & 9548 & 2008-3-31  & 49.6 & 6.1\\ 
{  } & {  } & \nustar{} & 60371003002 & 2017-12-23 & 49.1/48.9 & 3.3/2.3\\ 
NGC 3982 & 11:56:28.13, +55:07:30.86 & \chandra{} & 4845 & 2004-1-3 & 9.2 & 6.6\\ 
{  } & {  } & \nustar{} & 60375001002 & 2017-12-5 & 30.7/31.0 & 5.8/4.7\\ 
NGC 4501 & 12:31:59.161, +14:25:13.39 & \chandra{} & 2922 & 2002-12-9 & 17.1 & 11.7\\ 
{  } & {  } & \nustar{} & 60375002002 & 2018-1-28 & 58.0/59.4 & 4.2/3.4\\ 
{  } & {  } & - & 60375002004 & 2018-5-24 & 58.5/58.2 & 3.5/3.7\\ 
IC 3639 & 12:40:52.85, $-$36:45:21.11  & \chandra{} & 4844 & 2004-3-7 & 8.7 & 31.5\\ 
{  } & {  } & \nustar{} & 60001164002 & 2015-1-9  & 56.1/55.7 & 8.3/8.1\\ 
NGC 4922 & 13:01:24.90, +29:18:40.0 & \chandra{} & 4775 & 2004-11-2 & 3.8 & 11.8\\ 
{  } & {  } & - & 15065 & 2013-11-2  & 14.9 & 9.3\\
{  } & {  } & - & 18201 & 2016-3-6 & 5.8 & 10.7\\
{  } & {  } & \nustar{} & 60101074002 & 2015-11-9 & 20.2/20.1 & 4.2/2.8\\
NGC 5005 & 13:10:56.23, +37:03:33.14 & \chandra{} & 4021 & 2003-8-19 & 4.92 & 54.3\\
{  } & {  } & \xmm{} & 0110930501 & 2002-12-12 & 8.7/13.1/13.1 & 297.5/69.1/70.9\\
{  } & {  } & \nustar{} & 60001162002 & 2014-12-16 & 48.9/48.3 & 5.8/5.4\\
Mrk 463 & 13:56:02.87, +18:22:19.48 & \chandra{} & 4913 & 2004-6-11 & 49.3 &  24.3\tablenotemark{a}\\
{  } & {  } & - & 18194 & 2016-3-10 & 9.8 & 16.1\tablenotemark{a}\\ 
{  } & {  } & \nustar{} & 60061249002 & 2014-1-1 & 23.9/23.8 & 2.3/2.2\\ 
NGC 6890 & 20:18:18.10, $-$44:48:24.21 & \xmm{} & 0301151001 & 2005-9-29 & 0.9/7.5/7.8 & 131.1/26.3/28.3\\ 
{ } & { } & \swift{} & 00088188001 & 2018-3-6 & 1.7 & 11.13\\ 
{ } & { } & - & 00088188002 & 2018-5-25 & 2.0 & 20.1\\ 
{  } & {  } & \nustar{} & 60375003002 & 2018-5-25 & 34.6/34.5 & 59.5/56.2\\
\enddata 
\tablecomments{Net count rates for \chandra{} data are for the AGN core only.  Exposure times and net count rates for \nustar{} observations are FPMA/FPMB. Exposure times and net count rates for \xmm{} observations are pn/MOS1/MOS2.}
\tablenotetext{a}{For the brighter, eastern component of this merger system (see \S~3.8).}
\end{deluxetable}

\section{The Individual Galaxies}
\label{sec:gal}

We now discuss each of the nine galaxies in our sample individually, providing notes about each one and then details of the X-ray observations and analysis.

\subsection{NGC 1386}
NGC 1386 is a barred spiral galaxy in the Fornax Cluster \citep{1989AJ.....98..367F} with  prominent dust lanes, a ring of \ion{H}{2} regions, and AGN-ionized gas plumes visible in {\it Hubble} imagery of its central regions \citep{2000ApJS..128..139F}. It is optically classified as a Seyfert 2 galaxy \citep[e.g.,][]{1980ApJ...235..761P, 2011MNRAS.414.3084B} but it has also been classified as a S1i by \citet{2006A&A...455..773V} on the basis of a broad Paschen-beta (Pa$\beta$) component evident in its near-infrared (NIR) spectrum. 
\citet{2014MNRAS.438.3434R} did not find polycyclic aromatic hydrocarbon (PAH) features in its {\it Spitzer} nuclear spectrum, likely attributable to ionization by the AGN. The AGN is a water megamaser source \citep{1997AAS...19110402B}; such sources typically show higher levels of obscuration \citep[e.g.,][]{2006A&A...450..933Z, 2016A&A...589A..59M}.

\citet{2015ApJ...806...84L} reports that NGC 1386 has a mass outflow rate of $>1M_{\odot}\, {\rm yr}^{-1}$ and shows complex gas kinematics at its center, likely caused by an ionization cone intersecting the galactic disk at an angle. \citet{2017MNRAS.470.2845R} found even stronger outflows, comparable to that of a strong AGN, with a mass loss rate of $11 M_{\odot}\, {\rm yr}^{-1}$. The outflow takes the form of two expanding shells of gas that are coincident with the axis of the radio emission, implying they are likely powered by a radio jet rather than simply by the AGN radiation.  Between the broad Pa$\beta$ emission line and the radio maser activity, the broadband properties of NGC~1386 suggest a currently active, obscured Seyfert~2 galaxy. 

In the X-rays, \citet{2005MNRAS.356..295G} concluded the  \xmm{} spectrum was best fit by either scattering and transmission components, or by thermal and reflection components. \citet{2006AA...448..499B} confirmed a reflection-dominated model was the best fit based on \chandra{} data, but concluded spectral lines visible in the soft X-ray EPIC observations were more likely due to scattering off of photoionized plasma rather than thermal emission. \citet{2012ApJ...758...82L} presented a joint analysis of \chandra{} and \xmm{} data in the 0.5-2.0 keV range and found that it was best fit with a two-temperature APEC model, indicating the presence of two thermal gas components, one with $kT \sim$ 0.13~keV and one with $kT \sim$ 0.67~keV. They noted this was similar to X-ray observations of starburst galaxies \citep[e.g.,][]{1998ApJS..118..401D, 2004ApJS..151..193S}. In addition to two APEC components, their model also contained two powerlaw components with spectral indices tied together, each subject to both Galactic absorption and absorption at the source. The latter was found to be $N_{\rm H}\:=\:3.14\times10^{23}\:{\rm cm}^{-2}$, and \citet{2012ApJ...758...82L} measured the AGN contribution to the 0.5-2.0 keV X-ray luminosity to be $\mathrm{\approx70\%}$. Recently, \citet{2021ApJ...910...19J} reported \chandra\ detection of extended hard X-ray emission across the ionization cones of NGC~1386.

\citet{2011MNRAS.413.1206B} identified NGC 1386 as Compton-thick on the basis of its \xmm{} data, which shows a strong Fe K-alpha line ($\mathrm{EW_{6.4}=1710}$~eV in their model). They confirmed it was reflection-dominated, and measured a hydrogen column density of $N_{\rm H}\:=\:1.51\times10^{24}\:{\rm cm}^{-2}$.  Adding data taken by \nustar{} to the existing \xmm{} spectra, \citet{2015ApJ...805...41B} found a slightly higher column density, $N_{\rm H}\:=\:5.61\times10^{24}\:{\rm cm}^{-2}$. \citet{2016A&A...589A..59M} found similar results using a combination of a MyTORUS model \citep{2009MNRAS.397.1549M} and an emission line component at 6.5~keV.

\subsubsection{X-ray Observations and Data Extraction}

\begin{figure}
    \centering
    \plotone{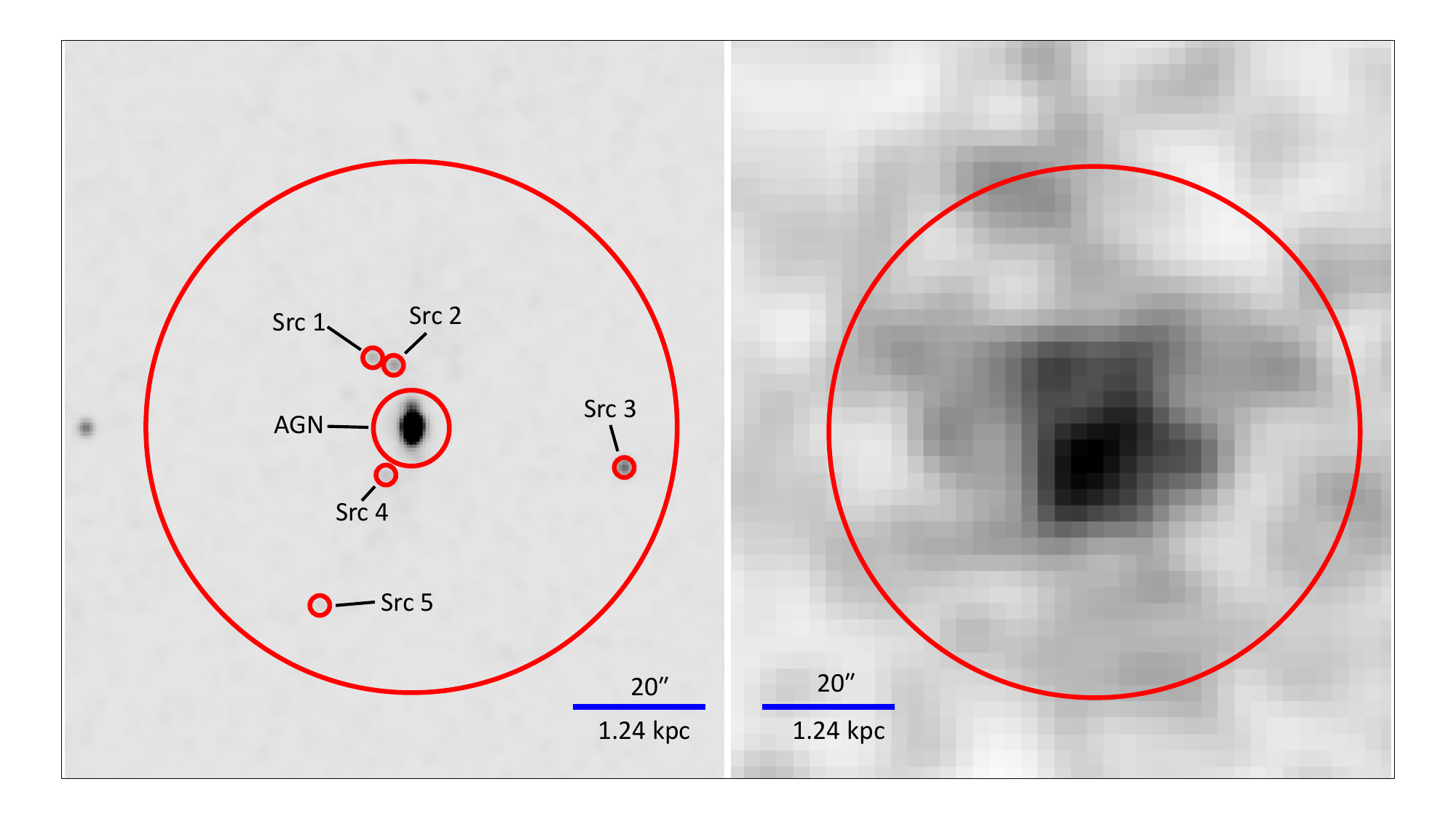}
    \caption{\chandra{} (ObsID: 13257) and \nustar{} (ObsID: 60201024002) FPMA images of NGC 1386. The larger, 40\arcsec{} radius circle denotes the \nustar{} extraction region, while the smaller circles denote the \chandra{} extraction regions. Five extra-nuclear point sources were visible in the \chandra{} observations, though all were sufficiently faint (i.e., $< 10$\% the flux of the nucleus) to be ignored in the AGN spectral analysis.}
    \label{fig:NGC1386_image}
\end{figure}

NGC~1386 was observed twice by \nustar\ and four times by \chandra; details, including observation dates and exposure times, are in Table~\ref{tab:xraydata}.  Figure~\ref{fig:NGC1386_image} presents the third \chandra{} observation and the second \nustar{} FPMA observation with the extraction regions overlaid.

The AGN \chandra{} spectrum was extracted with a circular source region 5.7\arcsec{} in radius. In addition to the AGN core, five extranuclear point sources in the \nustar{} beam were present in all four \chandra{} images. They were extracted using circular source regions 1.5\arcsec{} in radius. Since the count rates for all these sources were less than 10\% that of the core, they were ignored in the X-ray spectral fitting.

\subsubsection{X-ray Spectral Fitting}

\begin{figure}
    \plotone{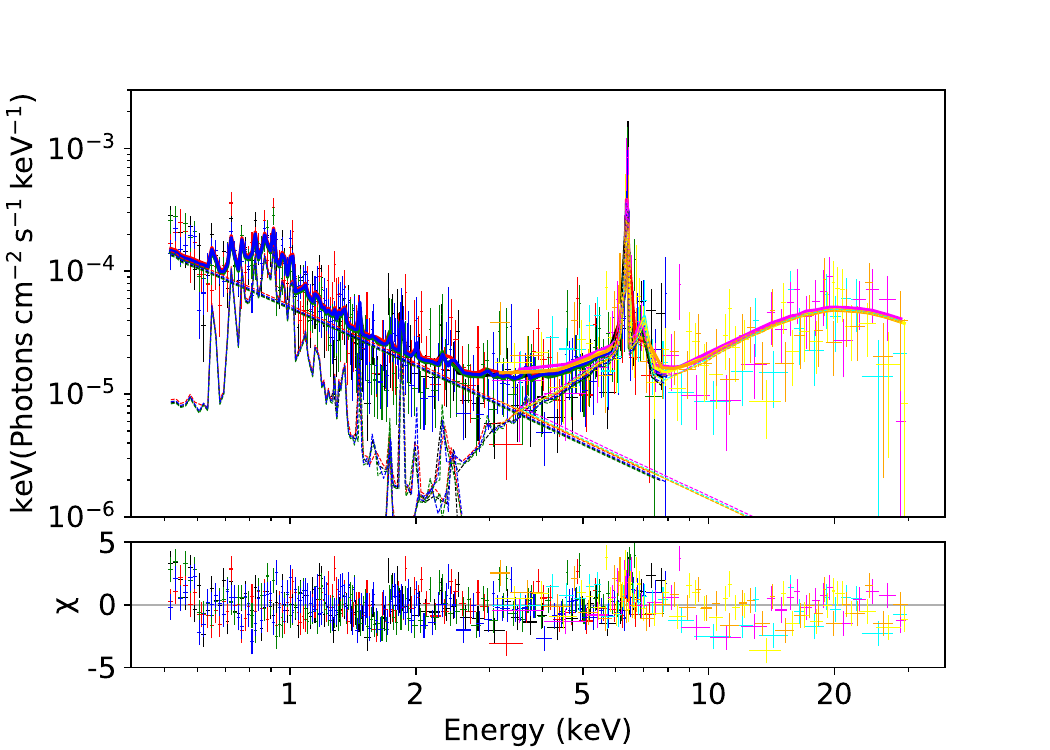}
    \caption{Unfolded spectrum and best-fit model for NGC 1386. Black, red, green, and blue denote \chandra{} data (ObsIDs 4076, 12289, 13185, and 13257). Cyan and magenta denote FPMA and FPMB data for \nustar{} observation 60001063002. Yellow and orange denote FPMA and FPMB for \nustar{} observation 60201024002.}
    \label{fig:NGC1386_uf}
\end{figure}

We first modeled the spectrum with TBABS*POWERLAW, which yielded a C-stat/d.o.f.\ of 2980.46/1698. A strong Fe~K-alpha emission line is evident in the unfolded spectrum (Figure \ref{fig:NGC1386_uf}), as well as a prominent Compton hump at 10-20 keV. We added a BORUS component to the original TBABS*POWERLAW fit to account for these reflection features, fixing the spectral index of the BORUS component to that of the POWERLAW component. This improved C-stat/d.o.f.\ to 2067.40/1694. Strong residuals above the power law component were present at energies 0.5-2.0 keV so an APEC component was added, resulting in C-stat/d.o.f.\ = 1679.64/1692. While this is a statistically good fit, $N_{\rm H}$ remains unconstrained. We therefore opted to freeze $\cos(\theta_{\rm inc})$ at its model value of 0.45 before refitting. The final fit had C-stat/d.o.f.\ = 1681.69/1693 and the parameters of the best-fit model are presented in Table \ref{tab:NGC1386params}. The 90\% confidence interval for the BORUS parameter $\log N_{\rm H}$ was unconstrained at the upper end, so it is listed as $\geq 24.5$. The powerlaw spectral index hit the upper bound of 2.6 in the model, so it is listed as $\geq 2.6$. The best-fit model is plotted over the unfolded spectrum in Figure \ref{fig:NGC1386_uf}. The logarithm of the 2-10 keV luminosity (in units of $\mathrm{erg\:s^{-1}}$) measured from the model is $42.29\pm{0.05}$.

\begin{deluxetable}{@{\extracolsep{10pt}}lccccccc@{}}
\tablecaption{Parameters for best-fit NGC 1386 model.}\label{tab:NGC1386params}
\tablewidth{2pt}
\tablehead{\multicolumn{2}{c}{APEC} & \multicolumn{4}{c}{BORUS} & \multicolumn{2}{c}{POWERLAW}\\ \cline{1-2}
\cline{3-6}  \cline{7-8} \colhead{$kT$} & \colhead{Norm} & \colhead{$\log({N_{\rm H}}$)\tablenotemark{a}} & \colhead{$\mathrm{CF_{Tor}}$\tablenotemark{b}} & \colhead{$\mathrm {\cos(\theta_{inc})}$\tablenotemark{c}} & \colhead{Norm} & \colhead{$\Gamma$} &\colhead{Norm}\\
 \colhead{(keV)} & \colhead{($10^{-5}$ cts $\mathrm{s^{-1}\:keV^{-1}}$)} & \colhead{} & \colhead{} & \colhead{} & \colhead{(cts $\mathrm{s^{-1}\:keV^{-1}}$)} & \colhead{} & \colhead{($10^{-5}$ cts $\mathrm{s^{-1}\:keV^{-1}}$)}
}
\startdata
$0.82\pm{0.03}$ & $2.28^{+0.44}_{-0.38}$ & $\geq{24.5}$ & $0.49\pm{0.01}$& $=0.45$\tablenotemark{d} & $0.09\pm{0.01}$ & $\geq 2.6$ & $5.65^{+0.98}_{-0.38}$
\enddata
\tablecomments{Error bars represent 90\% confidence intervals. The \chandra{} normalization constant values were $0.91^{+0.16}_{-0.13}$ (ObsID: 4076), $0.97^{+0.17}_{-0.14}$ (ObsID: 12289), $0.90^{+0.15}_{-0.10}$ (ObsID: 13185), and $0.53^{+0.16}_{-0.08}$ (ObsID:13527). The second \nustar{} FPMA and FPMB normalization constants were $0.99^{+0.15}_{-0.11}$ and $0.98^{+0.16}_{-0.11}$ (ObsID: 60201024002).}
\tablenotetext{a}{$N_{\rm H}$ in units of $\mathrm{cm^{-2}}$.}
\tablenotetext{b}{Covering factor of torus, equivalent to cosine of torus opening angle.}
\tablenotetext{c}{Cosine of torus inclination angle.}
\tablenotetext{d}{Frozen at this value.}
\end{deluxetable}

\subsection{NGC 3627}

NGC 3627 (also known as Messier 66) is a barred spiral galaxy in the Leo triplet of galaxies, along with NGC 3623 and NGC 3628, and is undergoing tidal interactions with them \citep{1987ApJ...320..145S, 1991ApJ...370..176H, 1993ApJ...418..100Z, 1996A&A...306..721R}. It exhibits low-luminosity nuclear activity, though its status as a true SMBH-powered AGN (as opposed to simply having a nuclear starburst) has been the subject of debate in the literature. Its optical activity type has been variously characterized as a transition object \citep[e.g.][]{2005ApJ...620..113D}, Seyfert 2 \citep[e.g.][]{2011MNRAS.414.3084B}, incapable of being distinguished between the transition object and Seyfert 2 classes \citep{1997ApJS..112..315H}, or simply a LINER \citep{2006A&A...455..773V}.

NGC 3627 presents a complex profile in the MIR, with diffuse emission across the entire galaxy \citep{2014MNRAS.439.1648A} from which a compact nuclear source cannot be clearly separated. 

In the X-rays, NGC 3627 was first detected by {\it ASCA} and {\it ROSAT}. \citet{2001MNRAS.324..737R} examined these observations and found that the spectrum was described well by a soft thermal component (0.5-1 keV) and a powerlaw component (2-5 keV). They measured the flux ratio between these components to be 0.56,
and argued that this indicated the two spectral components likely had a common, non-AGN origin. They noted this flux ratio was very similar to the {\it ASCA} flux ratio in the same energies for the starburst galaxy NGC 253. Therefore, they argued that NGC 3627 was unlikely to be a true AGN.

The first \chandra{} observation of NGC 3627, a  1.3~ks snapshot exposure, was initially published by \citet{2001ApJ...549L..51H}, who did not detect a dominant unresolved point source in the galaxy's core, only a group of sources. They therefore concluded that NGC 3627 was not a true AGN. Some later papers also suspected NGC 3627 not to be a true AGN, partially on this basis \citep[e.g.][]{2006A&A...455..173P, 2009A&A...506.1107G}; \citet{2006A&A...455..173P} put an upper limit of $L_{2-10}<7.6\times10^{37}\:{\rm erg}\:{\rm s}^{-1}$ on the nuclear 2-10~keV luminosity. In contrast, and based on the same observations, \citet{2009ApJ...699..281Z} argued the \chandra{} image does show a dominant central point source within 1\arcsec{} of the galaxy's center, and they report a significantly higher 0.3-8~keV X-ray luminosity of $L_{0.3-8}=9.1\times10^{39}\:{\rm erg}\:{\rm s}^{-1}$.

In NGC 3627's sole \xmm{} observation, \citet{2006A&A...455..173P} observed a point source at the galaxy nucleus, but noted it was equal in brightness to a second point source 10\arcsec{} away. Indeed, both \citet{2006A&A...455..173P} and \citet{2013A&A...556A..47H} agree the \xmm{} data is heavily contaminated by emission from sources other than the galaxy core. \citet{2009A&A...506.1107G} failed to find a unresolved point source in the harder bands observed by \xmm{} (4.5-8.0 keV). Their estimate of the 2-10 keV luminosity is $L_{2-10} \sim 10^{39}\:{\rm erg}\:{\rm s}^{-1}$ based on the \xmm{} data, assuming a powerlaw index of $\Gamma = 1.8$ and Galactic absorption. They nonetheless identified NGC 3627 as a Compton-thick AGN candidate on the basis of its $L_{2-10}/L_{\rm{[OIII]}}$ ratio \citep{2009ApJ...704.1570G}. In contrast, \citet{2011MNRAS.413.1206B} measured an ionized hydrogen column density of $5.01 \times10^{21}\, \mathrm{cm^{-2}}$ in the \xmm{} spectrum, which would clearly place it in the Compton-thin regime. \citet{2011MNRAS.413.1206B} modeled the \xmm{} observation of NGC 3627 with a soft thermal emission component and ionized absorber component in addition to Galactic absorption and powerlaw components.

A second, deeper (50.3~ks) \chandra{} observation of NGC 3627 was taken in 2008
\citep{2011ApJ...731...60G}. In this observation, one can see an unresolved nuclear point source embedded in diffuse emission \citep{2013ApJ...776...50C}.

\citet{2020ApJ...905...29E} fit the \nustar{} data for NGC 3627 with a partial covering absorber that included Galactic absorption. They measured an absorbing hydrogen column density of $1.8\times10^{24}\, \mathrm{cm^{-2}}$, which would put the AGN in the Compton-thick category. After correction for absorption they classified NGC 3627 as an AGN in the early stages of fading based on it being under-luminous in X-rays compared to the MIR. In their interpretation, NGC 3627 is observed at the beginning of the fading arc of the AGN duty cycle. 

\citet{2011MNRAS.414.3084B} plot NGC 3627 on several Baldwin, Phillips, and Terlevich (BPT) diagrams \citep{1981PASP...93....5B}. The position of NGC 3627 on the BPT diagram is an AGN if using [\ion{O}{3}]/H$\mathrm{\beta}$ vs. [\ion{N}{2}]/H$\mathrm{\alpha}$, a LINER if it is using [\ion{O}{3}]/H$\mathrm{\beta}$ vs. [\ion{S}{2}]/H$\mathrm{\alpha}$, and a Seyfert 2 if using [\ion{O}{3}]/H$\mathrm{\beta}$ vs. [\ion{O}{1}]/H$\mathrm{\alpha}$. We therefore adopt its optical classification as an AGN based on previous work.

\subsubsection{X-ray Observations and Data Extraction}
NGC~3627 has been observed by \chandra{} twice, for 1.3~ks on 1999 November 3 (ObsID: 394) and for 50.3~ks on 2008 March 31. Given that the first, significantly shorter exposure does not clearly detect any sources at the galaxy center, we ignore those data in our analysis.  Table~\ref{tab:xraydata} presents details of the latter \chandra\ observation, as well as the single \nustar\ observation of this galaxy to date.

\begin{figure}
    \centering
    \plotone{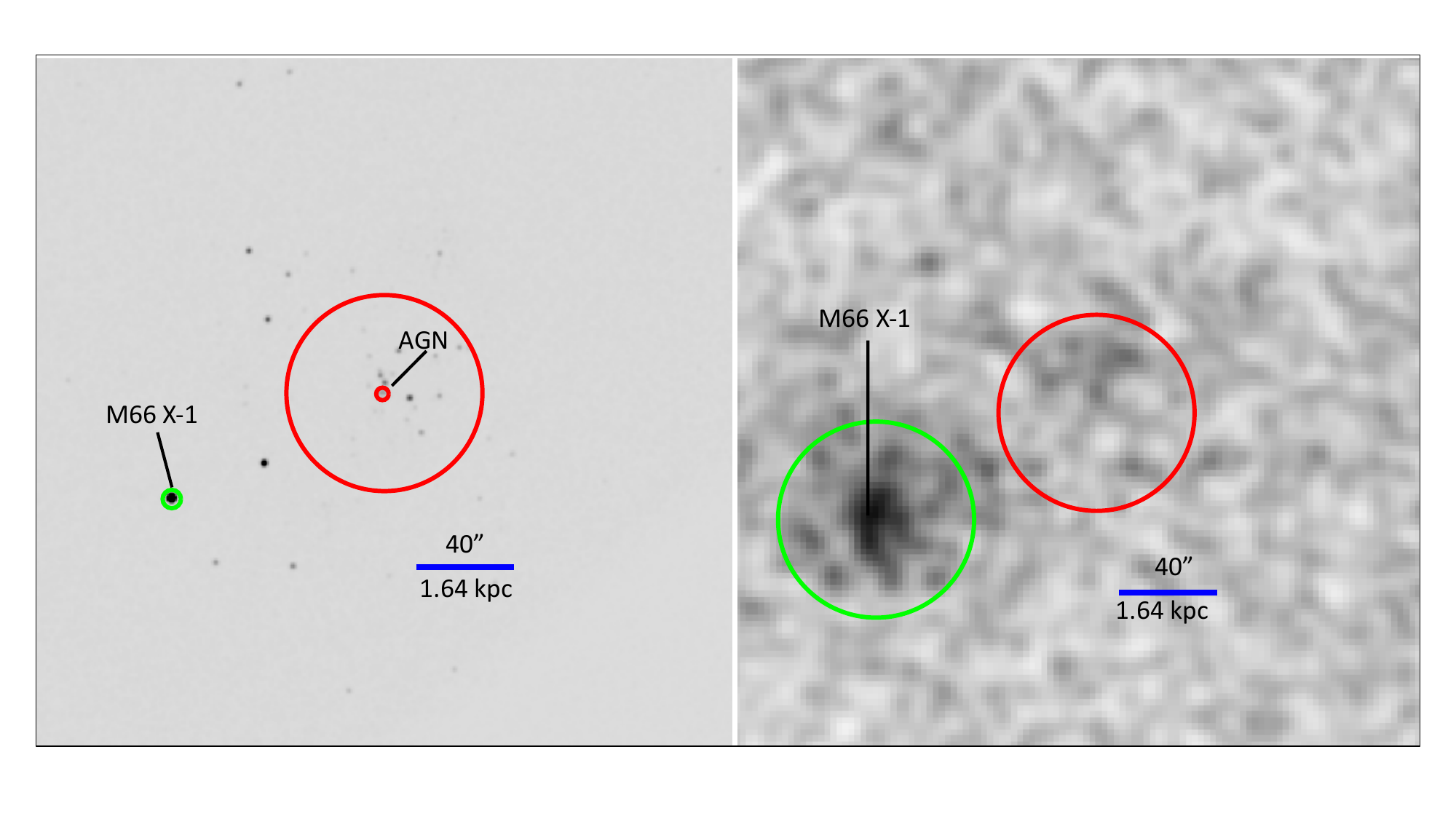}
    \caption{\chandra{} (left; ObsID: 9548) and \nustar{} FPMA (right) images of NGC 3627. The larger, 40\arcsec{} radius red circle denotes the \nustar{} extraction region for the AGN, while the smaller red circle denotes the \chandra{} extraction region for the AGN. The ULX M66~X-1 is highlighted with a green circle (3.75\arcsec{} diameter in \chandra; 40\arcsec{} radius in \nustar{}). M66~X-1 dominates the \nustar{} image, while the AGN is not clearly detected by \nustar{}.}
    \label{fig:NGC3627_zoomout_image}
\end{figure}

\begin{figure}
    \centering
    \plotone{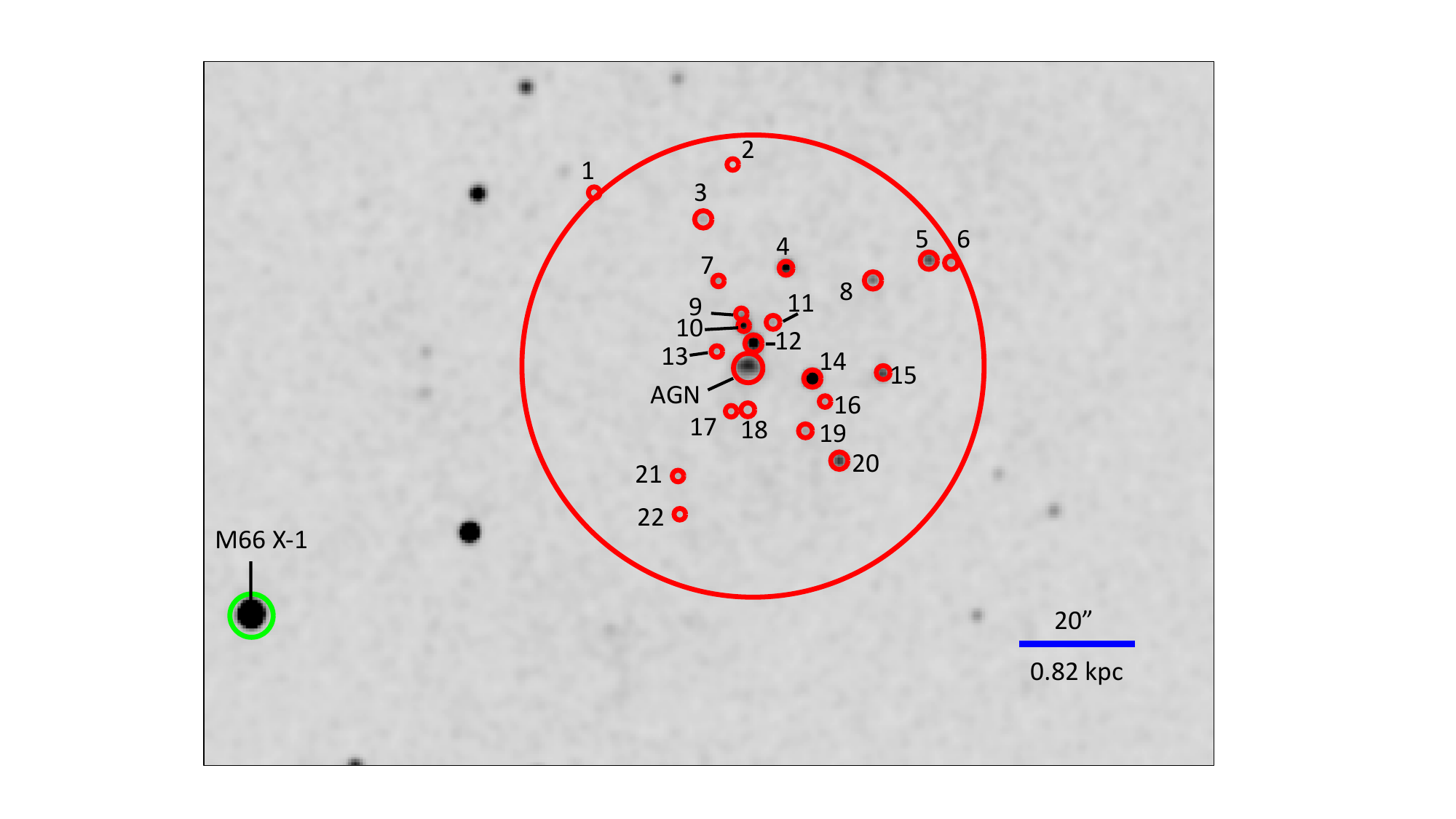}
    \caption{Zoomed-in and re-scaled \chandra{} image of NGC 3627 highlighting and labeling the plethora of off-nuclear point sources (small red circles) within the larger 40\arcsec{} radius  \nustar{} beam. The ULX M66~X-1 is visible to the southeast (green circle).}
    \label{fig:NGC3627_zoomin_image}
\end{figure}

The \chandra{} and \nustar{} FPMA images of NGC 3627 are shown in Figure~\ref{fig:NGC3627_zoomout_image}. There are a large number of sources visible in the \chandra{} image, with one diffuse, irregularly shaped source associated with the nucleus.  In addition, there is a bright point source approximately 1.5\arcmin\ to the southeast whose brightness dwarfs that of the nucleus as well as the numerous point sources within the \nustar\ beam (Figure~\ref{fig:NGC3627_zoomin_image}).
This source, associated with the ultraluminous X-ray source (ULX) M66~X-1 \citep{2011MNRAS.416.1844W}, dominates in the \nustar{} image, while the AGN, in contrast, is not clearly visible. Indeed, we used the \chandra-derived astrometric offset between the ULX and the AGN to place the AGN extraction aperture in the \nustar\ image.

Figure \ref{fig:NGC3627_zoomin_image} presents a zoomed-in version of the \chandra{} image, highlighting the 22 off-nuclear point sources visible within the \nustar{} beam. The \chandra{} AGN spectrum was extracted using a circular source region 2.55\arcsec{} in diameter. The off-nuclear point sources were extracted using circular source regions 1.5\arcsec{} in diameter for sources 3, 5, 8, 12, 14, and 20; 1.2\arcsec{} in diameter for sources 4, 11, 15, 18, and 19; and 1\arcsec{} in diameter for the remaining sources. Because the nucleus is so faint, 15 \chandra\ point sources within the \nustar{} beam are brighter than 10\% of its count rate.  For all other galaxies in our sample, we do joint fitting of the AGN and all off-nuclear point sources within the \nustar\ beam above that threshold.  However, fitting this many sources jointly would be prohibitive and most of the \chandra\ flux within the \nustar\ beam comes from the brightest of these off-nuclear sources.  Therefore, only the ten brightest point sources are included in the spectral fitting (i.e., sources 4, 5, 7, 8, 9, 10, 12, 14, 15, and 20).

\subsubsection{X-ray Spectral Fitting}
Due to the large number of point sources present in the \nustar{} beam, we first fit the off-nuclear point sources with their \chandra{} data alone. We then initially fit this galaxy with all parameters for the off-nuclear sources frozen based on their \chandra\ data, thereby avoiding having too many free parameters which can lead to parameter values being implausibly high or low.  

We started with a simple model consisting of a CONSTANT, a TBABS component frozen to the Galactic hydrogen column density, and 11 POWERLAW components, one for the AGN and one for each of the 10 brightest point sources, where the latter were frozen to the best-fit values from \chandra. This yielded a C-stat/d.o.f.\ of 1422.72/1347.  
However, this fit substantially overestimated the brightness of the \nustar{} data, likely because several of the off-nuclear point sources had hard spectra over the \chandra\ range that overestimated their brightness at the higher energies of  \nustar. 

We therefore decided to change the POWERLAW component in the extra point sources to a CUTOFFPL model. We started with the high-energy cutoffs frozen at 500 keV for all the sources, and tested whether thawing each one would decrease C-stat or not. Out of all the sources, only thawing the cutoffs on Sources 5, 8, 12, and 14 improved the fit. This fit had a C-stat/d.o.f.\ of 1182.16/1318. 

\begin{figure}
    \plotone{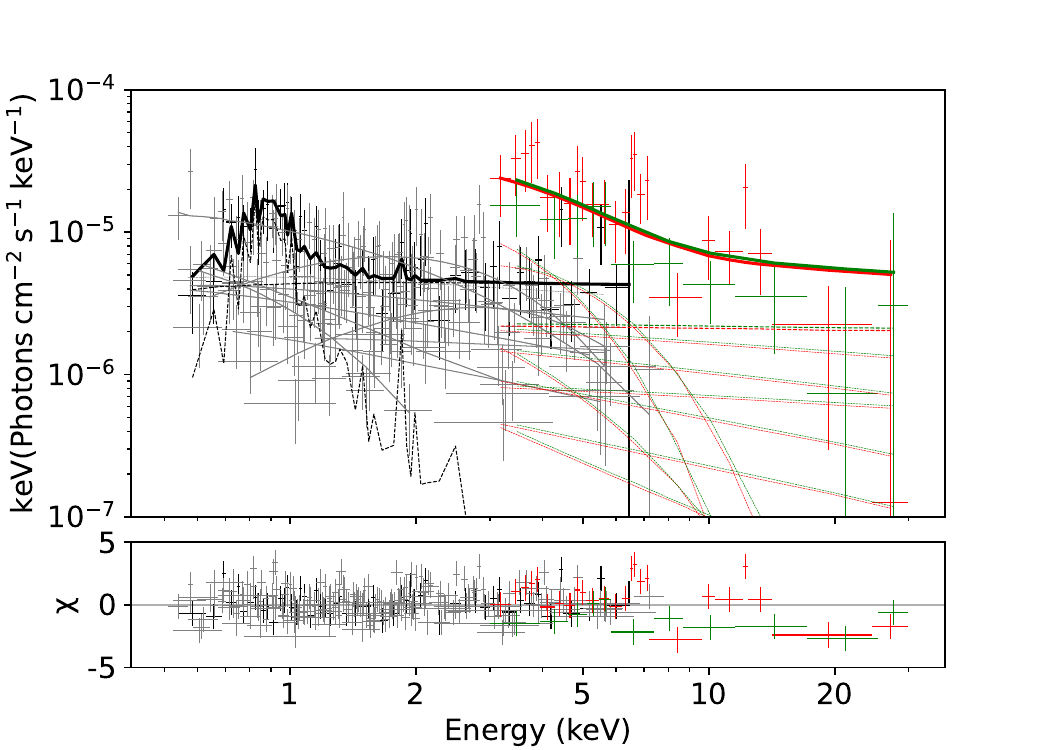}
    \caption{Unfolded spectrum and best-fit model for NGC 3627. Black denotes \chandra{} data and model for the AGN core. Red and green denote \nustar{} FPMA and FPMB data and models. The \chandra{} data and models for the off-nuclear sources are depicted in light grey.}
    \label{fig:NGC3627_uf}
\end{figure}

We fit the initial model with Sources 5, 8, 12 and 14's high-energy cutoffs thawed, then froze the high-energy cutoffs before refitting. We then added an APEC component to the model, as there is an excess between 0.5 and 2 keV. This led to a C-stat/d.o.f.\ of 1154.65/1320. The final parameter values are tabulated in Table \ref{tab:NGC3627params}. The spectrum and best fit final model are plotted in Figure \ref{fig:NGC3627_uf}. The logarithm of the 2-10 keV luminosity (in units of $\mathrm{erg\:s^{-1}}$) measured from the model is $38.38^{+0.16}_{-0.10}$.

\begin{deluxetable}{@{\extracolsep{10pt}}lccccc@{}}
\tablecaption{Parameters for best-fit NGC 3627 model.}\label{tab:NGC3627params}
\tablewidth{2pt}
\tablehead{\colhead{} & \multicolumn{2}{c}{APEC} &  \multicolumn{3}{c}{CUTOFFPL}\\ \cline{2-3}
\cline{4-6} \colhead{Source} & \colhead{$kT$} & \colhead{Norm}  & \colhead{$\Gamma$} & \colhead{Cutoff} & \colhead{Norm}\\
& \colhead{(keV)} & \colhead{($10^{-6}$ cts $\mathrm{s^{-1}\:keV^{-1}}$)} & \colhead{} & \colhead{(keV)} &  \colhead{($10^{-6}$ cts $\mathrm{s^{-1}\:keV^{-1}}$)}
}
\startdata
AGN & $0.83^{+0.12}_{-0.14}$ & $1.31^{+0.70}_{-0.43}$ & $1.04\pm{0.23}$ & -- & $2.29^{+1.54}_{-0.57}$\\
Src 4 & {} & {} & $1.31^{+0.22}_{-0.21}$ & 500\tablenotemark{a} & $2.2^{+2.3}_{-0.4}$\\
Src 5 & {} & {} & $-0.97\pm{0.32}$ & 1.40\tablenotemark{a} & $5.77^{+2.00}_{-4.68}$\\
Src 7 & {} & {} & $1.61^{+0.51}_{-0.46}$ & 500\tablenotemark{a} & $0.9^{+4.0}_{-0.2}$\\
Src 8 & {} & {} & $0.9^{+0.65}_{-0.64}$ & 0.51\tablenotemark{a} & $26.2^{+26.3}_{-17.6}$\\
Src 9 & {} & {} & $1.14^{+0.46}_{-0.41}$ & 500\tablenotemark{a} & $1.0^{+2.3}_{-0.4}$\\
Src 10 & {} & {} & $1.18^{+0.22}_{-0.20}$ & 500\tablenotemark{a} & $2.6^{+1.9}_{-0.5}$\\
Src 12 & {} & {} & $-0.56\pm{0.23}$ & 1.07\tablenotemark{a}& $26.8^{+7.4}_{-20.5}$\\
Src 14 & {} & {} & $1.39^{+0.19}_{-0.18}$ & 2.76\tablenotemark{a} & $7.7^{+26.2}_{-0.8}$\\
Src 15 & {} & {} & $2.26^{+0.39}_{-0.37}$ & 500\tablenotemark{a} & $1.84^{+6.89}_{-0.32}$\\
Src 20 & {} & {} & $1.54^{+0.27}_{-0.26}$ & 500\tablenotemark{a} & $2.7^{+3.4}_{-1.3}$\\
\enddata
\tablecomments{The AGN was fit using a POWERLAW model (i.e., not a CUTOFFPL model). The CUTOFFPL normalizations for sources 4, 7, 9, 10, and 20 were estimated with the STEPPAR command. The \chandra{} instrumental normalization constants on each of the sources could not be constrained.}
\tablenotetext{a}{Frozen at this value.}
\end{deluxetable}

\subsection{NGC 3982}
NGC 3982 is a barred spiral galaxy, classified as a Seyfert~1.9 since it possesses broad H$\mathrm{\alpha}$ but lacks broad H$\mathrm{\beta}$ in its optical spectrum \citep[e.g.][]{1997ApJS..112..315H, 2006A&A...455..773V}. Seyfert~1.9 galaxies are  believed to be highly obscured, and are often lumped together with Seyfert~2 AGN in population studies \citep[e.g.][]{2001ApJ...554L..19T}. The nucleus of NGC~3982 is surrounded by a partial ring of star formation, at a radius of approximately 500 pc  \citep{2017MNRAS.469.3405B}. At MIR wavelengths, NGC~3982 is a compact source with extended emission of unclear origin \citep{2014MNRAS.439.1648A}. \citet{2010ApJ...709.1257T} concluded that 81\% of the 19 $\mathrm{\mu m}$ emission originates from the AGN. \citet{2020ApJ...905...29E} identify NGC~3982 as a candidate fading AGN.

In the X-rays, NGC 3982 was first observed with {\it ASCA} \citep{2001ApJ...556L..75M} and was later serendipitously observed with \chandra{} as part of the \chandra{} Deep Field North survey \citep{2003AJ....126..539A}. The \chandra{} spectrum was first analyzed by \citet{2005AA...444..119G}, where the low number of counts hampered attempts to fit the spectrum to a Compton-thick model, though they did report a hydrogen column density $N_{\rm H} >1.6 \times 10^{24}\, {\rm cm}^{-2}$ and a very high Fe K-alpha equivalent width (8 keV based on their ``local'' fit).  These values suggest, though do not confirm, a Compton-thick nature for NGC~3982. Its \chandra{} spectrum was later re-analyzed by \citet{2007ApJ...656..105G} in an attempt to determine whether it was a `true' Seyfert~2, but the low number of counts prevented them from making a robust fit to the spectrum. However, because they did find evidence of photoelectric absorption, they concluded the `true' Seyfert~2 explanation for its 2-10 keV faintness seemed unlikely.

\citet{2007ApJ...657..167S} presented a joint fit of \chandra\ and \xmm\ spectra of NGC~3982, where they measured $N_{\rm H} > 10^{24}\, {\rm cm}^{-2}$ and the Fe~K-alpha equivalent width to be 6.31 keV. They therefore classified the AGN as Compton-thick. \citet{2009A&A...500..999A} also analyzed these \xmm{} data and measured somewhat less extreme values, finding $N_{\rm H} = 4.32 \times 10^{23}\, {\rm cm}^{-2}$ and an Fe~K-alpha equivalent width of 0.8 keV.  \citet{2011ApJ...729...52L} attempted to update the NGC~3982 Fe K-alpha properties using  archival \chandra{} data and a ZGAUSS model, but they were unable to constrain the parameters. \citet{2012ApJ...758...82L} fit the 0.5-2 keV spectrum with a single APEC and two powerlaw components with the goal of measuring the relative contributions of star formation (APEC) and the AGN (powerlaw) to the soft X-ray luminosity. The two powerlaws had their spectral indices tied together but separate absorption column densities, representing a partial covering geometry where some of the transmitted X-ray emission is absorbed and the rest is scattered along the line of sight. Adopting a absorption column density of $N_{\rm H} = 4.03 \times\ 10^{23}\, {\rm cm}^{-2}$ for the second powerlaw component, they estimated that 15\% of the soft X-ray emission was from the AGN.

Most recently, \citet{2020ApJ...901..161K} fit NGC 3982's \xmm{} and \nustar{} spectra using a PEXMON model and three variants of a MyTORUS model modified according to the procedures in \citet{2012MNRAS.423.3360Y}. The first variant was the standard MyTORUS model, while the other two were decoupled versions where the torus viewing angle was fixed to 90 degrees and the two sides of the torus were modeled separately. One model treated the torus as uniform and the other modeled it as patchy. While the PEXMON model resulted in a Compton-thin column density of $N_{\rm H} = 6 \times 10^{23}\, {\rm cm}^{-2}$, the decoupled MyTORUS models implied significantly higher, Compton-thick values of $N_{\rm H} = 5.3\, \times 10^{24}\, {\rm cm}^{-2}$ for a uniform torus and
$N_{\rm H} = 4.5 \times 10^{24}\, {\rm cm}^{-2}$ for a patchy torus.

\subsubsection{X-ray Observations \& Data Extraction}
\begin{figure}
    \centering
    \plotone{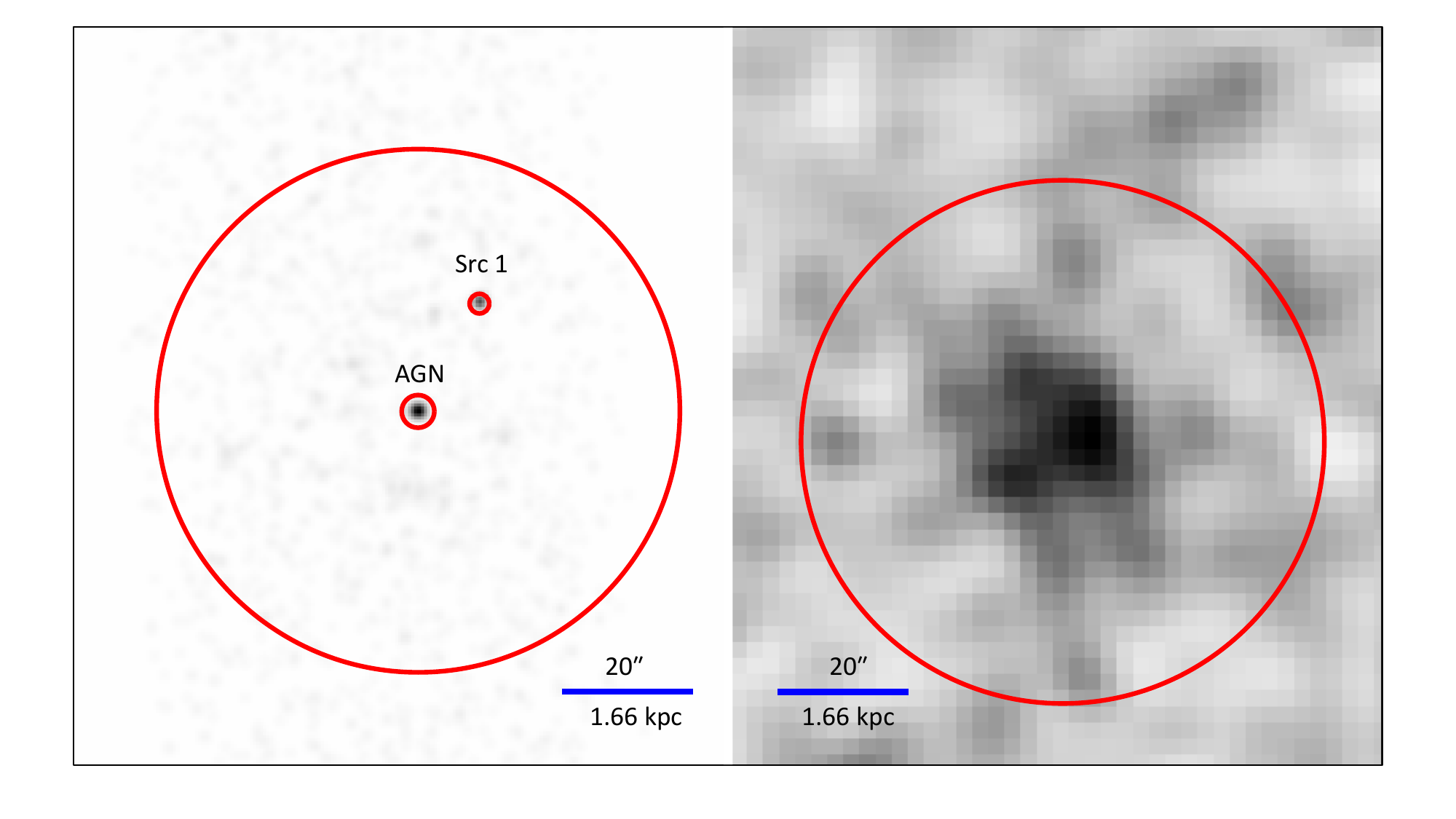}
    \caption{\chandra{} and \nustar{} FPMA images of NGC 3982. The larger, 40\arcsec{} radius circle denotes the \nustar{} extraction region, while the smaller circles denote the \chandra{} extraction regions. An off-nuclear point source (Src 1) is visible in the \chandra{} image, and it was bright enough that it had to be accounted for in the spectral fitting.}
    \label{fig:NGC3982_image}
\end{figure}

NGC 3982 was observed once by \chandra{}, on 2004 January 1 (ObsID: 4845), and once by \nustar{}, on 2017 December 5 (ObsID: 60375001002). The net exposure times were 9.20~ks and 61.67~ks, respectively. In addition to the AGN, Figure~\ref{fig:NGC3982_image}, which presents these images, shows a bright, off-nuclear \chandra\ point source (Source 1) within the \nustar\ beam.  For the \chandra\ data, we used a 2.5\arcsec{} radius circular aperture to extract the AGN, and a 1.5\arcsec{} radius circular aperture to extract Source~1. Since the Source 1 net count rate was more than 10\% that of the AGN, we included it in the X-ray spectral analysis.

\subsubsection{X-ray Spectral Fitting}

\begin{figure}
    \plotone{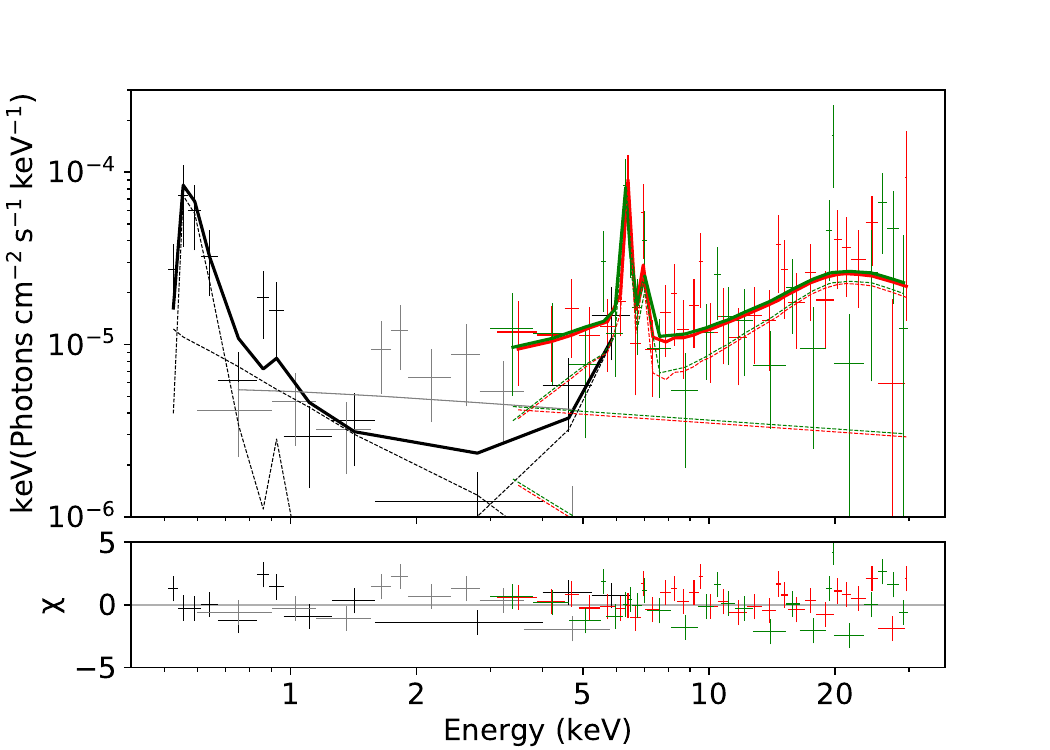}
    \caption{Unfolded spectrum and best-fit model for NGC 3982. Black denotes \chandra{} data of the AGN core. Red and green denote FPMA and FPMB data from the \nustar{} observation of NGC3982. The \chandra{} data of Src 1 is depicted in light grey.}
    \label{fig:NGC3982_uf}
\end{figure}

We started by fitting Source 1's \chandra{} spectrum alone with a simple POWERLAW model, finding best-fit values for its POWERLAW spectral index of $\Gamma = 1.17$ and normalization of $5.36\: \times \: 10^{-6}\, {\rm cts}\, {\rm s}^{-1}\, {\rm keV}^{-1}$. 
We then fit the AGN and Source 1 jointly, freezing the spectral parameters of Source 1 to the best-fit values from \chandra{}. We started with TBABS*(POWERLAW+POWERLAW)  and found C-stat/d.o.f.\ = 479.68/463. 
We then added BORUS (C-stat/d.of.\ = 398.96/459) and APEC (C-stat/d.o.f.\ = 382.89/457) components to the AGN. 
$\mathrm{CF_{tor}}$ was  unconstrained in this fit, but $\mathrm{\cos(\theta_{inc})}$ was  constrained. We froze $\mathrm{cos(\theta_{inc})}$ to its best-fit value of 0.86 before refitting, which allowed us to place a lower limit on $\mathrm{CF_{tor}}$. The final C-stat/d.o.f.\ is 396.15/458. The parameters of this final fit are tabulated in Table \ref{tab:NGC3982params} and the model is plotted over the X-ray data in Figure \ref{fig:NGC3982_uf}. The resulting logarithm of the 2-10 keV luminosity (in units of $\mathrm{erg\:s^{-1}}$) measured from this model is $42.83^{+0.13}_{-0.08}$.

\begin{deluxetable}{@{\extracolsep{10pt}}lcccccccc@{}}
\tablecaption{Parameters for best-fit NGC 3982 model.}\label{tab:NGC3982params}
\tablewidth{2pt}
\tablehead{\colhead{} & \multicolumn{2}{c}{APEC} & \multicolumn{4}{c}{BORUS} & \multicolumn{2}{c}{POWERLAW}\\ \cline{2-3}
\cline{4-7} \cline{8-9} \colhead{Source} & \colhead{$kT$} & \colhead{Norm} & \colhead{log($N_{\rm H}$)} & \colhead{$\mathrm{CF_{Tor}}$} & \colhead{$\mathrm {\cos(\theta_{inc})}$} & \colhead{Norm} & \colhead{$\Gamma$} &\colhead{Norm}\\
 \colhead{} & \colhead{(keV)} & \colhead{($10^{-5}$ cts $\mathrm{s^{-1}\:keV^{-1}}$)} & \colhead{} & \colhead{} & \colhead{} & \colhead{(cts $\mathrm{s^{-1}\:keV^{-1}}$)} & \colhead{} & \colhead{($10^{-6}$ cts $\mathrm{s^{-1}\:keV^{-1}}$)}
}
\startdata
AGN & $0.16\pm{0.03}$ & $2.64^{+1.32}_{-1.07}$ & $\geq{25.3}$ & $\geq{0.92}$& $=0.86$\tablenotemark{a} & $0.15^{+0.05}_{-0.02}$ & $2.48^{+0.06}_{-0.29}$ & $9.64^{+12.0}_{-5.72}$\\
Src 1 & {} & {} & {} & {} & {} & {} & 1.18\tablenotemark{a} & 5.36\tablenotemark{a}
\enddata
\tablecomments{The instrumental normalization constant for the \chandra{} AGN data  $0.53^{+0.41}_{-0.11}$. The normalization constant for the \chandra{} data of Src 1 was $1.07^{+0.30}_{-0.26}$.}
\tablenotetext{a}{Frozen at this value.}
\end{deluxetable}

\subsection{NGC 4501}

NGC 4501 (also known as Messier 88) is a spiral galaxy in the Virgo Cluster \citep{1982A&AS...47..505K}. In the optical, it has been classified as a Seyfert 2 \citep[e.g.,][]{1993ApJS...89....1R, 2006A&A...455..773V} but has occasionally been labeled a LINER \citep[e.g.,][]{1999RMxAA..35..187C,2017MNRAS.469.3405B}. The galaxy has a concurrent starburst based on its MIR spectra \citep{2011MNRAS.414..500H}, though the central regions of the galaxy seem to consist only of evolved stars \citep{2017MNRAS.464..293R,2017MNRAS.469.3405B}. The galaxy is approaching the center of the Virgo cluster and has already become depleted of neutral hydrogen due to ram-pressure stripping \citep[e.g.,][]{2008A&A...483...89V, 2009A&A...502..427V, 2016A&A...587A.108N}.

NGC 4501 is radio-loud \citep{2013ApJS..204...23V} and its  powerlaw SED across the 1-10 $\mathrm{\mu m}$ range of its {\it Spitzer} spectrum could include a contribution from  synchrotron emission from a jet \citep{2013ApJ...764..159L}. \citet{2010ApJ...709.1257T} report that 70\% of its 19 $\mathrm{\mu m}$ emission comes from the AGN. While these results seem to indicate a strong AGN MIR component, the AGN was barely detectable in subarcsecond MIR images from \citet{2014MNRAS.439.1648A}, and it was not detected in the {\it M}-band ($\mathrm{\lambda_{c}=4.66\, \mu m}$) by the Very Large Telescope (VLT) Infrared Spectrometer and Array Camera \citep[ISAAC;][]{2021ApJ...910..104I}. 

In the X-ray, NGC 4501 was first detected by {\it ASCA} \citep{2000ApJ...539..161T}, where its spectrum showed no evidence of heavy absorption or Fe K-alpha emission. \citet{2005ApJ...633...86S} analyzed the \chandra{} observation of NGC 4501 and found it contained multiple X-ray components of equal brightness instead of a dominant hard X-ray component. On this basis they classified NGC 4501 as a non-AGN LINER, though they did note a lack of a dominant hard X-ray component could be caused by absorbing column densities of $\geq 10^{24}\,\mathrm{cm^{-2}}$.  \citet{2012ApJ...758...82L} estimated that  approximately 15\% of the soft (0.5-2 keV) X-ray emission in NGC 4501 was from the AGN. The \xmm{} observations of NGC 4501 were first analyzed in detail by \citet{2006A&A...446..459C}, who found its 0.5-10~keV spectrum could be fit well with a soft thermal component and a powerlaw component. They concurred with \citet{2000ApJ...539..161T} that there was no evidence of heavy absorption.

In contrast to these researchers' conclusions, \citet{2008MNRAS.390.1241B} argued that NGC 4501's \chandra{} observation does indeed show a hard X-ray component coincident with the galaxy's optical nucleus. They fit this hard component using a PEXMON model. Using the hard X-ray component to estimate the bolometric luminosity of the AGN, they concluded that the AGN was more likely heavily obscured than intrinsically faint. They also noted that previous studies using \xmm{} data had been hampered by contamination from extranuclear point sources. 

\subsubsection{X-ray Observations and Data Extraction}

\begin{figure}
    \centering
    \plotone{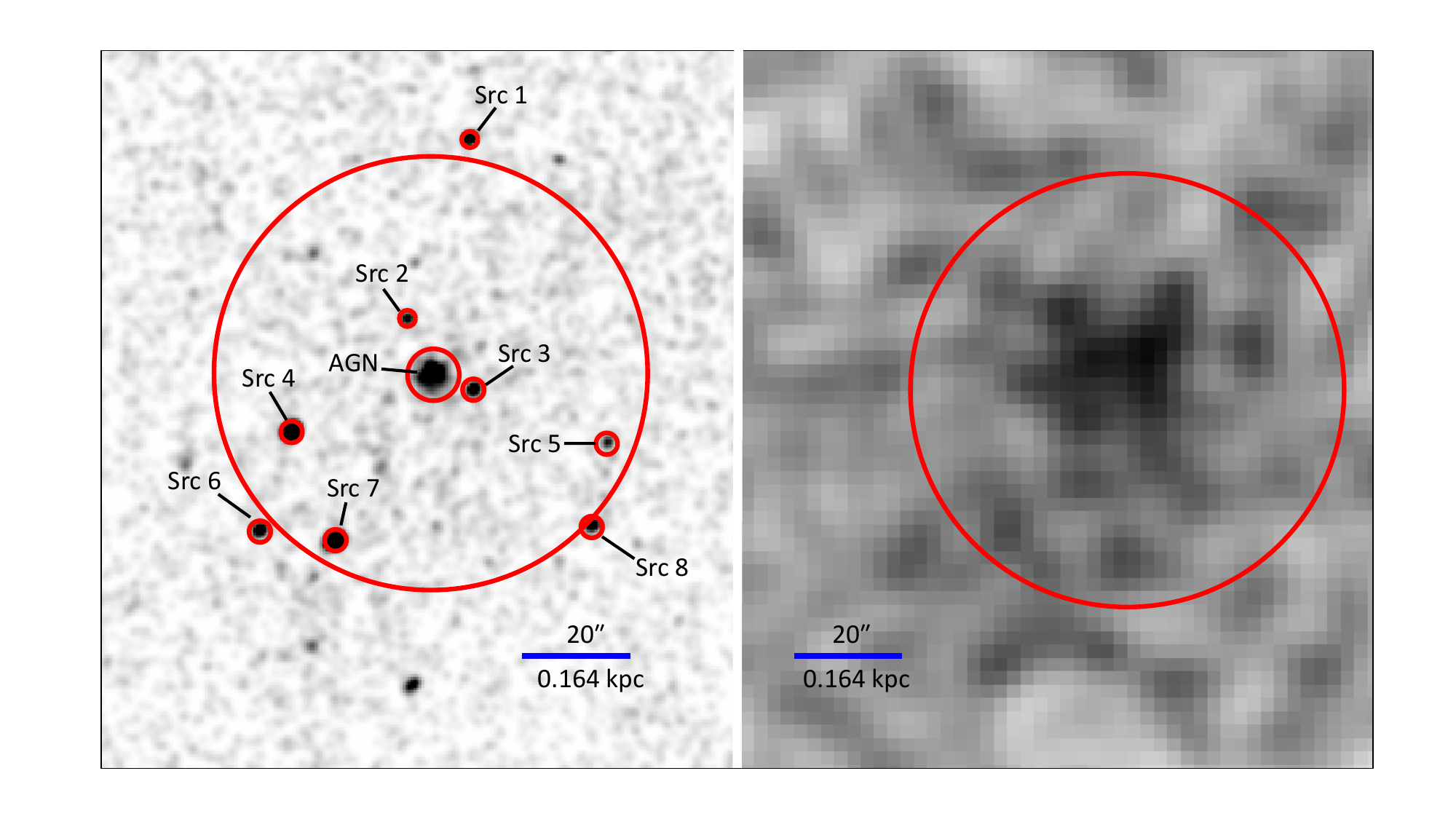}
    \caption{\chandra{} and \nustar{} FPMA (ObsID: 60375002002) images of NGC 4501. The larger, 40\arcsec{} radius circle denotes the \nustar{} extraction region, while the smaller circles denote the \chandra{} extraction regions. Eight off-nuclear point sources are visible in the \chandra{} image;  all but Src 5 are sufficiently bright that they are included in the X-ray spectral fitting.}
    \label{fig:NGC4501_image}
\end{figure}

NGC~4501 was observed twice by \nustar\ and once by \chandra; details, including observation dates and exposure times, are in Table~\ref{tab:xraydata}. 
The \chandra{} and \nustar{} images of NGC~4501 are
presented in Figure~\ref{fig:NGC4501_image}, with  extraction regions overlaid.  The \chandra{} AGN spectrum was extracted with a circular source region 4.78\arcsec{} in radius.  In addition to the AGN core, 8 extra-nuclear sources in the \nustar{} beam are visible in the \chandra{} image. These were extracted with circular source regions 2\arcsec{} in radius from the \chandra\ data, other than the second and eighth sources which were extracted with circular source regions 1.5\arcsec{} in radius. Of these eight point sources, all but the fifth source (Source~5 in the labeled image) had greater than 10\% the count rate of the AGN core, and so they were included in the X-ray spectral fitting.

\subsubsection{X-ray Spectral Fitting}
Because the large number of extra point sources would create too many free parameters for XSPEC to fit, we repeated the procedure we initially attempted in NGC 3627 for the off-nuclear point sources. That is, we fit each off-nuclear source \chandra{} spectrum individually to find the best-fit parameters for its model components. Then, in the joint-fitting step with the \nustar{} data, model parameters were frozen to their best-fit values from \chandra{}. The only free parameter present for each off-nuclear source in the final fitting was its normalization constant.

In the preliminary \chandra{} fitting, most of the off-nuclear sources were best fit by a simple TBABS*POWERLAW model. The exceptions were Source 1 and Source 4, which both required an additional ZTBABS component, and Source 7, which required an additional APEC component.

We began the joint-fitting with a simple fit consisting of a normalization constant, TBABS, and eight powerlaws, one for the AGN and the rest for the off-nuclear sources. The resulting C-stat/d.o.f.\ was 1409.95/1489. We then added the other point-source model components and refit each time: the ZTBABS component on Source 4 (C-stat/d.o.f.\ = 1377.45/1489), the APEC component on Source 7 (C-stat/d.o.f.\ = 1360.06/1489), and the ZTBABS component on Source 1 (C-stat/d.o.f.\ = 1345.19/1489).

\begin{figure}
    \plotone{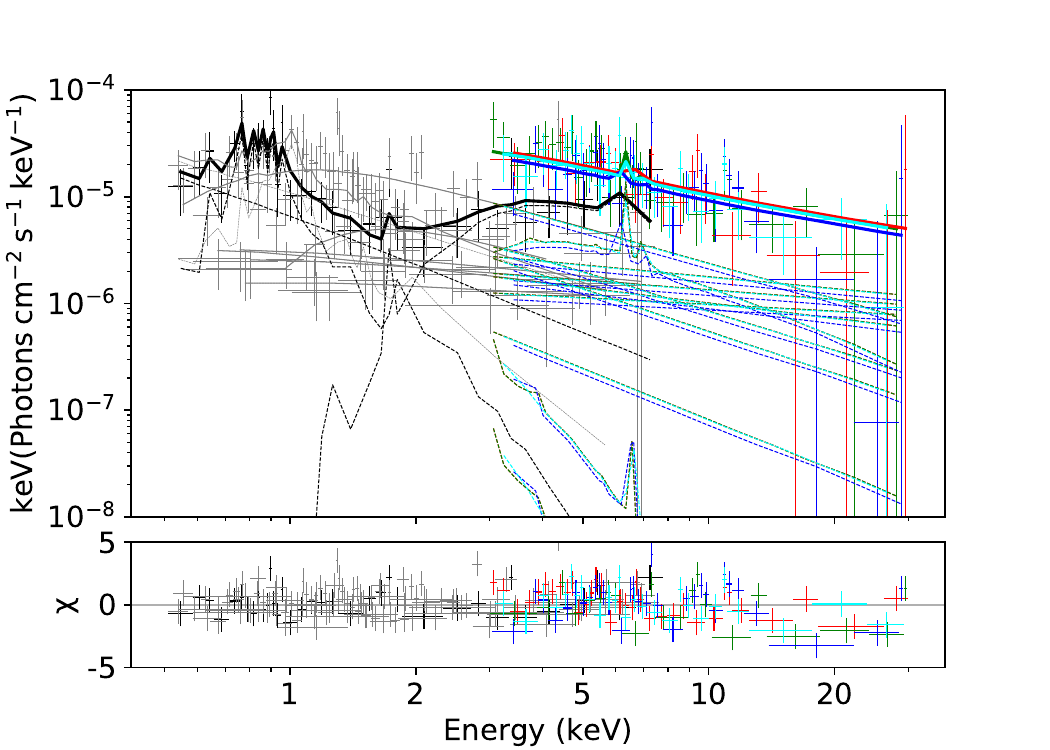}
    \caption{Unfolded spectrum and best-fit model for NGC 4501. Black denotes \chandra{} data and model for the AGN core. Red and green denote FPMA and FPMB data and models for \nustar{} observation 60375002002, while blue and cyan denote FPMA and FPMB data and models for \nustar{} observation 60375002004. The \chandra{} data and models for the extra point sources are depicted in light grey.}
    \label{fig:NGC4501_uf}
\end{figure}

There is a prominent hard component that rises towards 5 keV in the unfolded \chandra{} spectrum of the AGN (Figure \ref{fig:NGC4501_uf}), as would be expected for an Fe K-alpha line created by an obscuring torus along the line of sight. However this hard component is not seen in the \nustar{} data taken 12 years later. This raises two intriguing possibilities. It is possible the AGN has become less luminous in the intervening decade. Modeling all the sources as simple powerlaws, the total \chandra{} 3-8 keV flux within the \nustar\ beam was $1.9\times10^{-13}$ $\mathrm{erg\:cm^{-2}\:s^{-1}}$ with the AGN included, and $1.5\times10^{-13}$ $\mathrm{erg\:cm^{-2}\:s^{-1}}$ without the AGN. The 3-8 keV flux for the \nustar{} observations ranged from $1.4\times 10^{-13}$ $\mathrm{erg\:cm^{-2}\:s^{-1}}$ to $1.6\times10^{-13}$ $\mathrm{erg\:cm^{-2}\:s^{-1}}$. As the 3-8 keV \chandra{} flux without the AGN was always closer to the \nustar{} fluxes than with it included, this raised the potential for luminosity variation in NGC 4501. It is is also possible that the obscuration of NGC 4501 has changed in the intervening time; if it became very heavily obscured, then even the hard X-ray component could be blocked. Neither possibility is out of the question, as AGN are known to sometimes vary in both luminosity \citep[e.g.][]{2015ApJ...800..144L,2017ApJ...835..144G} and obscuration \citep[e.g.][]{2014ApJ...788...76W,2015ApJ...804..107R} over the timescales in question. However, given that 8 point sources other than the AGN are visible in the \nustar{} beam it is also possible that the \nustar{} spectrum is simply contaminated by them, washing out the AGN's hard X-ray component. 

To test the first possibility (that the AGN varied in luminosity) we allowed the normalization of the AGN to freely vary. The AGN spectrum shows a clear soft excess around 1 keV, so we first added an APEC model. We then added a BORUS model to account for the hard component. This rendered $kT$ implausibly large, however, so we set a lower limit of 0.1 keV and an upper limit of 2.0 keV on $kT$. Because the \nustar{} data do not show a reflection/torus component, the inclination angle and covering factor of the AGN torus cannot be measured with much accuracy. For this reason we froze the BORUS $\mathrm{CF_{Tor}}$ parameter to 0.5, and the BORUS $\mathrm {\cos(\theta_{inc})}$ parameter to 0.17 (corresponding to an inclination angle of 80 deg). The final fit had a C-stat/d.o.f.\ of 1296.11/1485. The parameters of this fit are tabulated in Table \ref{tab:NGC4501params}. The cross-calibration coefficient of the AGN in this fit was $2.66^{+1.33}_{-0.95}$, which includes 1.71 within its 90\% confidence interval. This is not an extreme value for this coefficient to take. As such the claim that the AGN decreased in luminosity cannot be made with confidence. However, this still leaves open the possibility of the obscuration varying between the time of the \chandra{} observation and the time of the \nustar{} observations.

To test this second possibility (that the AGN varied in obscuration), we untied the \chandra{} and \nustar{} values of the BORUS parameter $N_{\rm{H}}$ from each other, but did not allow the AGN to vary in luminosity between the \chandra{} and \nustar{} data. After refitting with these changes, the C-stat/d.o.f. was 1296.11/1485. The value of log($N_{\rm{H}}/\rm{cm}^{-2}$) in this model changed from $22.91^{+0.28}_{-0.21}$ in the \chandra{} observation to $22.43^{+0.28}_{-0.24}$ in the \nustar{} observations, too similar to explain the lack of appearance of a hard component in the \nustar{} data. Considering this and the negligible improvement in C-stat/d.o.f.\ if we let the obscuration vary instead of the luminosity, we conclude that there is no evidence in our data of NGC 4501 obscuration variability. 

Given that we have no strong evidence of either luminosity or obscuration variability in this AGN, the most parsimonious explanation for the lack of a hard component in the \nustar{} data is contamination from the eight extra point sources. It should be noted, however, that the possibility of variability cannot be ruled out with this data. The best fit with the cross normalization constant on the AGN left to freely vary is plotted in Figure \ref{fig:NGC4501_uf}. The logarithm of the resulting 2-10 keV luminosity (in units of $\mathrm{erg\:s^{-1}}$) measured from the model is $41.50^{+0.25}_{-0.11}$.

\begin{deluxetable}{@{\extracolsep{10pt}}lccccccccc@{}}
\tablecaption{Parameters for best-fit NGC4501 model.}\label{tab:NGC4501params}
\tablewidth{2pt}
\tablehead{\colhead{} & \colhead{ZTBABS} & \multicolumn{2}{c}{APEC} & \multicolumn{4}{c}{BORUS} & \multicolumn{2}{c}{POWERLAW}\\ 
\cline{2-2} \cline{3-4} \cline{5-8} \cline{9-10} \colhead{Source} & \colhead{$N_{\rm H}$} & \colhead{$kT$} & \colhead{Norm} & \colhead{log($N_{\rm H}$)} & \colhead{$\mathrm{CF_{Tor}}$} & \colhead{$\mathrm {\cos(\theta_{inc})}$} & \colhead{Norm}  & \colhead{$\Gamma$} &\colhead{Norm}\\
 \colhead{} & \colhead{($\mathrm{10^{22}\:cm^{-2}}$)} & \colhead{(keV)} & \colhead{} & \colhead{} & \colhead{} & \colhead{} & \colhead{} & \colhead{} & \colhead{}
}
\startdata
AGN & {} & $0.75\pm{0.11}$ & $2.65^{+2.56}_{-1.48}$ & $22.87^{+0.25}_{-0.15}$ & 0.5\tablenotemark{a} & 0.17\tablenotemark{a} & $2.03^{+1.60}_{-0.44}$ & $\geq1.98$ & $3.24^{+3.30}_{-1.83}$\\
Src 1 & 0.72\tablenotemark{a} & {} & {} & {} & {} & {} & {}  & 2.25\tablenotemark{a} & 15.2\tablenotemark{a}\\
Src 2 & {} & {} & {} & {} & {} & {} & {} & 1.21\tablenotemark{a} & 1.59\tablenotemark{a}\\
Src 3 & {} & {} & {} & {} & {} & {} & {} & 1.48\tablenotemark{a} & 3.05\tablenotemark{a}\\
Src 4 & 0.21\tablenotemark{a} & {} & {} & {} & {} & {} & {} & 2.12\tablenotemark{a} & 31.4\tablenotemark{a}\\
Src 6 & {} & {} & {} & {} & {} & {} & {} & 1.37\tablenotemark{a} & 4.17\tablenotemark{a}\\
Src 7 & {} & 1.09\tablenotemark{a} & 7.25\tablenotemark{a} & {} & {} & {} & {} &  2.33\tablenotemark{a} & 11.7\tablenotemark{a}\\
Src 8 & {} & {} & {} & {} & {} & {} & {} & 1.30\tablenotemark{a} & 2.68\tablenotemark{a}
\enddata
\tablecomments{The instrumental normalization constants for the \chandra{} data, in the order of sources from the table, are $2.66^{+1.33}_{-0.95}$, $1.03^{+0.27}_{-0.23}$, $1.07^{+0.43}_{-0.34}$, $1.01^{+0.31}_{-0.26}$, $1.00^{+0.12}_{-0.11}$, $0.78^{+0.23}_{-0.19}$, $1.00^{+0.13}_{-0.12}$, and $1.01^{+0.32}_{-0.27}$.
The \nustar{} normalization constants for ObsID 60375002004A are $0.88^{+0.15}_{-0.14}$ for FPMA and $0.98^{+0.18}_{-0.16}$ for FPMB. The normalizations for the model components are in units of $10^{-6}$ cts $\mathrm{s^{-1}\:keV^{-1}}$ for APEC, $10^{-3}$cts $\mathrm{s^{-1}\:keV^{-1}}$ for BORUS, and $10^{-6}$ cts $\mathrm{s^{-1}\:keV^{-1}}$ for POWERLAW.}
\tablenotetext{a}{Frozen at this value.}
\end{deluxetable}

\subsection{IC 3639}
IC 3639 is a barred spiral galaxy containing a Seyfert 2 nucleus, as well as a nuclear starburst within the central 80~pc of the galaxy \citep{1998ApJ...505..174G, 2018A&A...611A..46F}. It is part of a compact group of galaxies, though it lacks features indicative of recent mergers or interactions \citep{2001MNRAS.324..859B}. IC~3639 has polarized broad H$\mathrm{\alpha}$ emission, though the nature of that emission is uncertain:  some researchers considerate it as indicative of a hidden broad-line region \citep{1997Natur.385..700H, 2001MNRAS.327..459L, 2003ApJS..148..353T}, while others claim it is a kinematic feature of the narrow-line region \citep{2007ApJ...656..105G}.  MIR interferometry reveals a compact, sub-arcsecond, unresolved nuclear point source \citep{2014MNRAS.439.1648A, 2016ApJ...822..109A} surrounded by a halo of MIR emission associated with the compact nuclear starburst \citep{2018A&A...611A..46F}. The starburst contributes 70\% of the observed MIR flux.

The first published X-ray observations of IC~3639 suggested that it possessed a very high hydrogen column density and a strong Fe K$\mathrm{\alpha}$ line \citep{1999MmSAI..70...73R}. A more detailed analysis of \chandra{}, \suzaku{}, and \nustar{} data by \citet{2016ApJ...833..245B} confirmed it has a hydrogen column density of $\mathrm{10^{25}\,cm^{-2}}$ and an extreme Fe K$\mathrm{\alpha}$ equivalent width of 2.29 keV. They also found it has a 2-10 keV luminosity well below the expected value based on the luminosity of its [O III] line, assuming the relations of \citet{2006A&A...455..173P} and \citet{2015MNRAS.454.3622B}. Overall, \citet{2016ApJ...833..245B} conclude that IC 3639 is a Compton-thick AGN possessing an active central engine generating a strong reflection component in its X-ray spectrum. 

\subsubsection{X-Ray Observations \& Data Extraction}

\begin{figure}
    \centering
    \plotone{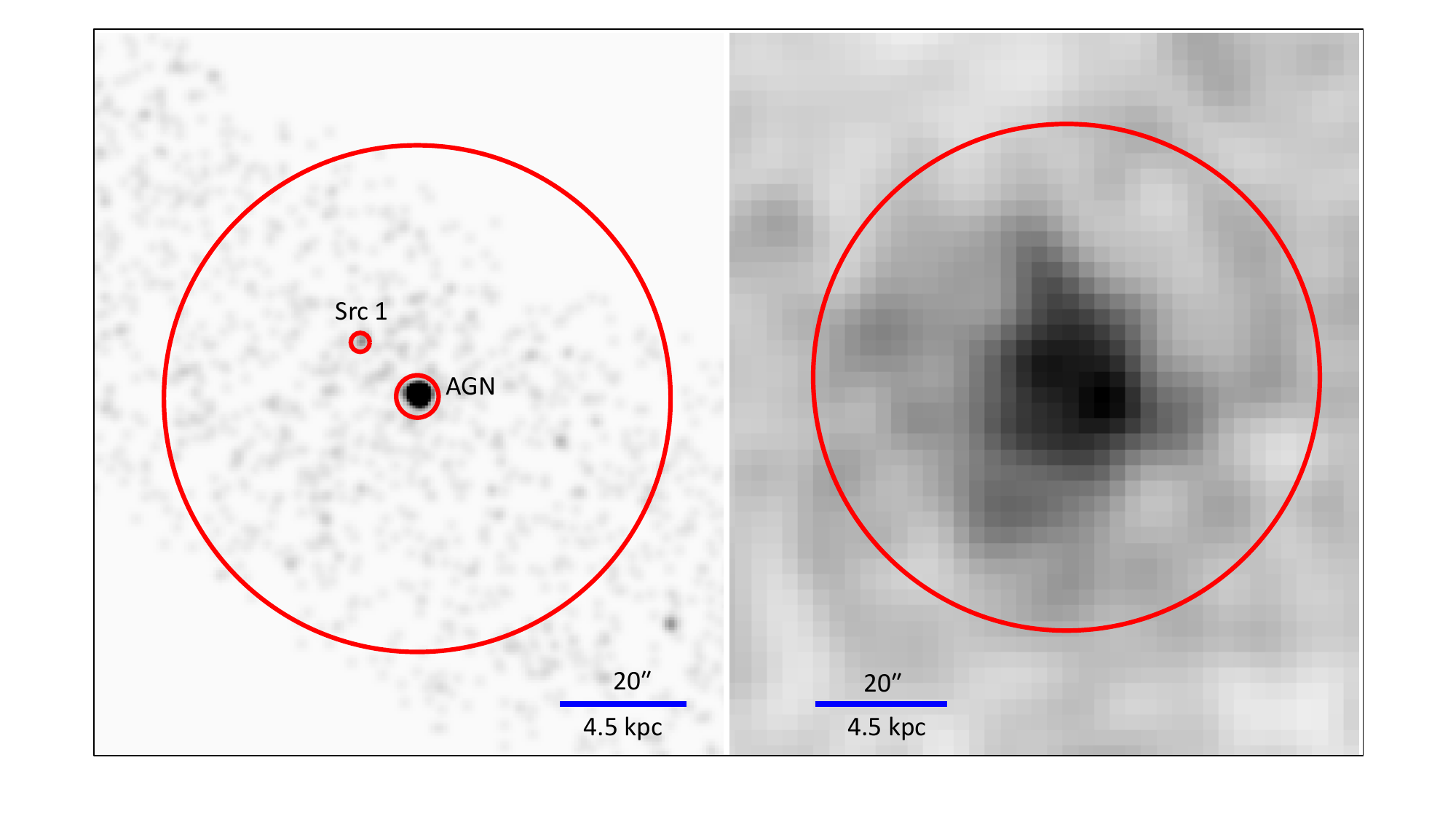}
    \caption{\chandra{} and \nustar{} FPMA images of IC 3639. The larger, 40\arcsec{} radius circle denotes the \nustar{} extraction region, while the smaller circles denote the \chandra{} extraction regions. An off-nuclear point source (Src 1) is visible in the \chandra{} image, but is sufficiently faint to be ignored in the X-ray spectral fitting.}
    \label{fig:IC3639_image}
\end{figure}

IC~3639 was observed once by both \nustar\ and \chandra. The observation dates and exposure times are in Table~\ref{tab:xraydata} and Figure~\ref{fig:IC3639_image} presents the images. The higher resolution \chandra{} data reveal a faint, off-nuclear point source (labeled ``Src 1'' in Figure \ref{fig:IC3639_image}) as well as the AGN in the 40\arcsec\ radius \nustar{} beam. The \chandra\ AGN spectrum was extracted with a circular region of 3.35\arcsec{} radius, while Source 1 was extracted with a 1.5\arcsec{} radius region. Since Source~1 has $\leq$10\% the net count rate of the AGN, its spectrum is not used in the spectral fitting.

\subsubsection{X-Ray Spectral Fitting}
We first fit the \chandra{} and \nustar{} data jointly with a simple absorbed powerlaw (TBABS*POWERLAW) model. The resulting C-stat/d.o.f.\ was 1062.90/830, indicating a poor fit. 

\begin{figure}
    \plotone{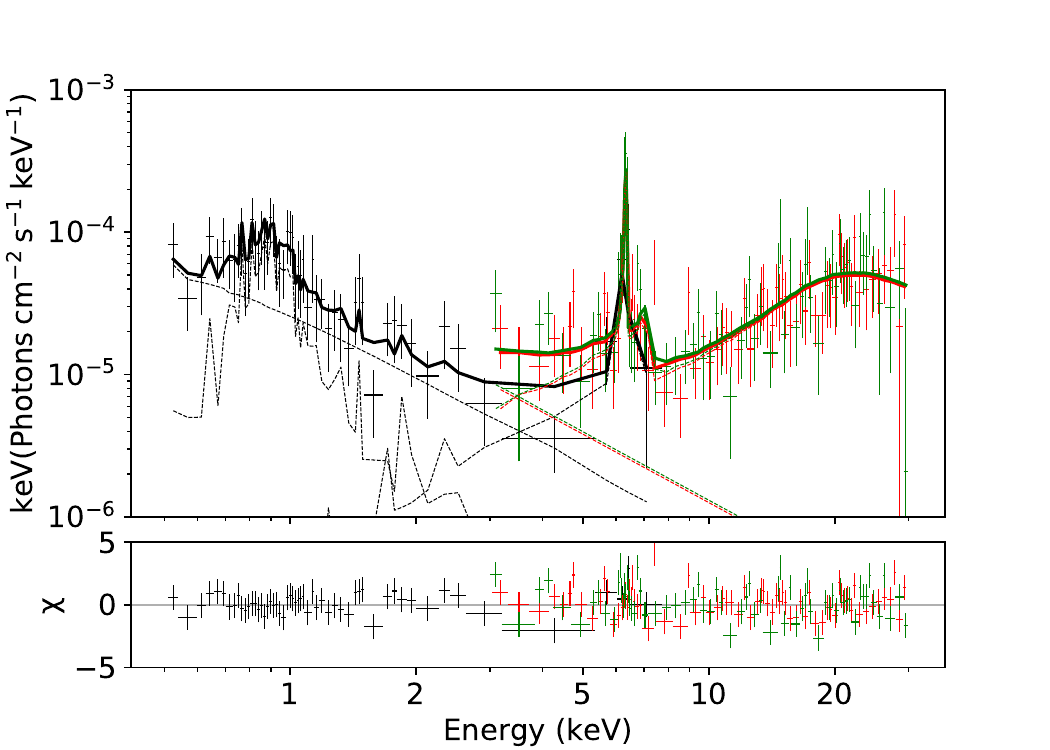}
    \caption{Unfolded spectrum and best-fit model for IC 3639. Black denotes \chandra{} data, green denotes \nustar{} FPMA data, and red denotes \nustar{} FPMB data.}
    \label{fig:IC3639_uf}
\end{figure}

Looking at the unfolded spectrum for IC 3639 (Figure \ref{fig:IC3639_uf}), an extremely strong Fe K$\mathrm{\alpha}$ line can be seen around 6.4 keV. The unfolded spectrum also shows a substantial rise from 10-20 keV, with a pronounced Compton hump at 20 keV. We therefore added a BORUS component to the initial TBABS*POWERLAW model.

Prominent residuals remained at 0.5-2.0 keV, so we also added an APEC component. 
The resulting best fit model has C-stat/d.o.f.\ = 606.54/824. The parameter values for the best fit model are tabulated in Table \ref{tab:IC3639params}. Since the upper error bar for $\mathrm{CF_{tor}}$, the lower error bar for $\mathrm{\cos(\theta_{inc})}$, and the lower error bar for the BORUS normalization were less than 0.005 in value, we have rounded them up to 0.01.  The best fit model is plotted over the unfolded spectrum in Figure \ref{fig:IC3639_uf}.
The logarithm of the 2-10 keV luminosity (in units of $\mathrm{erg\:s^{-1}}$) measured from the model is $43.07^{+0.18}_{-0.12}$.

\begin{deluxetable}{@{\extracolsep{10pt}}lccccccc@{}}
\tablecaption{Parameters for best-fit IC 3639 model.}\label{tab:IC3639params}
\tablewidth{2pt}
\tablehead{
\multicolumn{2}{c}{APEC} & \multicolumn{4}{c}{BORUS} & \multicolumn{2}{c}{Powerlaw}\\
\cline{1-2} \cline{3-6} \cline{7-8} \colhead{$kT$} & \colhead{Norm} & \colhead{$\log({N_{\rm H}})$} & \colhead{$\mathrm{CF_{Tor}}$} & \colhead{$\mathrm{\cos(\theta_{inc})}$} & \colhead{Norm} & \colhead{$\Gamma$} &\colhead{Norm}\\
\colhead{(keV)} & \colhead{($10^{-5}$ cts $\mathrm{s^{-1}\:keV^{-1}}$)} & \colhead{} & \colhead{} & \colhead{} & \colhead{(cts $\mathrm{s^{-1}\:keV^{-1}}$)} & \colhead{} & \colhead{($10^{-5}$ cts $\mathrm{s^{-1}\:keV^{-1}}$)}
}
\startdata
$0.85^{+0.12}_{-0.12}$ & $3.06^{+0.69}_{-0.64}$ & $25.00^{+0.06}_{-0.26}$ & $0.87^{+0.01}_{-0.12}$ & $0.77^{+0.07}_{-0.01}$ & $0.04^{+0.02}_{-0.01}$ & $\geq2.4$ & $5.01^{+2.59}_{-0.81}$
\enddata
\tablecomments{Error bars shown are for 90\% confidence intervals. The \chandra{} instrumental normalization constant value was $0.59^{+0.24}_{-0.06}$.}

\end{deluxetable}

\subsection{NGC 4922}
NGC 4922 is a pair of galaxies in the late stages of a merger \citep{2017MNRAS.468.1273R}. The northern galaxy has been classified as a luminous infrared galaxy \citep{2010ApJ...723..993D} and a Seyfert~2 \citep{2010ApJ...709..884Y}. It is also a water megamaser \citep{2004ApJ...617L..29B}. The southern galaxy is an elliptical galaxy with no obvious signs of activity \citep{1999MNRAS.302..561A}.

In the X-rays, NGC 4922 was first studied in detail with \rosat{}, which detected extended soft X-ray emission across the entire merging pair \citep{1999MNRAS.302..561A}. Further observations by \citet{2017MNRAS.468.1273R} revealed the northern galaxy is brighter in X-rays, with the southern galaxy's nucleus only detectable in the 0.3-2 keV band by \chandra{}, and it was not detected by\nustar{}.  Based on joint analysis of \chandra{} and \nustar{} observations, \citet{2017MNRAS.468.1273R} reported the northern galaxy to be Compton-thick, with $N_{\rm H}>4.27\:\times\:10^{24}\:{\rm cm}^{-2}$.

\subsubsection{X-Ray Observations and Data Extraction}
\begin{figure}
    \centering
    \plotone{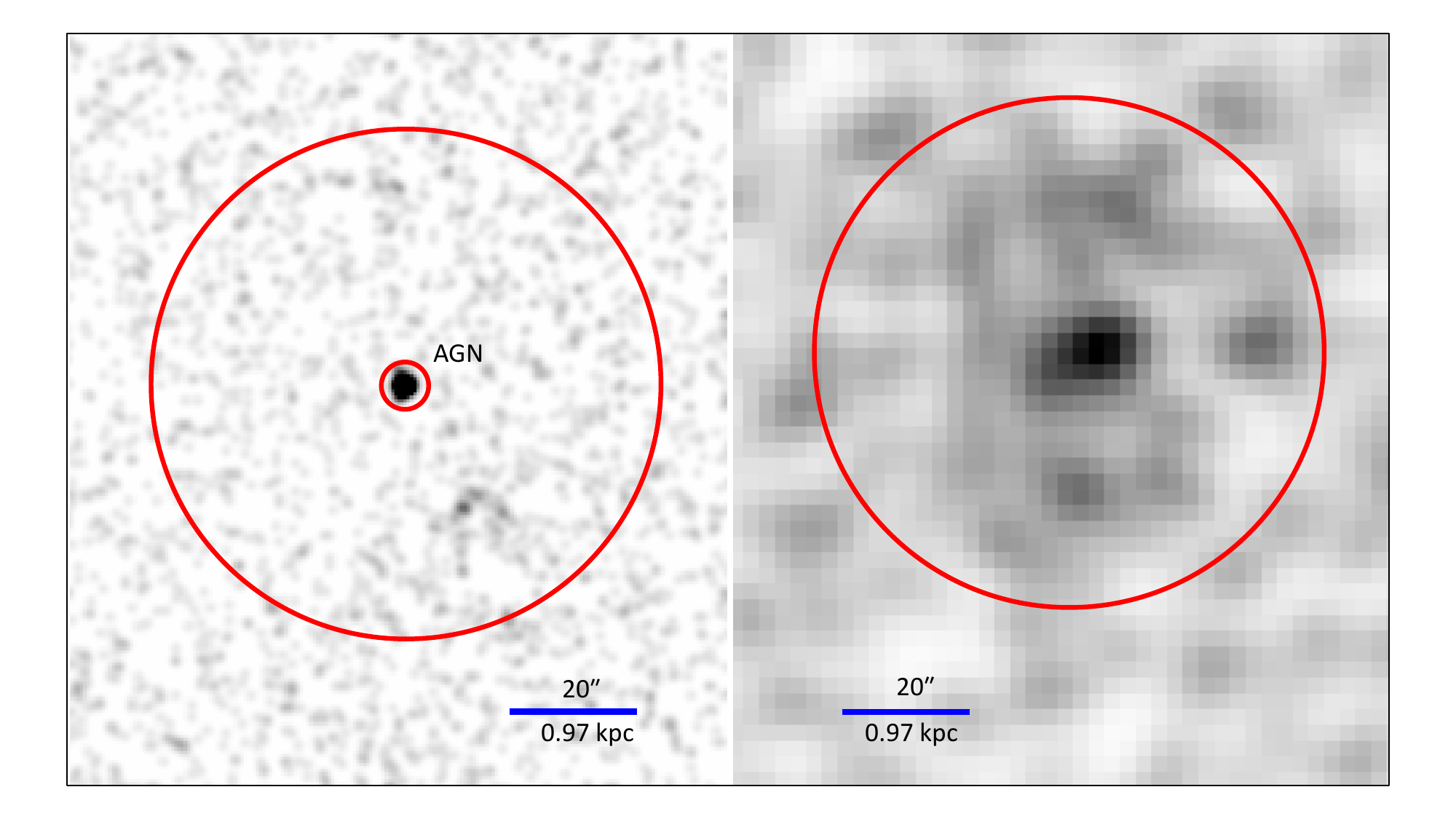}
    \caption{\chandra{} (ObsID: 15065) and \nustar{} FPMA images of NGC 4922. The larger, 40\arcsec{} radius circle denotes the \nustar{} extraction region, while the smaller circle denotes the \chandra{} extraction region.}
    \label{fig:NGC4922_image}
\end{figure}

NGC~4922 was observed once by \nustar\ and three times by \chandra; the observation dates and exposure times are in Table~\ref{tab:xraydata}.  The \chandra{} spectra were extracted using circular source regions 3.7\arcsec{} in radius. Figure \ref{fig:NGC4922_image} shows the second \chandra\ observation and the \nustar\ FPMA observation with the extraction regions overlaid.

\subsubsection{X-Ray Spectral Fitting}
We first fit the \chandra{} and \nustar{} data jointly with a simple TBABS*POWERLAW fit, using a Galactic hydrogen column density of $N_{\rm{H}}^{\rm{Gal}}\rm{=1.06\times{}10^{20}\,cm^{-2}}$. The resulting C-stat/d.o.f.\ was 407.61/423. 

\begin{figure}
    \plotone{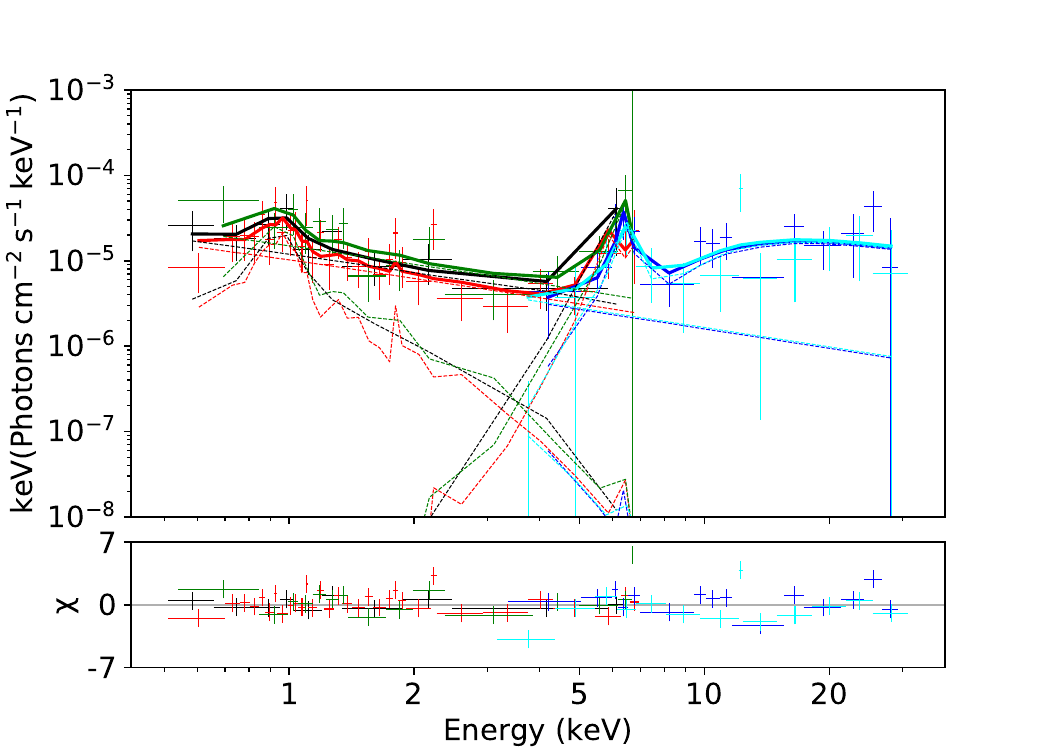}
    \caption{Unfolded spectrum and best-fit model for NGC 4922. Black, red and green denote \chandra{} data (ObsIDs 4775, 15065, 18201), while blue and cyan denote FPMA and FPMB data.}
    \label{fig:NGC4922_uf}
\end{figure}

The unfolded spectrum (Figure \ref{fig:NGC4922_uf}) shows a less prominent Compton rise than some of the other galaxies in the sample (e.g., NGC 1386), but it is present. A presumed Fe K$\mathrm{\alpha}$ line is also present at 6.4 keV. While the signal-to-noise is lower than in the aforementioned galaxies, NGC 4922 nonetheless shows the features typical of Compton-thick AGN. We therefore added a BORUS component to the initial TBABS*POWERLAW fit, fixing the BORUS spectral index to the POWERLAW spectral index. This resulted in a C-stat/d.o.f.\ of 338.44/414. An excess of soft X-ray emission was present from 0.5-2.0 keV, so an APEC component was added, resulting in C-stat/d.o.f.\ = 324.51/412. Because $\mathrm{\cos(\theta_{inc})}$ was completely unconstrained with these model components, its value was frozen at 0.17 (or $\mathrm{\theta_{inc}}\:\approx\:80\degree$), representing a near edge-on line of sight. We then refit the model. The final C-stat/d.o.f.\ was 321.07/412.
The resulting values for each of the model parameters are shown in Table \ref{tab:NGC4922params}. The model is plotted over the \chandra{} and \nustar{} data as the solid lines in Figure \ref{fig:NGC4922_uf}. The logarithm of the 2-10 keV luminosity (in units of $\mathrm{erg\:s^{-1}}$) measured from the model is $42.29^{+0.12}_{-0.47}$.

\begin{deluxetable}{@{\extracolsep{10pt}}lccccccc@{}}
\tablecaption{Parameters for best-fit NGC 4922 model.}\label{tab:NGC4922params}
\tablewidth{2pt}
\tablehead{\multicolumn{2}{c}{APEC} & \multicolumn{4}{c}{BORUS} & \multicolumn{2}{c}{POWERLAW}\\
\cline{1-2} \cline{3-6} \cline{7-8} \colhead{$kT$} & \colhead{Norm} & \colhead{
$\log({N_{\rm H}}$)} & \colhead{$\mathrm{CF_{Tor}}$} & \colhead{$\mathrm {\cos(\theta_{inc})}$} & \colhead{Norm} & \colhead{$\Gamma$} &\colhead{Norm}\\
 \colhead{(keV)} & \colhead{($10^{-6}$ cts $\mathrm{s^{-1}\:keV^{-1}}$)} & \colhead{} & \colhead{} & \colhead{} & \colhead{($10^{-4}$ cts $\mathrm{s^{-1}\:keV^{-1}}$)} & \colhead{} & \colhead{($10^{-6}$ cts $\mathrm{s^{-1}\:keV^{-1}}$)}
}
\startdata
$1.06^{+0.34}_{-0.21}$ & $4.71^{+6.02}_{-2.71}$ & $23.89^{+0.11}_{-0.17}$ & $\geq0.25$& $=0.17$\tablenotemark{a} & $3.96^{+1.02}_{-2.50}$ & $1.75\pm{0.34}$ & $9.04^{+6.72}_{-3.67}$
\enddata
\tablecomments{The \chandra{} instrumntal normalization constant values were $1.35^{+0.75}_{-0.52}$ (ObsID: 4775), $1.17^{+0.56}_{-0.41}$ (ObsID: 15065), and $1.69^{+0.87}_{-0.62}$ (ObsID: 18201).}
\tablenotetext{a}{Frozen at this value.}
\end{deluxetable}

\subsection{NGC 5005}

NGC 5005 is a weakly barred spiral galaxy with a nucleus that is heavily shrouded in dust \citep{2000ApJ...532..323P}. Its AGN is known to be variable over timescales of months \citep{2012A&A...539A.104Y}. NGC 5005's optical classification has been ambiguous. \citet{1981ApJ...250...55S} were able to identify H$\mathrm{\alpha}$, [\ion{S}{2}], [\ion{O}{2}], and [\ion{O}{3}] emission lines in its nuclear spectrum, but no others. They did not specify a classification for it but regarded it as unlikely to be a Seyfert 2. Later papers in the literature have classified it as a LINER \citep[e.g.,][]{1992ApJ...393...90H, 1997ApJS..112..315H, 2006A&A...455..773V}, a Seyfert 2 \citep[e.g.,][]{2017MNRAS.464.2139A}, or both a LINER and a Seyfert 2 at the same time \citep[e.g.,][]{2006ApJS..166..498S, 2017ApJ...846..102M}.  Palomar spectra for NGC 5005 show a broad H$\alpha$ component blended with narrow H$\alpha$ and [\ion{N}{2}] emission \citep{1996ApJ...471..190R, 1997ApJS..112..315H}, suggesting NGC 5005 is an unobscured AGN. However, \citet{2014A&A...563A.119B} were unable to find a broad H$\alpha$ component in later {\it Hubble} spectroscopy when using the [\ion{O}{1}] line as a template for deblending, and therefore concluded that either the broad H$\alpha$ detection in the Palomar data was spurious, or NGC~5005 is a changing-look AGN. \citet{2015ApJ...814..149C} did, in contrast, identify a broad H$\alpha$ line in the {\it Hubble} spectra when using the [\ion{S}{2}] line as a template for deblending, measuring a broad H$\alpha$ component with FWHM of $2610\, {\rm km}\, {\rm s}^{-1}$.  A detailed analysis of new ground-based spectra as well as the archival {\it Hubble} spectra for NGC 5005 was published by \citet{2018MNRAS.480.1106C}, who found a broad H-alpha component in the {\it Hubble} spectra, blended with [\ion{S}{2}] and [\ion{N}{2}].  The broad H$\alpha$ component had a FWHM of $2152\, {\rm km}\, {\rm s}^{-1}$, was very weak, and was not visible in their ground-based spectra.

NGC 5005's core is embedded in extended MIR emission that appears to trace out its spiral structure \citep{2014MNRAS.439.1648A}. Based on {\it Spitzer} data, \citet{2010ApJ...709.1257T} estimated that only 44\% of its 19$\mathrm{\mu m}$ emission is from an AGN. Based on the NIR [\ion{Fe}{2}] and [\ion{P}{2}] forbidden line flux ratios, \citet{2016ApJ...833..190T} found that, unusually, NGC~5005's narrow line region seems to be predominantly shock-ionized rather than UV-ionized.

In the X-rays, NGC 5005 was first detected by {\it ASCA}, where its spectrum was analyzed by \citet{1999ApJ...522..157R}. They reported a hydrogen column density of $N_{\rm H}>10^{24}\,\rm{cm}^{-2}$, implying a Compton-thick AGN. Further evidence of NGC 5005's Compton-thick nature comes from the unusually low ratio between its observed 2-10~keV X-ray and [\ion{O}{3}] luminosities. \citet{1999ApJ...522..157R} note, however, that NGC~5005 showed no evidence of a reflection component in its {\it ASCA} spectrum, with an upper limit of 0.9~keV on the equivalent width of the Fe K$\mathrm{\alpha}$ line. They concluded that the hydrogen column density was so thick that the soft X-ray emission from the AGN was completely absorbed, leaving only extended emission from a concurrent starburst to create the {\it ASCA} spectrum. 

Observations by \chandra{} and \xmm{} revealed new features of NGC~5005's X-ray emission. The AGN core was found to be embedded in a background of extended X-ray emission that follows the contours of the galaxy \citep{2005MNRAS.356..295G}, and that might be responsible for a large soft excess observed in its 0.6-1~keV X-ray spectrum \citep{2006MNRAS.365..688G}. \citet{2005MNRAS.356..295G} concluded the X-ray spectrum was unlikely to be dominated by an inverse Compton component, and placed an upper limit on the equivalent width of an Fe K$\mathrm{\alpha}$ line of $\leq 0.24$~keV. In contrast to \citet{1999ApJ...522..157R}, they measured $N_{\rm H} \simeq 1.5\: \times\: 10^{20}\: {\rm cm}^{-2}$. Furthermore, their search of the available literature at the time \citep[e.g.,][]{1981ApJ...250...55S, 1988ApJS...67..249D, 1997ApJS..112..315H} revealed a wide range of reported [\ion{O}{3}] fluxes for NGC 5005, some of which were not overluminous compared to the X-ray flux. They therefore claimed NGC 5005 was misidentified as a Compton-thick AGN.  These conclusions were further reinforced by later analyses of the \chandra{} and \xmm{} observations, with values of $N_{\rm H}$ closer to $\mathrm{10^{20}\:cm^{-2}}$ \citep{2011MNRAS.414.3084B} or $\mathrm{10^{21}\:cm^{-2}}$ \citep{2012A&A...539A.104Y} than to Compton-thick column densities.  In summary, the latest analyses of optical and X-ray observations of NGC 5005 suggest that it might be intrinsically underluminous rather than heavily obscured.

\subsubsection{X-ray Observations and Data Extraction}

\begin{figure}
    \centering
    \plotone{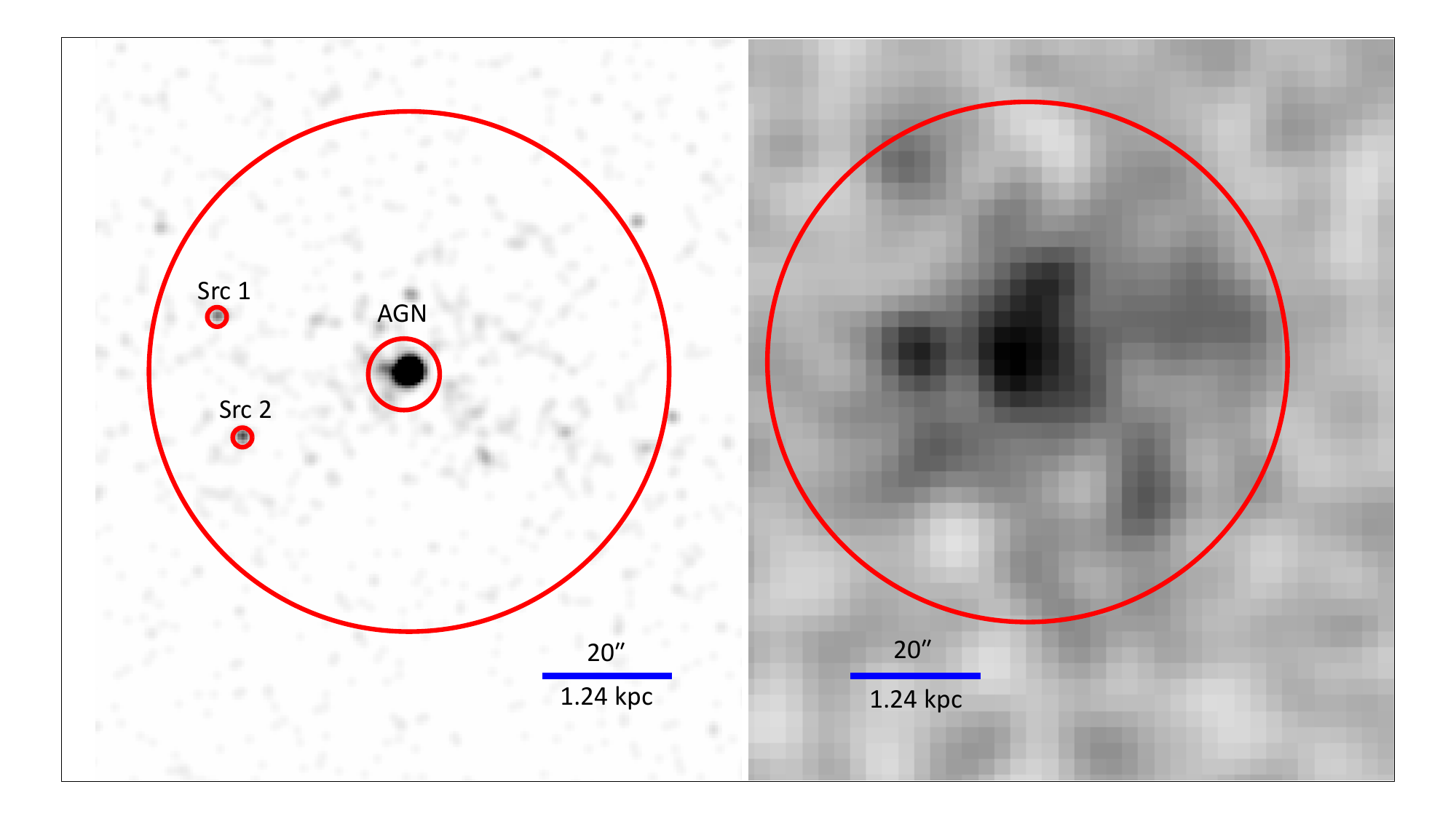}
    \caption{\chandra{} and \nustar{} FPMA images of NGC 5005. The larger, 40\arcsec{} radius circle denotes the \nustar{} extraction region, while the smaller circles denote the \chandra{} extraction regions. Two off-nuclear point sources (Src 1 and Src 2) are visible in the \chandra{} image, but were faint enough to be ignored in the spectral fitting.}
    \label{fig:NGC5005_image}
\end{figure}

NGC 5005 has been observed once each by \nustar, \chandra, and \xmm; details of the observations, including the observation dates and exposure times are in Table~\ref{tab:xraydata}.  \chandra{} and \nustar{} images of the galaxy are presented in Figure~\ref{fig:NGC5005_image}. The \chandra{} AGN spectrum was extracted from a 5.5\arcsec{} radius circular source region. Two extranuclear \chandra\ point sources (labeled ``Src 1'' and ``Src 2'') are visible within the \nustar{} beam, which we extracted with  1.5\arcsec{} radius circular source regions. Since their count rates were less than 10\% the count rate of the AGN, they were ignored in the X-ray spectral fitting. 

Since no background flares were evident in the \xmm\ 10-12~keV lightcurve of NGC 5005, the EPIC pn spectrum was extracted from the full dataset. We used 30\arcsec\ circular source regions with 60\arcsec{} radius background regions. For the MOS data, we filtered out times with high background, defined as times when the 10-12~keV count rate was $> 0.35\, {\rm ct}\, {\rm s}^{-1}$. Using patterns 0-12, we extracted the MOS source spectra with 30\arcsec{} radius circular regions and 50-80\arcsec\ annular background regions.

\subsubsection{X-ray Spectral Fitting}
\begin{figure}
    \plotone{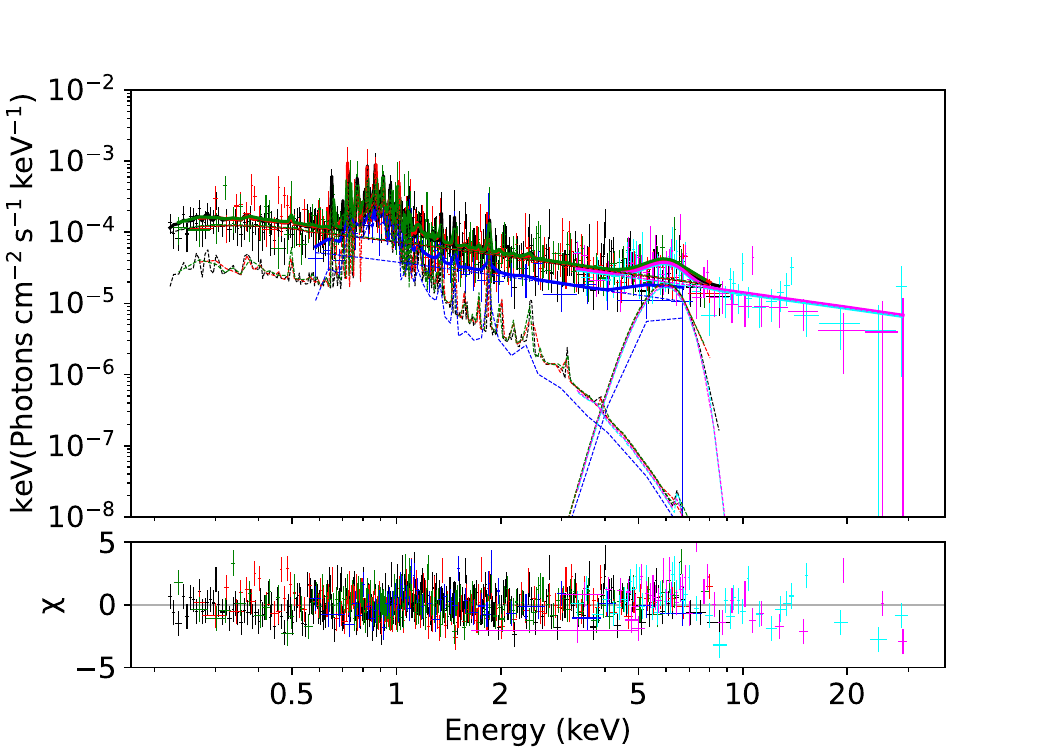}
    \caption{Unfolded spectrum and best-fit model for NGC 5005. Black denotes \xmm{} pn data, red denotes \xmm{} MOS1 data, green denotes \xmm{} MOS2 data, blue denotes \chandra{} data, cyan denotes FPMA data, and magenta denotes FPMB data.}
    \label{fig:NGC5005_uf}
\end{figure}

We initially began our analysis with the \chandra{} and \nustar{} data only. We started with a simple TBABS*POWERLAW fit, with the Galactic hydrogen column density set to $N_{\rm{H}}^{\rm{Gal}}\rm{=1.17\times{}10^{20}\,cm^{-2}}$. The C-stat/d.o.f.\ for this fit was 550.23/584, indicating that a powerlaw model captures most of this AGN's spectrum. Next we added an APEC component to account for the soft excess visible from 0.5-2.0 keV. This reduced C-stat/d.o.f.\ to 515.15/582, so the APEC component was kept. We then added a BORUS component, as would be appropriate for a Compton-thick AGN, which brought C-stat/d.o.f.\ down to 504.28/578. However, looking at the unfolded spectrum of NGC 5005 (Figure \ref{fig:NGC5005_uf}), it is unclear whether a BORUS component is truly justified. The hard X-ray emission does not possess the Compton hump characteristic of a reflection-dominated spectrum, but instead appears to be flat or even declining. It may possess a broad line component in the \nustar{} spectrum, visible as a bump of emission from 4-8 keV. Together, these facts suggest the AGN spectrum might be better fit with just a ZGAUSS component rather than an entire BORUS component.

We added a ZGAUSS component to the  TBABS*(APEC+POWERLAW) model, fixing the line energy at 6.4 keV and fixing the line width at $10^{-3}$ keV. This did not significantly change C-stat/d.o.f., though freeing the line width to vary in the fitting brought C-stat/d.o.f.\ down to 497.25/580.

To further ascertain the nature of the unusual  bump at 4-8 keV we extracted the \xmm{} observation of NGC 5005. The bump from 4-8 keV seen in the \nustar{} spectrum is not clearly seen in its \xmm{} spectra; however the \xmm{} data were taken a decade earlier, so the lack of the line may simply be due to variability.  To determine whether the line was truly absent from the \xmm{} and \chandra{} data, we first fit the \nustar{} data alone to a TBABS*(ZGAUSS+POWERLAW) model to find the best fit parameters for the line. The C-stat/d.o.f.\ of this fit was 391.98/468; for comparison, the C-stat/d.o.f.\ for a TBABS*(BORUS+POWERLAW) fit to the \nustar{} data was 394.76/467. The resulting line parameters were an energy of 5.91 keV, a line with of 0.76 keV, and a normalization of $5.26\times 10^{-6}\ {\rm cts}\, {\rm s}^{-1}\, {\rm keV}^{-1}$.  We then fit the \xmm{} and \chandra{} data alone with a TBABS*(APEC+ZGAUSS+POWERLAW) model, with the ZGAUSS energy and width set to the values measured from the \nustar{} data alone. The normalizations were left to vary freely. The resulting normalizations were consistent with the \nustar{} data for both the \xmm{} and \chandra{} data.

We ran 10,000 Monte Carlo simulations to estimate the false alarm probability for the putative line. We simulated fake \nustar{} observations in XSPEC with the parameters of the best fit to the \nustar{} data using only a POWERLAW component, then tried fitting the data with both a POWERLAW model and a POWERLAW+ZGAUSS model. The normalization of the ZGAUSS component was left to freely vary, while the line width was fixed at the value measured from \nustar{}. We then stepped through the values in line energy and saved the best fit. The resulting decrease in C-stat was greater than the decrease for the real data only in 4 out of 10,000 runs. The same was true if we instead fit it with a POWERLAW+ZGAUSS model where the line was unresolved (width fixed at $3\times10^{-3}$ keV). We therefore estimate the false positivity rate as 0.04\%. This is a  $>$3.3 sigma detection, and we treat the line as real.

For the final fit, we froze the ZGAUSS parameters to the best-fit values from the \nustar{} data alone. The resulting C-stat/d.o.f.\ was 1872.72/2130.
The parameters for this model are listed in Table \ref{tab:NGC5005params}. It is plotted over the \xmm{}, \chandra{} and \nustar{} data in Figure \ref{fig:NGC5005_uf}. The logarithm of the 2-10 keV luminosity (in units of $\mathrm{erg\:s^{-1}}$) measured from the model is $42.29^{+0.12}_{-0.47}$.

\begin{deluxetable}{@{\extracolsep{10pt}}lcccccc@{}}
\tablecaption{Parameters for best-fit NGC 5005 model.}\label{tab:NGC5005params}
\tablewidth{2pt}
\tablehead{\multicolumn{2}{c}{APEC} & \multicolumn{3}{c}{ZGAUSS} & \multicolumn{2}{c}{POWERLAW}\\
\cline{1-2} \cline{3-5} \cline{6-7} \colhead{$kT$} & \colhead{Norm} & \colhead{Line Energy} & \colhead{$\mathrm{\sigma}$} & \colhead{Norm} & \colhead{$\Gamma$} &\colhead{Norm}\\
 \colhead{(keV)} & \colhead{($10^{-5}$ cts $\mathrm{s^{-1}\:keV^{-1}}$)} & \colhead{(keV)} & \colhead{(keV)} & \colhead{($10^{-6}$ cts $\mathrm{s^{-1}\:keV^{-1}}$)} & \colhead{} & \colhead{($10^{-5}$ cts $\mathrm{s^{-1}\:keV^{-1}}$)}
}
\startdata
$0.79\pm{0.03}$ & $4.62^{+0.67}_{-0.59}$ & $5.91^{+0.60}_{-0.62}$\tablenotemark{a} & $0.74^{+0.47}_{-0.48}$\tablenotemark{a}& ${5.26}^{+4.39}_{-3.08}$\tablenotemark{a}& ${1.69}\pm{0.05}$ & $6.61^{+0.88}_{-0.79}$
\enddata
\tablecomments{The \xmm{} pn, MOS1, and MOS2 instrumental normalization constant values were $1.14^{+0.14}_{-0.13}$, $1.14^{+0.15}_{-0.13}$, and $1.17^{+0.15}_{-0.13}$ (ObsID: 0110930501). The \chandra{} normalization constant value was $0.59^{+0.10}_{-0.08}$ (ObsID: 4021). All parameters for APEC and POWERLAW components were measured using a fit with the ZGAUSS components fixed to the values from the \nustar{} data alone.}
\tablenotetext{a}{Measured from fit to \nustar{} data alone}
\end{deluxetable}

\subsection{Mrk 463}

Mrk 463 is a complex ongoing merger with two galactic nuclei and prominent tidal tails visible in optical light \citep{1989AJ.....97.1306H}. It has long been known to be an ultraluminous infrared galaxy and possess a Seyfert 2 AGN \citep{1988ApJ...328L..35S}. In fact, Mrk 463 possesses dual AGN \citep{2008MNRAS.386..105B}, with the eastern AGN more luminous than the western AGN. The eastern AGN possesses a hidden BLR in polarized light \citep{2001ApJ...554L..19T}. Two-sided conical [\ion{O}{3}] outflows extend from the eastern nucleus \citep{1995A&A...298..343C}, creating an extended emission line region to the south of the galaxy, similar to a voorwerp \citep{2018ApJ...854...83T}. The eastern nucleus and its ionization cones generate radio fluxes comparable to a radio-loud quasar or radio galaxy \citep{1991AJ....102.1241M}, which is highly unusual for a Seyfert AGN. Based on the amount of energy required to create the observed ionization and emission line features, \citet{2018ApJ...854...83T} argued the eastern AGN was $\sim$3-20 times more luminous $\sim$40,000 years ago. They argue that it might become a bona fide quasar in the future as the galaxy merger progresses.

Mrk 463 displays prominent photoionized metal lines in its \xmm{} spectra, including from almost fully stripped \ion{Fe}{26} \citep{2004AJ....127..758I} and the \ion{O}{7} radiation recombination continuum \citep{2008MNRAS.386..105B}. It also has a neutral Fe K$\mathrm{\alpha}$ line in its \xmm{} spectra \citep{2004AJ....127..758I}. Both \citet{2004AJ....127..758I} and \citet{2008MNRAS.386..105B} concluded that Mrk 463 is overall Compton thin. Using the \chandra{} data, \citet{2008MNRAS.386..105B} detected a strong Fe K$\mathrm{\alpha}$ line in the eastern nucleus (EW $\simeq$ 250 eV), while only an upper limit could be placed on the Fe K$\mathrm{\alpha}$ line from the western nucleus. The eastern nucleus was also more heavily absorbed. They therefore concluded the eastern nucleus is more obscured than the western nucleus, a claim that is also supported by NIR data.

\subsubsection{X-Ray Observations \& Data Extraction}
\begin{figure}
    \centering
    \plotone{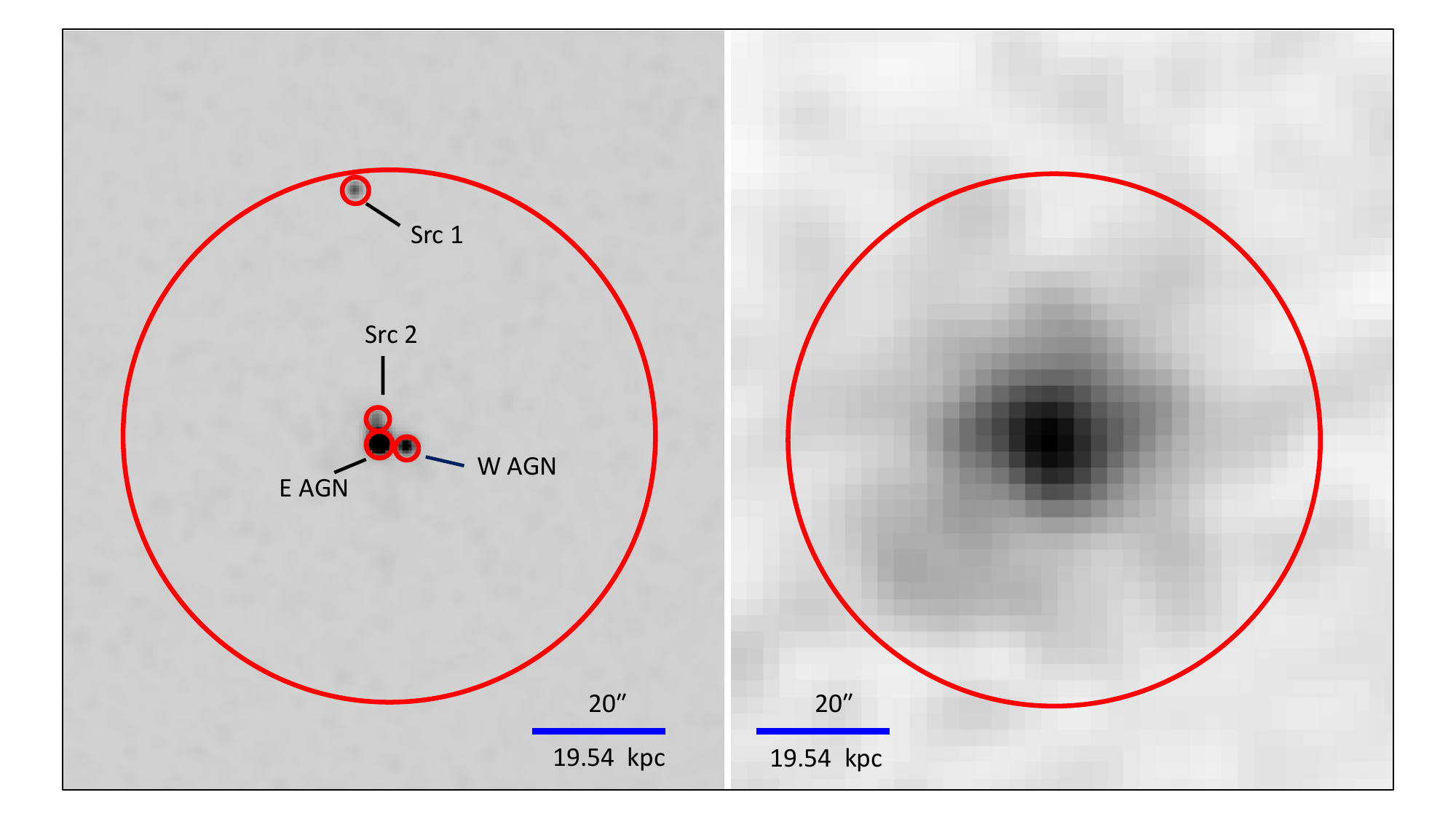}
    \caption{\chandra{} (ObsID: 4913) and \nustar{} FPMA images of Mrk 463. The larger, 40\arcsec{} radius circle denotes the \nustar{} extraction region, while the smaller circles denote the \chandra{} extraction regions. Two extra-nuclear point sources (Source 1 and Source 2) were visible in all \chandra{} observations and were used in the fitting process.}
    \label{fig:Mrk463_image}
\end{figure}

Mrk~463 has been observed once by \nustar{} and twice by \chandra; details of the observations, including the observation dates and exposure times are in Table~\ref{tab:xraydata}. The image from the first \chandra{} observation (ObsID: 4913) and the FPMA image from the \nustar{} observation are shown side by side in Figure \ref{fig:Mrk463_image}.  The higher resolution \chandra{} image clearly resolves the brighter eastern AGN and the fainter western AGN. An extra-nuclear point source (Source 1) is present in the \nustar{} beam. While at first glance the eastern AGN appears to be an elongated ellipse (and indeed was extracted as such by \citealp{2008MNRAS.386..105B}), closer inspection reveals the northern lobe of the ellipse is not part of the AGN, but rather an area of fainter, extended emission that is not detected above 2 keV in energy. It was therefore extracted as a separate source, labeled Source 2. Both extra-nuclear sources have more than 10\% the count rate of the fainter, western AGN in the 0.5-8.0 keV band, so they were ultimately used in the fitting.

The eastern and western AGNs were extracted with circular source regions of radius 2\arcsec{} and 1.76\arcsec{}, respectively, while Source 1 and Source 2 were extracted with circular source regions of radius 2\arcsec{} and 1.77\arcsec{}, respectively.

\subsubsection{X-ray Spectral Fitting}

\begin{figure}
    \plotone{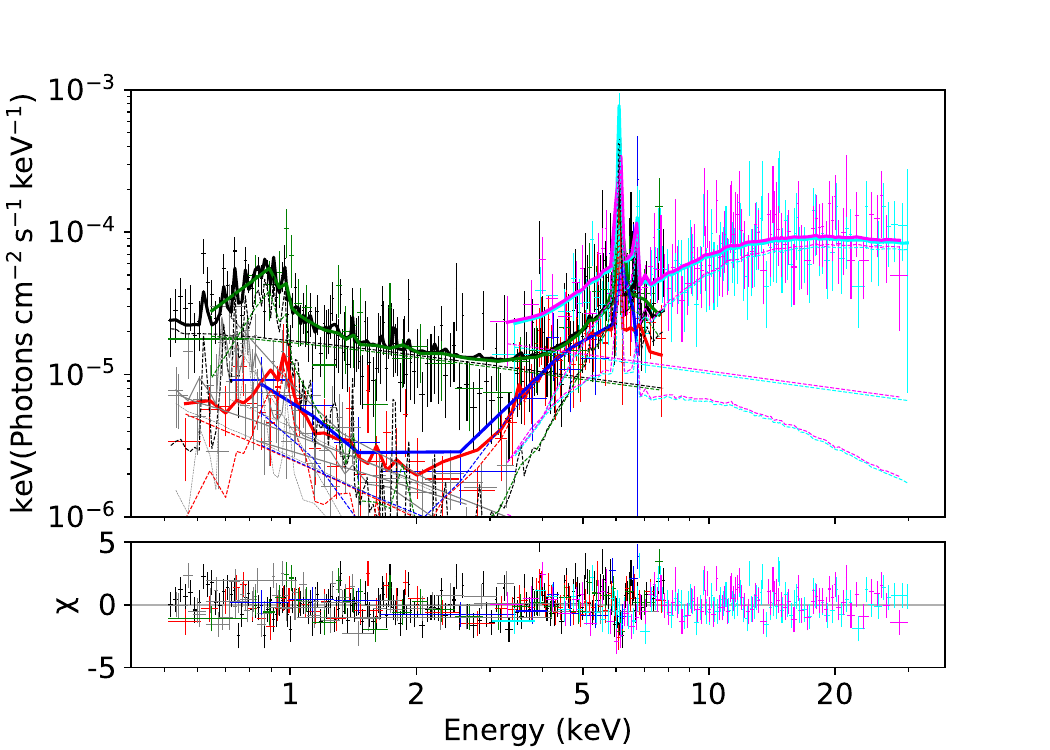}
    \caption{Unfolded spectrum and model for Mrk 463. The model shown is the fit with all the APEC parameters frozen (i.e. the fit that was used to recover the BORUS parameters). Black denotes \chandra{} observation 4913 of the east AGN. Red denotes \chandra{} observation 4913 of the west AGN. Green denotes \chandra{} observation 18194 of the east AGN. Blue denotes \chandra{} observation 18194 of the west AGN. Cyan and magenta represent \nustar{} FPMA and FPMB data respectively. The \chandra{} observations of Source 1 and Source 2 are depicted in light grey.}
    \label{fig:Mrk463_uf}
\end{figure}

Similarly to NGC 3627 and NGC 4501, we jointly fit the data freezing Source 1's and Source 2's parameters to the best-fit values from \chandra{} alone. We began with a fit that was simply four POWERLAW components. The soft X-ray spectra of the two AGN and Source 2 (see Figure \ref{fig:Mrk463_uf}) suggest the need for an APEC component, though this is not necessarily true for Source 1. We therefore added APEC components to all the sources but Source 1. The resulting C-stat/d.o.f.\ was 1673.97/1669.

The \nustar{} spectra (see Figure \ref{fig:Mrk463_uf}) show a prominent Fe K$\mathrm{\alpha}$ line and Compton hump, while the \chandra{} spectra for both AGN show a pronounced rise towards the Fe K$\mathrm{\alpha}$ line from 4-6 keV. We therefore added a BORUS component to both AGN. Adding it to the east AGN brought C-stat/d.o.f.\ down to 1517.85/1610, while adding it to the west AGN brought C-stat/d.o.f.\ down to 1504.83/1610. However, this caused the APEC $kT$ on the east AGN to become implausibly small ($\approx8\times10^{-3}$ keV), APEC $kT$ on the west AGN to be implausibly large ($\approx62$ keV), and the constant on the west AGN to be implausibly large ($\approx 4.65\pm{14}$). To resolve these issues we fixed the normalization constants of the west AGN to have an upper limit of 2 and a lower limit of 0.5. C-stat/d.o.f.\ was brought down to 1469.63/1606. Most of the BORUS parameters remained unconstrained, however, so we froze $\mathrm{\cos(\theta_{inc})}$ to 0.17 (i.e., $\mathrm{\theta_{inc}}$ = 80\degree). This provided some improvement in C-stat/d.o.f., but some of the BORUS parameters were still not converging. We therefore froze the APEC $kT$ and normalization of both AGN. The final estimates for each parameter are tabulated in Table \ref{tab:Mrk463params}. The best fit model is plotted over the data as the solid lines in Figure \ref{fig:Mrk463_uf}. The logarithm of the 2-10 keV luminosity (in units of $\mathrm{erg\:s^{-1}}$) measured from the model is $44.01^{+0.03}_{-0.10}$ for the eastern AGN and $43.57^{+0.19}_{-0.76}$ for the western AGN.

\begin{deluxetable*}{@{\extracolsep{10pt}}lcccccccc@{}}
\tablecaption{Parameters for best-fit Mrk463 model.}\label{tab:Mrk463params}
\tablewidth{2pt}
\tablehead{\colhead{} & \multicolumn{2}{c}{APEC} & \multicolumn{4}{c}{BORUS\tablenotemark{a}} & \multicolumn{2}{c}{POWERLAW}\\ \cline{2-3} \cline{4-7} \cline{8-9} \colhead{Source} & \colhead{$kT$} & \colhead{Norm} & \colhead{log(${N_{\rm{H}}}$)} & \colhead{$\mathrm{CF_{Tor}}$} & \colhead{$\mathrm{\cos(\theta_{inc})}$} & \colhead{Norm} & \colhead{$\Gamma$} &\colhead{Norm}\\
 \colhead{} & \colhead{(keV)} & \colhead{} & \colhead{} & \colhead{} & \colhead{} & \colhead{} & \colhead{} & \colhead{} 
}
\startdata
E AGN & $0.91^{+0.10}_{-0.08}$ & $12.8^{+3.7}_{-8.2}$ & $23.86^{+0.12}_{-0.07}$ & $0.27^{+0.09}_{-0.07}$& $0.17$\tablenotemark{b} & $2.73^{+0.18}_{-0.59}$ & $\leq1.54$ & $27.8^{+5.8}_{-15.9}$\\
W AGN & $1.11^{+0.23}_{-0.20}$ & $1.42^{+7.34}_{-0.66}$ & $23.50^{+0.10}_{-0.22}$ & ${0.28}^{+0.43}_{-0.03}$ & 0.17\tablenotemark{b} & $4.50^{+2.39}_{-3.73}$ & $\geq{1.58}$ & $1.30^{+6.00}_{-0.54}$\\
Src 1 & {} & {} & {} & {} & {} & {} & 2.17\tablenotemark{b} & 4.03\tablenotemark{b}\\
Src 2 & 0.55\tablenotemark{b} & 2.6\tablenotemark{b} & {} & {} & {} & {} & 2.49\tablenotemark{b} & 2.76\tablenotemark{b}
\enddata
\tablecomments{The instrumental  normalization constants for \chandra{} observations 4913 and 18194 were $0.66^{+2.96}_{-0.08}$ and $0.65^{+2.68}_{-0.11}$ for the East AGN, $1.00^{+0.16}_{-0.14}$ and $0.71^{+0.45}_{-0.32}$ for Source 1, and $1.01^{+0.13}_{-0.12}$ and $1.61^{+0.69}_{-0.54}$ for Source 2. No errors on the normalization constants for the West AGN could be calculated because a hard lower limit of 0.5 and a hard upper limit of 2.0 were placed on them. The normalizations of the model components are in units of $10^{-6}$ cts $\mathrm{s^{-1}\:keV^{-1}}$ for APEC, $10^{-3}$cts $\mathrm{s^{-1}\:keV^{-1}}$ for BORUS, and $10^{-6}$ cts $\mathrm{s^{-1}\:keV^{-1}}$ for POWERLAW. }
\tablenotetext{a}{Parameters for this component recovered by freezing APEC component and refitting.}
\tablenotetext{b}{Frozen at this value}
\end{deluxetable*}

\subsection{NGC 6890}
NGC 6890 is a spiral galaxy. Its optical activity has traditionally been classified as Seyfert 2 \citep[e.g.,][]{1996ApJ...471..190R}, but it has also been more specifically classified as a S1.9 \citep{2006A&A...455..773V}.

NGC 6890's MIR spectrum is dominated by a red continuum suggestive of cool dust and polycyclic aromatic hydrocarbon features \citep{2006AJ....132..401B}, where the latter is indicative of star formation \citep{2009ApJ...705...14D}. Based on its {\it Spitzer} IRS spectrum, \citet{2010ApJ...709.1257T} argued roughly 90\% of the 19$\mathrm{\mu m}$ emission is due to the AGN. Its MIR morphology is circular and centered on the nucleus \citep{2014MNRAS.439.1648A} but might be somewhat extended \citep{2016ApJ...822..109A}.

The \xmm{} observations of NGC~6890 were first analyzed by \citet{2007ApJ...657..167S}, who fitted it with an unabsorbed powerlaw. However, since its 2-10 keV X-ray flux was significantly depressed compared to its [\ion{O}{3}] flux, they still regarded it as a Compton-thick AGN. In contrast, \citet{2011ApJ...729...52L} found that the spectrum was best fit with two absorbed powerlaws. They also detected an Fe K$\mathrm{\alpha}$ line at the 93\% confidence level. The equivalent width was 1.21 keV if they used a global fit, and 0.93 keV if they used a local fit.

\citet{2011MNRAS.413.1206B} presented a more detailed analysis of the \xmm{} data, including fits with PEXMON and TORUS models to account for a reflection component. They measured a hydrogen column density of $\mathrm{10^{21}\:cm^{-2}}$ for this reflection component, which would put it outside the Compton-thick regime.

\subsubsection{X-Ray Observations \& Data Extraction}

NGC~6890 has not been observed by \chandra, but has been observed by \xmm\ once, by \swift\ twice, and by \nustar\ once. The \nustar\ observation was concurrent with the second \swift\ observation.  Details of these observations, including their observation dates and exposure times, are in Table~\ref{tab:xraydata}.

For the \xmm{} data, we filtered out times with high background, defined as when the count rate in the 10–12 keV range was $\rm{>}$0.4 cts $\rm{s^{-1}}$ for the pn and $\rm{>}$0.35 cts $\rm{s^{-1}}$ for the MOS cameras. We extracted the source spectra with 30\arcsec{} radius circular regions, and background spectra from an annulus of 50–80\arcsec{} with pattern 0–4 for the pn and pattern 0–12 for the MOS cameras. 

\begin{figure}
    \centering
    \plotone{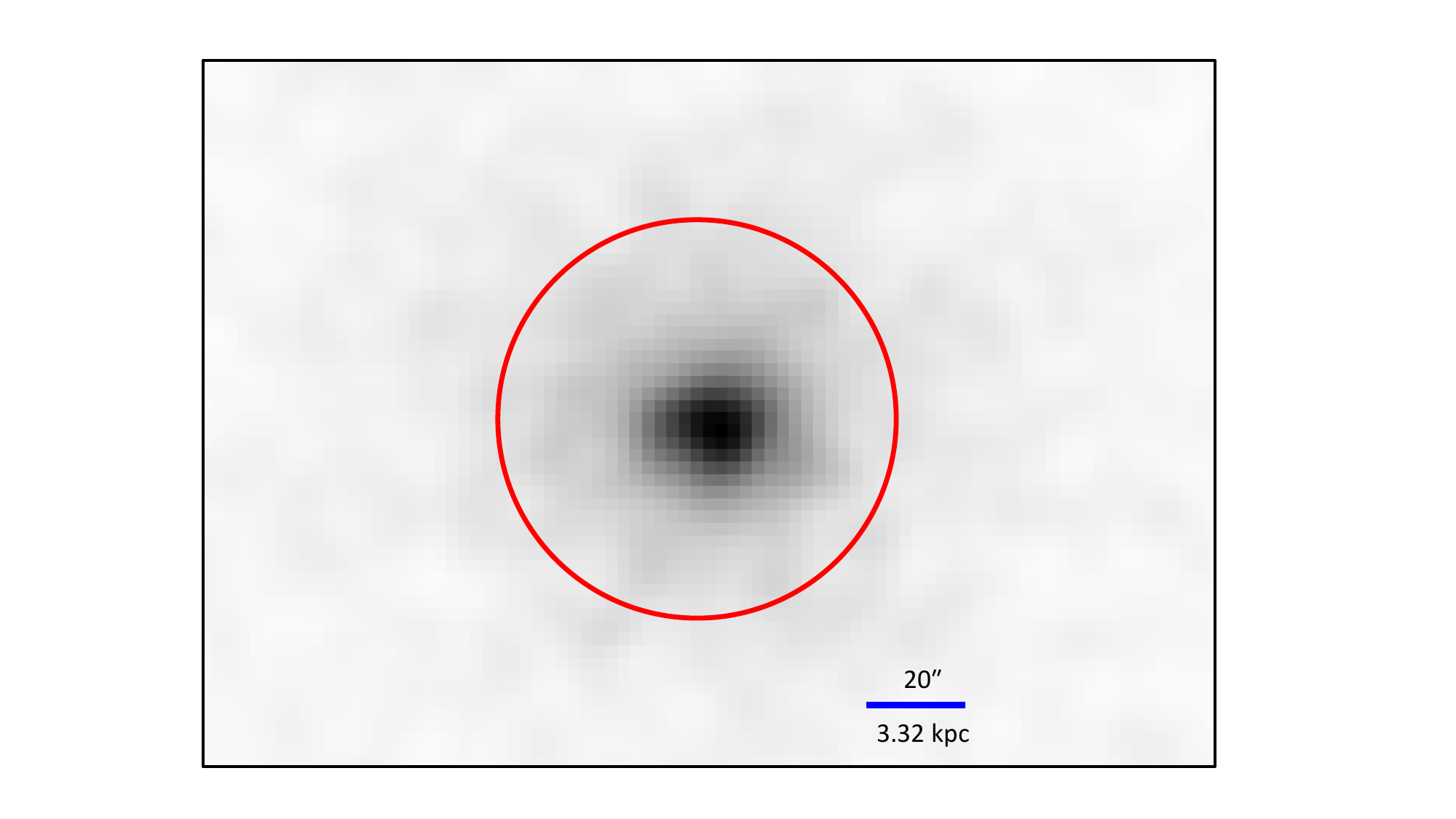}
    \caption{\nustar{} FPMA image of NGC 6890. The \xmm{} data, which have lower anguler resolution than \chandra, did not detect any off-nuclear point sources within the \nustar{} beam (shown in red).}
    \label{fig:NGC6890_image}
\end{figure}

\subsubsection{X-ray Spectral Fitting}
\begin{figure}
    \plotone{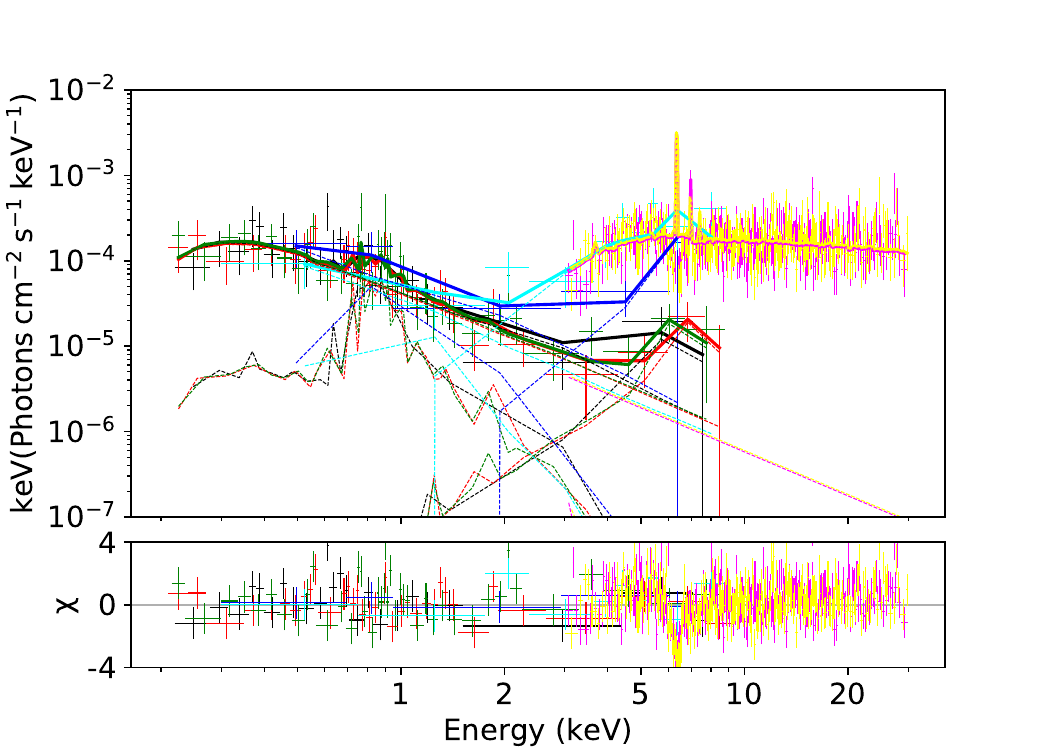}
    \caption{Unfolded spectrum and best-fit model for NGC 6890. Black denotes \xmm{} pn data, red denotes \xmm{} MOS1 data, green denotes \xmm{} MOS2 data, blue denotes data from \swift{} observation 00088188001, cyan denotes data from \swift{} observation 0008818800, and magenta and yellow denote \nustar{} FPMA and FPMB data, respectively.}
    \label{fig:NGC6890_uf}
\end{figure}

We began with a simple CONSTANT*TBABS*POWERLAW model. The C-stat/d.o.f.\ for this fit was 1951.31/1580, implying room for improvement. A BORUS component improved C-stat/d.o.f.\ to 1736.27/1576. We then added an APEC component, which improved C-stat/d.o.f.\ to 1724.84/1574. Looking at the unfolded spectrum (Figure \ref{fig:NGC6890_uf}), the BORUS component seems to have a higher intensity in the \swift{} and \nustar{} data than in the \xmm{} data. The APEC component of the \xmm{} data is of similar flux density to the \nustar{} data, but energies in the \xmm{} data above 1 keV do not match up with the \nustar{} data. 

To test the possibility that the BORUS component was varying, we created fits where $N_{\rm{H}}$ and the BORUS normalization varied between each observation. When these parameters were left free to vary, their values for the \xmm{} MOS1 and MOS2 data were tied to the EPIC pn value, and their \nustar{} FPMB and FPMA values were tied together. The values for the \swift{} observations were left to freely vary independently. We set the cross-normalization constants all to 1. For the case of $N_{\rm{H}}$ varying, C-stat/d.o.f.\ was 1886.20/1576. For the case of the BORUS normalization varying, C-stat/d.o.f.\ was 1828.16/1576. And for the case of both $N_{\rm{H}}$ and the BORUS normalization varying, C-stat/d.o.f.\ was 1836.111/1573. 

The best fit seemed to be the one where only the BORUS normalization varied. However, adding $N_{\rm{H}}$ variability should not have made the fit worse than the fit with the normalization varying alone. We therefore freed $N_{\rm{H}}$. We then reset log($N_{\rm{H}}/\rm{cm}^{-2}$) to be 25.5 for \xmm{} and 23 for \swift{} and \nustar{} and refit. This led to a C-stat/d.o.f. = 1810.31/1573.  
However, the POWERLAW component was underestimating the \xmm{} data. We therefore untied the POWERLAW spectral index from the BORUS spectral index. This allowed the scattered powerlaw to differ from the intrinsic powerlaw input to the BORUS model. This fit had a C-stat/d.o.f.\ of 1711.07/1572.

We next tied the values of  $N_{\rm{H}}$ and the BORUS normalization for the second \swift{} observation and \nustar{} together, since these observations were contemporaneous. Lastly we thawed the cross-normalization constants, setting limits of 0.5-2.0 on all of them. 
The final C-stat/d.o.f.\ was 1699.97/1569, the parameters of the final fit are in Table \ref{tab:NGC6890params}, and the fit is plotted over the data in Figure \ref{fig:NGC6890_uf}. The logarithms of the 2-10 keV luminosities (in units of $\mathrm{erg\:s^{-1}}$) measured from the model from each epoch are $42.25^{+0.89}_{-0.24}$ for September 2009, $42.73^{+0.33}_{-0.48}$ for March 2018, and $43.66^{+0.09}_{-0.01}$ for May 2018.

\begin{deluxetable}{@{\extracolsep{10pt}}lccccccccc@{}}
\tablecaption{Parameters for best-fit NGC 6890 model.}\label{tab:NGC6890params}
\tablewidth{0pt}
\tablehead{\colhead{Observation} & \multicolumn{2}{c}{APEC} & \multicolumn{5}{c}{BORUS} & \multicolumn{2}{c}{POWERLAW}\\ \cline{2-3} \cline{4-8} 
\cline{9-10} \colhead{} & \colhead{$kT$} & \colhead{Norm} & \colhead{$\Gamma$} & \colhead{log($N_{\rm{H}}$)} & \colhead{$\mathrm{CF_{Tor}}$} & \colhead{$\mathrm {\cos(\theta_{inc})}$} & \colhead{Norm} & \colhead{$\Gamma$} &\colhead{Norm}\\
\colhead{} & \colhead{(keV)} & \colhead{} &\colhead{} & \colhead{} & \colhead{} & \colhead{} & \colhead{} & \colhead{} & \colhead{}
}
\startdata
\xmm{} & $0.73^{+0.14}_{-0.15}$ & $7.14^{+9.64}_{-3.54}$ & $\leq1.41$ & $24.12^{+0.29}_{-1.10}$ & $\leq0.10$ & $\leq0.10$ & $1.89^{+13.1}_{-0.78}$ & $2.71^{+0.15}_{-0.20}$ & $2.95^{+4.28}_{-0.27}$\\
1st \swift{} & {} & {} & {} & $23.40^{+0.86}_{-0.88}$ & {} & {} & $5.81^{+6.67}_{-3.88}$  & {} & {}\\
2nd \swift{}/\nustar{} & {} & {} & {} & $23.01^{+0.02}_{-0.12}$ & {} & {} & $49.5^{+11.4}_{-1.3}$ & {} & {}\\
\enddata
\tablecomments{The instrumental normalization constant for \xmm{} MOS2 was $1.46^{+0.19}_{-0.18}$. The rest were pegged at the upper limit of 2. The BORUS $\Gamma$ and $\mathrm{CF_{Tor}}$ constraints were derived by freezing the BORUS log($N_{\rm{H}}$) and Normalization at their best fit values in all observations. The normalizations are in units of $10^{-6}$ cts $\mathrm{s^{-1}\:keV^{-1}}$ for APEC, $10^{-3}$ cts $\mathrm{s^{-1}\:keV^{-1}}$ for BORUS, and $10^{-5}$ cts $\mathrm{s^{-1}\:keV^{-1}}$ for POWERLAW. }
\end{deluxetable}

\section{Discussion}

In this section we discuss the implications of our results. \S~4.1 discusses how the intrinsic luminosities of the AGN were derived. \S~4.2 compares these luminosities to those expected from scaling relations. \S~4.3 discusses the obscuration levels measured from the X-ray spectral fits. \S~4.4 describes how Eddington ratios were computed and whether there are any correlations observed with Eddington ratio. Lastly \S~4.5 explains in further detail special features observed in the individual galaxies.

\begin{deluxetable*}{llccccccc}
\tablecaption{Summary of AGN Properties.\label{tab:summary}}
\tablewidth{1pt}
\tablecolumns{9}
\tablehead{
\colhead{Object} &
\colhead{Type} &
\colhead{$\log{(M_{\rm BH}}$)} &
\colhead{$\log{(N_{\rm H})}$}&
\colhead{$\log{(L_{\rm 2-10})}$} &
\colhead{$\log{(L_{\rm [OIII]})}$} &
\colhead{$\log{(L_{\rm MIR})}$} &
\colhead{${L_{\rm bol}/L_{\rm Edd}}$} &
\colhead{Refs}\\
\colhead{} &
\colhead{} &
\colhead{(${M_{\odot}}$)} &
\colhead{($\mathrm{cm^{-2}}$)} &
\colhead{$\mathrm{(erg\:s^{-1})}$} &
\colhead{$\mathrm{(erg\:s^{-1})}$} &
\colhead{$\mathrm{(erg\:s^{-1})}$} & 
\colhead{} &
\colhead{}
}
\startdata 
NGC 1386 & S1i & 7.24 & $\geq24.5$ & $42.29\pm{0.05}$ & 40.16 & $42.39\pm{0.08}$ & $1.38\times10^{-2}$ & 1,2,6,10\\
NGC 3627 & S3 & 6.93 & - & $38.38^{+0.16}_{-0.10}$ & 39.40 & $40.60\pm{0.11}$ & $3.67\times10^{-6}$ & 1,3,6,10\\
NGC 3982 & S1.9 & 6.95 & $\geq25.3$ & $42.83^{+0.13}_{-0.08}$ & 39.87 & $41.56\pm{0.06}$ & $9.50\times10^{-2}$ & 1,4,6,10\\
NGC 4501 & S2 & 7.30 & $22.87^{+0.25}_{-0.15}$ & $41.50^{+0.25}_{-0.11}$ & 39.86 & $40.56\pm{0.06}$ & $1.93\times10^{-3}$ & 1,3,6,10\\
IC 3639 & S1h & 7.01 & $25.00^{+0.06}_{-0.26}$ & $43.07^{+0.18}_{-0.12}$ & 42.0 & $43.52\pm{0.04}$ & 0.146 & 1,4,7,10\\
NGC 4922 & S2 & {} & $23.89^{+0.11}_{-0.17}$ & $42.29^{+0.12}_{-0.47}$ & 42.3 & {} & {} & 1,8\\
NGC 5005 & S3b\tablenotemark{a} & 8.27 & - & $40.17^{+0.04}_{-0.05}$ & 39.03 & $40.78\pm{0.12}$ & $9.67\times10^{-6}$ & 1,4,6,10\\
Mrk 463 & E: S1h & {} & $23.86^{+0.12}_{-0.07}$ & $44.01^{+0.03}_{-0.10}$ & 42.62\tablenotemark{b} & 44.83 & {} & 1,9,11\\
{} & W: S2 & {} & $23.50^{+0.10}_{-0.22}$ & $43.57^{+0.19}_{-0.76}$ & & \nodata & {} & \\
NGC 6890\tablenotemark{c} & S1.9 & 7.07 &  &  & 42.02 & $42.60\pm{0.09}$ & & 1,5,6,10 \\
--- 2009 Sep & & & $24.12^{+0.29}_{-1.10}$ & $42.25^{+0.89}_{-0.24}$ & & & $1.86\times10^{-2}$ & \\
--- 2018 Mar & & & $23.40^{+0.86}_{-0.88}$ & $42.73^{+0.33}_{-0.48}$ & & & $5.70\times10^{-2}$ & \\
--- 2018 May & & & $23.01^{+0.02}_{-0.12}$ & $43.66^{+0.09}_{-0.01}$ & & & 0.530 & \\
\enddata
\tablenotetext{a}{Broad component detected in H$\alpha$, no others.}
\tablenotetext{b}{Combined $\log{(L_{\rm [OIII]})}$ for E and W components of Mrk~463, not corrected for intrinsic dust extinction.}
\tablenotemark{c}{In temporal order: 2009 Sep = \xmm{} observation; 2018 Mar = first \swift{} observation; 2018 May = second \swift\ observation + \nustar{} observation.}
\tablecomments{S1i indicates a Seyfert 2 with broad lines detected in the infrared, S1h indicates a Seyfert 2 with a hidden BLR detected in polarized light, S1.9 denotes a Seyfert with broad H$\mathrm{\alpha}$ but no broad H$\mathrm{\beta}$, and S3 indicates a LINER.  ${L_{\rm 2-10}}$ is the intrinsic absorption-corrected X-ray luminosity. MIR luminosities are at 12 $\mu$m. Bolometric luminosities were computed using $K_{X}(L_{X})$ from Table 1 of \citet{fred}. Error bars on luminosities are given if available. Ref is references for optical classification, ${M_{\rm BH}}$, ${L_{\rm [OIII]}}$, and ${L_{\rm MIR}}$.}
\tablerefs{(1) \citet{bob}, (2) \citet{2002ApJ...579..530W}, (3) \citet{2016ApJ...818...47S}, (4) \citet{2016ApJ...831..134V} \& references therein, (5) \citet{2010MNRAS.406..493M}, (6) \citet{2011MNRAS.414.3084B} and references therein, (7) \citet{2016ApJ...833..245B}, (8) \citet{2021ApJ...908..221L} and references therein, (9)\citet{2005ApJ...634..161H} and references therein, (10) \citet{2015MNRAS.454..766A}, (11) \citet{2016MNRAS.463.2405A}.}
\end{deluxetable*}

\subsection{Intrinsic Luminosities}

The 9 AGN in the sample have low observed \xmm{} 2-10 keV luminosities. Recall this can mean one of two things: that the AGN is heavily obscured, or that the AGN is recently deactivated.  Observations in the hard X-ray band from \nustar{} are essential for distinguishing between the two scenarios.  With hard X-ray data, we can model the spectrum more accurately, and from that model we can estimate the true intrinsic 2-10 keV luminosity. In the case of an obscured AGN, we would expect to see additional flux at higher energies, where the photons have enough energy to penetrate the obscuring torus. This would lead to a modeled intrinsic X-ray luminosity that is higher than that originally derived from observed 2-10 keV fluxes. If an AGN has recently deactivated, we will not observe additional X-rays from the AGN at higher energies. This means the 2-10 keV band will capture most of the AGN's X-rays, and so the intrinsic luminosity inferred from the model will be similar to that inferred with the 2-10 keV observed fluxes alone.

For most of the AGN in the sample, there was a jump by several orders of magnitude between the observed and intrinsic X-ray luminosities. This indicates that they are obscured AGN, as the hard X-ray data indicate their modeled spectra have to be more luminous in the 2-10 keV band than directly observed. For NGC 3627 and NGC 5005 however, the change in X-ray luminosity between observed and intrinsic was within an order of magnitude. This indicates that they are not as heavily obscured as the other AGN in the sample.

\subsection{Scaling Relations}

\begin{figure}
    \centering
    \plotone{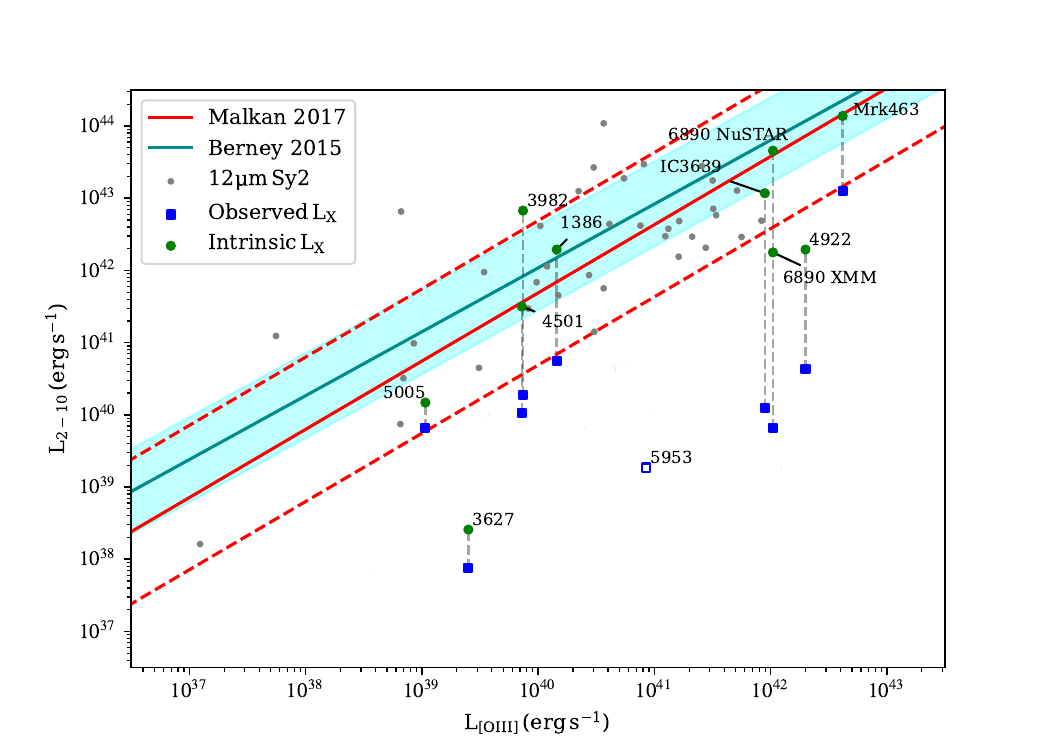}
    \caption{Intrinsic 2-10 keV X-ray luminosities versus updated [\ion{O}{3}] luminosities for the galaxies in our sample. The instrinsic luminosities are plotted alongside their former positions from Figure \ref{fig:Proposal_Plot}. The Mrk 463 2-10 keV luminosity is the combined luminosity of the eastern and western AGNs. IC 3639 has been moved slightly to the left to better distinguish it from NGC 6890.}
    \label{fig:Proposal_plot_updated_Lx}
\end{figure}

\begin{figure}
    \centering
    \plotone{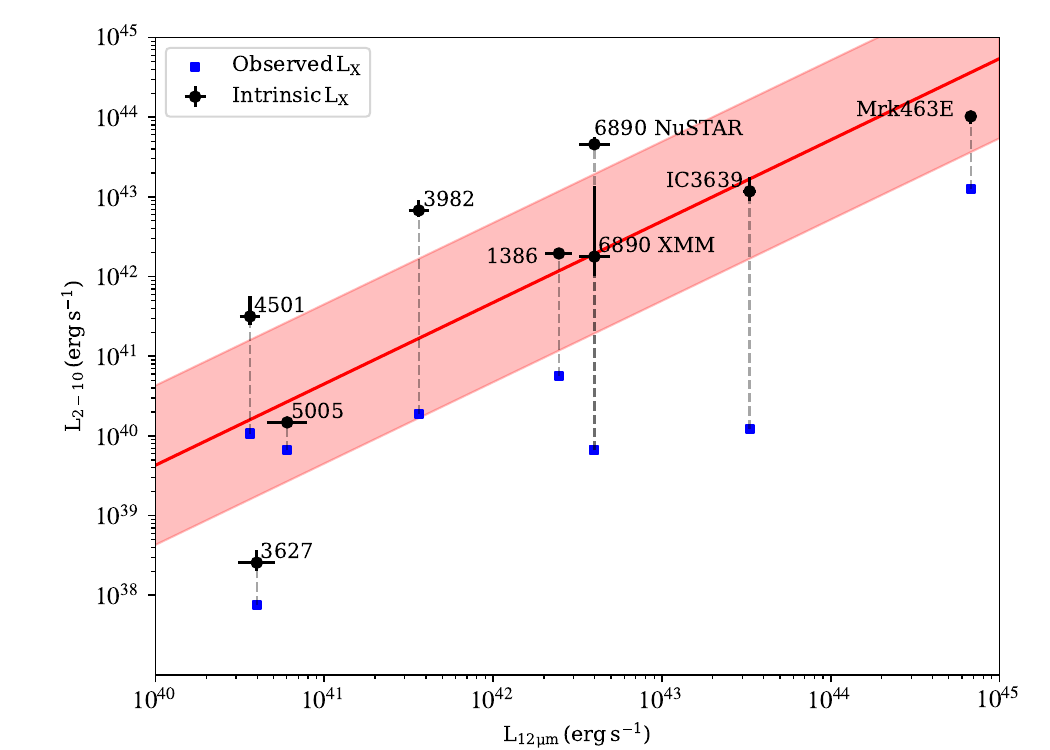}
    \caption{Intrinsic 2-10 keV X-ray luminosities versus $12\mu {\rm m}$ luminosities for the galaxies in our sample. The blue squares are the galaxies plotted with observed 2-10 keV luminosities. The black points with errorbars use the intrinsic 2-10 keV luminosities. The red line is the mean $L_{2-10}$ vs $L_{\rm 12\mu m}$ relation for the complete reliable sample in \citet{2015MNRAS.454..766A}. The scatter of this relation is 0.33 dex which is depicted as the light red shaded region. $L_{2-10}$ errors were derived from our measurements as explained in Section 4. Errors on  $L_{\rm 12\mu m}$ are derived from the literature.}
    \label{fig:Lx_vs_LMIR}
\end{figure}

The hard X-ray continuum from an AGN is believed to originate from the corona, located very close to the central black hole \citep[$\leq$~0.1pc][]{2017ApJ...844...21I}, which translates to a light-crossing time of the corona of less than 3 months, and it is known from observations that the corona can vary in timescales of days \citep[e.g.,][]{2014MNRAS.443.2746W,2015ApJ...806..149K,2020ApJ...898L...1R}. In contrast, the MIR emission from the torus and the [\ion{O}{3}]
emission from the NLR originate from much further out  could only vary {\it much} more slowly. Therefore, when an AGN deactivates, the corona will fade out well before the torus and NLR do, on timescales far shorter than theirs (i.e. days vs. decades and centuries). We therefore expect a recently deactivated AGN to have an intrinsic X-ray luminosity well below that which is expected based on its [\ion{O}{3}] and MIR luminosities.  If, in contrast, the AGN is merely heavily obscured, we would expect to find an intrinsic X-ray luminosity consistent with its [\ion{O}{3}] and MIR luminosities.

In Figure \ref{fig:Proposal_plot_updated_Lx} we  plot the intrinsic 2-10 keV luminosities of our sample on \ref{fig:Proposal_Plot}. The red line is the mean $L_{\rm [OIII]}$ vs  $L_{2-10}$ relation for Seyfert 1 galaxies in the 12 $\mathrm{\mu m}$ sample, and the dashed red lines represent one dex above and below it. It is derived from the observed luminosities in \citet{2017ApJ...846..102M}. Most of the galaxies lie within 1 dex of the mean relation when the intrinsic 2-10 keV luminosity is considered. NGC 3982 does lie more than one dex above it. However, it is still placed within the scatter of the other Seyfert 2 galaxies in the 12 $\mathrm{{\mu m}}$ sample. NGC 4922 is more than 1 dex below the mean relation. But by far the furthest outlying galaxy is NGC 3627, located more than 2 dex below from the mean correlation. This makes it the strongest candidate for its corona having faded relative to the luminosity inferred from the NLR.

In Figure \ref{fig:Lx_vs_LMIR}, we plot the observed and intrinsic 2-10 keV luminosities of our sample versus their $\mathrm{12\mu m}$ luminosities. The $\mathrm{12\mu m}$ luminosities are derived mostly from \citet{2015MNRAS.454..766A}, which had the subarcsecond resolution necessary to resolve the nuclear MIR emission and separate it from the overall host galaxy emission. The exceptions to this are NGC 4922, for which no $\mathrm{12\mu m}$ luminosities could be found in the literature, and Mrk 463, for which the $\mathrm{12\mu m}$ luminosity was taken from \citet{2016MNRAS.463.2405A}. 

The red line is from \citet{2015MNRAS.454..766A} and represents their estimate of the ${L_{\rm 12\mu m}}$ vs ${L_{2-10}}$ correlation using their entire reliable sample. This relation has an intrinsic scatter of 0.33 dex, which is shown as the light red shaded region. With the original observed estimates of the 2-10 keV luminosity, all the AGN except for NGC 4501 are located more than 0.33 dex below the mean relation. With the absorption-corrected intrinsic 2-10 keV luminosity, NGC 3982, NGC 4501, and NGC 6890 lie more than 1 dex above it. For NGC 6890 this is clearly due to the increase in luminosity observed in the X-ray data. Because the torus is located further out than the corona, this could imply the corona has gotten brighter in recent years for the other two X-ray overluminous AGN as well, while the torus has yet to respond to the increase in luminosity.

Once again, NGC 3627 is the furthest below the mean relation, more than 1 dex below it. This can be taken to represent deactivation of the corona while the torus continues to be emitting. According to \citet{2017ApJ...844...21I} the torus fading timescale is dominated by the light crossing time. This leads to a fading timescale of around 10-100 years based on typical sizes of the torus. This would imply that NGC 3627 deactivated no more than 10-100 years ago, since this galaxy still preserves torus emission. Indeed, the conclusion of \citet{2020ApJ...905...29E} was that NGC 3627 has faded on timescales of decades. However, NGC 3627's $\mathrm{12\mu m}$ emissions are distributed throughout the galaxy, so it is unclear how much of the nuclear $\mathrm{12\mu m}$ contribution is from an AGN torus as opposed to star formation.

\subsection{Obscuration}

Of the galaxies in our sample, all but two have the X-ray spectra typical of obscured AGN,  displaying the soft thermal emission from 0.5-2 keV, a hard reflection component and narrow Fe K$\alpha$ line, and the Compton hump at $\sim 20$~keV (for examples, see NGC 1386 or IC 3639). The hydrogen column densities of the AGN are summarized in Table \ref{tab:summary}. We replicate the result that NGC 1386 and IC 3639 are Compton-thick. NGC 3982 is also Compton-thick. NGC 4501, NGC 4922, and both AGNs in Mrk 463 are obscured, but not quite at the Compton-thick level. NGC 6890 was nearly Compton-thick during the time of its \xmm{} observations, but became definitively Compton-thin during later observations. This makes it a new X-ray changing-look AGN \citep[e.g.][]{2003MNRAS.342..422M}. NGC 3627 and NGC 5005 are unobscured.

It is noteworthy that our selection method (searching for AGN that are underluminous in soft X-rays relative to their [\ion{O}{3}] luminosity), which is a common method of selecting Compton thick AGN candidates, resulted in a sample where only 33\% of the objects were actually Compton thick at the time of their \nustar{} observations. Most (77\%) of the AGN are indeed heavily obscured ($N_{\rm H} > 10^{23} \,{\rm cm}^{-2}$).

\subsection{Eddington Ratios}
We computed the bolometric luminosities from the intrinsic 2-10 keV luminosities using the general expression for $\kappa_{X}=L_{\rm bol}/L_{X}$ from Table 1 of \citet{fred}. We then converted these to Eddington ratios using the most recent black hole masses available in the literature.

We find that the obscured AGN are accreting at a higher rate (i.e. $L_{\rm bol}/L_{\rm Edd}>10^{-3}$) than the two AGN that do not show evidence of obscuration (NGC 3627 and NGC5005; $L_{\rm bol}/L_{\rm Edd}\sim10^{-6}$). Of note is that the most heavily obscured AGN, IC 3639, has an Eddington ratio of 0.146, more typical of quasars than Seyfert galaxies, and that NGC 6890's Eddington ratio increased by an order of magnitude between its \xmm{} and \nustar{} observations, for a final ratio of 0.530. 

There is no clear correlation between Eddington ratio and the position of the AGN on the $L_{\rm [OIII]}$ vs  $L_{2-10}$ relation for the sample as a whole, as the high Eddington ratio AGN IC 3639 and NGC 6890 are near the mean relation (as are low Eddington ratio AGN NGC 1386, NGC 4501, and Mrk 463) while low Eddington ratio NGC 3982 lies more than 1 dex above it. The very lowest Eddington ratio AGN (NGC 3627 and NGC 5005) are located at two very different positions in the graph, with NGC 5005 being only 1 dex away from the mean relation, while NGC 3627 lies more than 2 dex away.

The same is true for the ${L_{\mathrm{12\mu m}}}$ vs ${L_{2-10}}$ relation, as low Eddington ratio NGC 3982 and NGC 4501 lie more than 1 dex above the mean relation like high Eddington ratio NGC 6890. NGC 5005 is within the intrinsic scatter of the relation, but NGC 3627 is not.

\subsection{Notes about Individual Galaxies}

\subsubsection{NGC 3627}
The \nustar{} data for NGC~3627 were recently analyzed and concluded to show evidence of a Compton-thick nature for this AGN \citep{2020ApJ...905...29E}. However, this is clearly not the case for our analysis of the data (\S~3.2), which shows no evidence for a reflection component, and thus no evidence for obscuration. This would seem to favor the fading AGN scenario for this galaxy, since it is well below the X-ray luminosity expected for its MIR luminosity. However, it is still possible that a stronger AGN could be hidden behind extremely high levels of obscuration (such that not even the hard X-rays are able to escape). This scenario can not be completely ruled out by \nustar{}; a more sensitive hard X-ray telescope could potentially do so. In the case of a strong, extremely obscured AGN, we would still expect strong MIR emission, as the dusty torus would still be heated. Given the AGN in NGC 3627 does not dominate above the MIR background of its host galaxy, it seems likely this AGN is intrinsically low-luminosity. If we accept the measured 2-10 keV luminosity of $\mathrm{10^{38.38}\,erg\,s^{-1}}$ as the true intrinsic luminosity, NGC 3627's luminosity is below the Eddington luminosity of a stellar black hole ($\mathrm{1.26\times10^{39}\,erg\,s^{-1}}$ for a 10 $\mathrm{M_{\odot}}$ black hole). It therefore might not even be a currently active AGN by some definitions,  even though it does present as one on the BPT diagram.

 Many correlations between AGN [\ion{O}{3}] luminosity and NLR size have been published. Following the prescription of \citet{2017ApJ...844...21I}, we use the correlation from \citet{2017ApJ...837...91B} which spatially separated AGN [\ion{O}{3}] emission from larger, star formation driven [\ion{O}{3}] emission  based on integral-field unit spectroscopy of nearby type 1 and type 2 AGN.  That prescription is:

\begin{equation} \label{eq:NLRsize}
\log(R_{
\rm NLR})=0.41\,\log(L_{\rm [OIII]})-14.00,
\end{equation}
where the NLR radius $R_{\rm NLR}$ is in units of parsec and the luminosity is in units of ${\rm erg}\, {\rm s}^{-1}$. 
Plugging in the non-reddening corrected [\ion{O}{3}] luminosity for NGC 3627, we get a NLR radius of 27.2~pc.

The overall size of the [OIII] emitting region can be estimated as well. We use the relation for Seyfert galaxies from \citet{2003ApJ...597..768S}:
\begin{equation} \label{eq:OIIIsize}
\log(R_{\rm{[OIII]}})=0.33\,\log(L^{\rm int}_{\rm [OIII]})-10.78,
\end{equation}
where the radius is in units of parsec and the luminosity is in units of ${\rm erg}\, {\rm s}^{-1}$ and is intrinsic (i.e. reddening-corrected). Plugging in our reddening-corrected [\ion{O}{3}] luminosity, this relation gives ${R_{\rm [OIII]}}$ = 67.3~pc. The resulting timescale for fading based on light-crossing times implies that the central AGN in NGC 3627 deactivated no earlier than 87 - 220 years ago.

\subsubsection{NGC 5005}
Lacking a prominent hard X-ray reflection component, NGC 5005 does not present a typical obscured AGN X-ray spectrum (\S~3.7). This could indicate that the AGN is currently inactive, and we are only seeing softer X-ray emission from star formation. However, since its optical spectrum exhibits a broad H$\alpha$ component, the simplest conclusion is that this AGN is actually unobscured. This is in contrast to many of its classifications in the literature, which refer to it as a Seyfert 2. It is hypothetically conceivable that the central black hole in this source has faded relative to the BLR, but the extreme rapidity of the timescale of which this would occur given the 10s-100s of light days size of the BLR makes this very unlikely.

As noted previously, \nustar{} data on NGC 5005 show a broad emission line centered on 5.91 keV, but this is not seen in the \xmm{} and \chandra{} data. Based on our MC simulations, this line is a real feature. In being broad it resembles the relativistic iron lines that have been observed in other AGN \citep[e.g.,][]{2006ApJ...652.1028B, 2013Natur.494..449R, 2020MNRAS.499.1480W}, and the centroid energy lower than the rest-frame energy transition at 6.4 keV of the line suggests it is gravitationally redshifted, like some of the relativistically broadened lines observed in  Seyfert galaxies \citep[e.g.,][]{2007MNRAS.382..194N}. 

\begin{figure}
    \centering
    \plotone{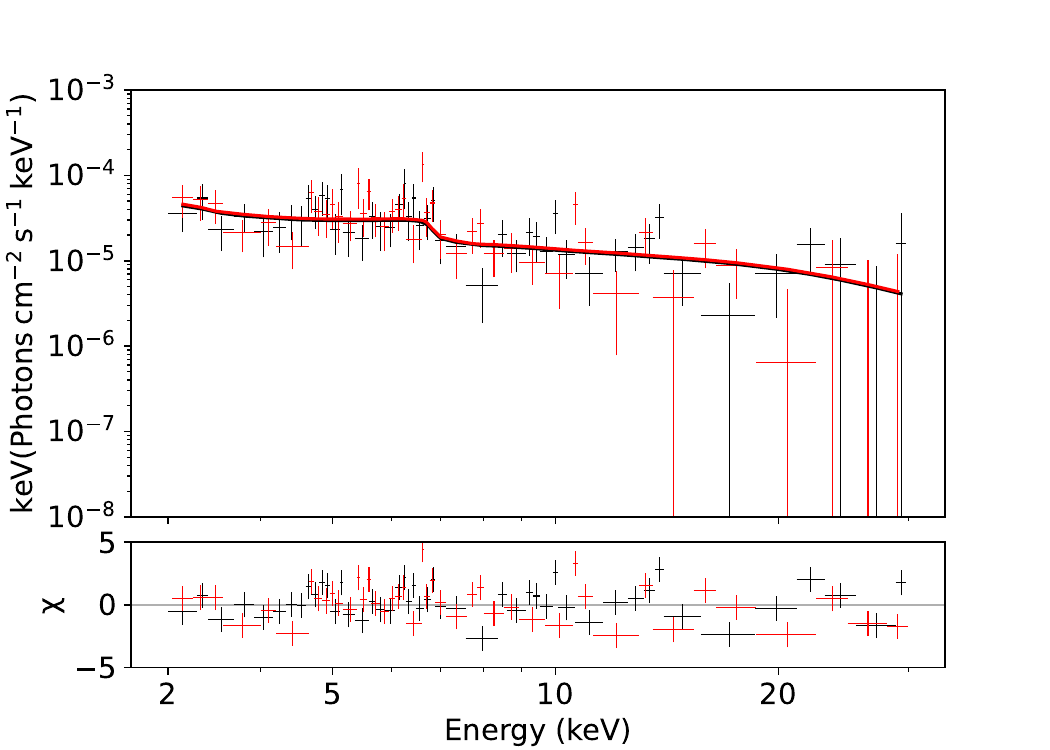}
    \caption{Unfolded spectrum and best-fit model for NGC 5005 \nustar{} data using a TBABS*RELXILL model and realistic reflection fraction values. Black is FPMA data, red is FPMB data. The spin in this case is a=0.998.}
    \label{fig:relxill}
\end{figure}

We fit the \nustar{} spectra of NGC 5005 with the relativistic reflection model RELXILL \citep[version 1.4.3;][]{2014ApJ...782...76G,2014MNRAS.444L.100D,2016A&A...590A..76D}. We first used all the default parameter values for the model except for the iron abundance and redshift, which we froze to solar and the redshift of the galaxy. The fit was better than both a BORUS+POWERLAW model fit just to the \nustar{} data, and the ZGAUSS*POWERLAW fit used in Section 3. However, the black hole spin parameter, $a$, could not be well constrained, and the reflection fraction was too high to be physical (i.e. reflection fraction $>10$ for spin $a<0.9$). The reflection fraction is defined as the ratio of the amount of observed radiation reflected off of the accretion disk to the amount of radiation directly transmitted to the observer from the corona. For a given spin value of the black hole there is a maximum possible value of this fraction \citep[see Figure 3 in][]{2014MNRAS.444L.100D}. We therefore fit the spectrum with the black hole spin fixed to a variety of values, and set the upper limit on the reflection fraction set to the upper limits from \citet{2014MNRAS.444L.100D}. The C-stat declined continuously as the spin increased, and the best fit was obtained with a near-maximum spin value ($a=0.998$). Given the strength of the broad line in NGC 5005, a high spin is clearly favored, as this allows a higher reflection fraction (in this case, $>12$). We have plotted that fit in Figure \ref{fig:relxill}. The C-stat/d.o.f.\ for this fit was comparable to that of the ZGAUSS+POWERLAW fit, but not lower. 

\subsubsection{NGC 6890}
NGC 6890 varied in both obscuration and luminosity between the time of its \xmm{} and \nustar{} observations (\S~3.9).  The observed change in luminosity makes NGC 6890 different from many other X-ray changing-look Seyfert galaxies, which have been traditionally interpreted as varying in obscuration alone \citep[e.g.][]{2002ApJ...571..234R}, the most famous of which is NGC 1365, which shows rapid variability in absorption levels  \citep{2002ApJ...571..234R,2014ApJ...788...76W,2015ApJ...804..107R}. However, other types of changing-look AGN, such as changing-look quasars \citep[which vary between optical classifications; e.g.][]{2020MNRAS.491.4925G}, are thought to indeed be due to physical changes in the accretion disk \citep[e.g.][]{2018ApJ...864...27S,2018MNRAS.480.4468R,2020ApJ...890L..29A} and/or accretion rate \citep[e.g.][]{2015ApJ...800..144L,2016MNRAS.455.1691R,2016MNRAS.457..389M,2017ApJ...835..144G,2018ApJ...858...49W}. NGC 6890's increase in luminosity by an order of magnitude implies a change in the central engine. This might make it more similar to optical-changing look AGN than to NGC 1365, and/or lend credence to the hypothesis that a decrease in the magnitude of an X-ray reflection component could also be caused by AGN brightening in addition to reduced obscuration \citep{2003MNRAS.342..422M}.

\subsection{\bf Implications for AGN Variability and Duty Cycle}
Previous attempts to search for fading AGN have often relied on finding extended emission line regions, such as the Voorwerpen 
\citep[e.g.,][]{2009MNRAS.399..129L, 2012AJ....144...66K, 2012MNRAS.420..878K, 2018MNRAS.474.2444S} and green beans \citep[e.g.,][]{2013ApJ...763...60S, 2015MNRAS.449.1731D}. These have probed fading timescales of $\mathrm{10^4-10^5}$ years. Because our sample, which is a subset of the {\it IRAS} $12~\mu$m Seyfert sample, identified AGN with strong MIR torus emission but apparently lacking coronal X-ray emission, we have probed much shorter fading timescales (decades and centuries).  While our sample does include one Voorwerp in Mrk~463E, it is distinctly different from the green beans sample, probing a lower [\ion{O}{3}] luminosity than that sample (i.e., $L_{\rm [OIII]} > 10^{43}\, {\rm erg}\, {\rm s}^{-1}$ for the green beans).

The \citet{2017ApJ...846..102M} X-ray and [\ion{O}{3}] data that was used as the initial starting point for our sample included 39 Seyfert~1 AGN and 47 Seyfert~2 AGN. We can presume the Seyfert 1 AGN are currently active, and of the 47 Seyfert 2 AGN, we found one candidate recently deactivated AGN. This is a rate of approximately 1\%. That we were able to find one recently deactivated AGN in a sample of 86 AGN suggests that recently deactivated AGN may be more common than previously thought.

A majority of our sources are mergers in progress (Mrk 463, NGC 4922) or located close to other galaxies and thereby potentially interacting (NGC 1386, NGC 3627, IC 3639). These results together could be understood in a context of merger/interaction-induced AGN activity \citep[e.g.,][]{2008ApJS..175..356H}.

The timescale of NGC 3627's fading is shorter than the $\mathrm{\sim 10^4-10^5}$ year viscous timescale for AGN but it is also longer than the $\mathrm{\sim}$10 year thermal timescales used to explain changing-look AGN \citep{2018ApJ...864...27S}. This could imply that AGN can vary on timescales intermediate between these two timescales, but it could also simply be we are observing the beginning of a viscous-timescale-related fading.

\section{Conclusions}
In this paper, we presented \nustar{} observations of 9 AGN underluminous in the 2-10 keV X-rays from the 12 $\mathrm{\mu m}$ galaxy sample. We combined these \nustar{} data with \chandra{}, \swift{}, and \xmm{} data as necessary to perform broad-band X-ray spectral fitting and determine whether these AGN were truly intrinsically underluminous and potentially deactivated, or simply heavily obscured. We find that all but NGC 5005 and NGC 3627 are obscured AGN, whereas NGC 5005 and NGC 3627 are intrinsically low luminosity. Of the two low-luminosity AGN, NGC 3627 appears to not be active. Since this galaxy preserves NLR [\ion{O}{3}] emission and nuclear MIR emission, we conclude that it is a candidate recently deactivated AGN, having turned off no more than 87-220 years ago.
\bigskip{}

The scientific results reported in this article are based on data obtained from the \chandra{} Data Archive. This work is based on on observations obtained with \xmm{}, an ESA science mission with instruments and contributions directly funded by ESA Member States and NASA. We acknowledge the use of public data from the \swift{} data archive. This research has
made use of data and/or software provided by the High Energy Astrophysics Science Archive Research Center (HEASARC), which
is a service of the Astrophysics Science Division at NASA/GSFC and the High Energy Astrophysics Division of the Smithsonian Astrophysical Observatory. This work has made use of data obtained from the \nustar{} mission, a project led by Caltech, funded by NASA and
managed by NASA/JPL. MLS wants to thank Lisbeth D. Jensen for her help in digging through the data used in \citet{2017ApJ...846..102M}, and Donaji Esparza-Arredondo for showing her the fitting results for NGC 3627 from \citet{2020ApJ...905...29E}.
J.A.G. acknowledges support from NASA grant 80NSSC21K1567  and from the Alexander von Humboldt Foundation.

\facilities{CXO, NuSTAR, XMM, Swift}

\software{HEASOFT \citep{2014ascl.soft08004N}, CIAO \citep{2006SPIE.6270E..1VF}, XMM-Newton SAS \citep{2004ASPC..314..759G}}

\begin{figure}
    \plotone{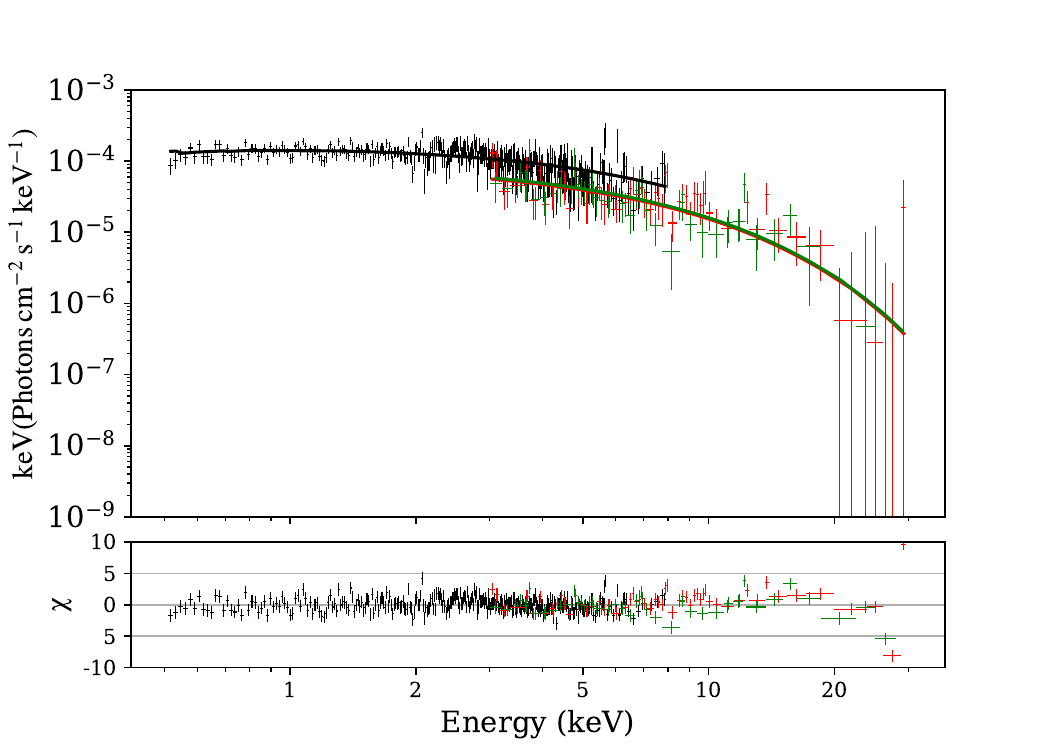}
    \caption{Unfolded spectrum and model for the ultraluminous X-ray source M66 X-1, which was observed along with NGC 3627. Black denotes \chandra{} data, while red and green represent \nustar{} FPMA and FPMB data.}
    \label{fig:M66X-1}
\end{figure}

\appendix
We present the spectrum of the ultraluminous X-ray source M66 X-1, the first reported spectrum for this source that uses \nustar{} data. The spectrum was best fit by a TBABS*CUTOFFPL model. The C-stat/d.o.f.\ was 885.78/912. The spectrum is plotted with the unfolded model in Figure \ref{fig:M66X-1}. The parameters of the best fit model are given in Table \ref{tab:M66X-1params}.

\section{M66 X-1 Spectrum}

\begin{deluxetable}{lcc}
\tablecaption{Parameters for best-fit M66 X-1 model.}\label{tab:M66X-1params}
\tablewidth{2pt}
\tablehead{\multicolumn{3}{c}{CUTOFFPL} \\
\colhead{$\Gamma$} & \colhead{Cutoff} & \colhead{
Norm}\\ \colhead{} &
 \colhead{(keV)} & \colhead{($10^{-5}$ cts $\mathrm{s^{-1}\:keV^{-1}}$)}
}
\startdata
$0.94^{+0.08}_{-0.09}$ & $5.10^{+1.05}_{-0.79}$ & $9.48^{+1.31}_{-1.16}$
\enddata
\tablecomments{ The \chandra{} normalization constant value was $1.92^{+0.22}_{-0.19}$.}
\end{deluxetable}

\bibliography{Seyfert2}{}
\nocite{*}
\bibliographystyle{aasjournal}

\end{document}